\documentclass[aps,rmp,reprint,amsmath,amssymb,graphicx,longbibliography]{revtex4-1}
\usepackage{graphicx}
\usepackage{bm}
\usepackage{braket}
\usepackage[english]{babel}
\usepackage[T1]{fontenc}
\usepackage{xcolor}
\usepackage{comment}

\usepackage{enumitem}

\newcommand{\hl}{\textcolor{black}}
\usepackage[colorlinks=true, allcolors=blue]{hyperref}
\usepackage{makecell}

\DeclareUnicodeCharacter{2009}{XXXX here is the little bugger XXXX}

\begin{document}

\title{Quantum cryptography beyond key distribution: theory and experiment}

\author{Mathieu Bozzio}
\affiliation{University of Vienna, Faculty of Physics, Vienna Center for Quantum Science and Technology (VCQ), \\1090 Vienna (Austria).}

\author{Claude Cr\'epeau}
\affiliation{École de Technologie Supérieure, Département Génie Logiciel et Technologie de l’Information, \\ H3C 1K3 Montréal (Canada).}

\author{Petros Wallden}
\affiliation{University of Edinburgh, School of Informatics, Quantum Software Lab, \\ EH8 9AB Edinburgh (United Kingdom).}

\author{Philip Walther}
\affiliation{University of Vienna, Faculty of Physics,\: Vienna Center for Quantum Science and Technology (VCQ),\\ 1090 Vienna (Austria).}\affiliation{QUBO Technology GmbH, 1090 Vienna (Austria).}

\date{\today{}}

\begin{abstract}
Owing to its fundamental principles, quantum theory holds the promise to enhance the security of modern cryptography, from message encryption to anonymous communication, digital signatures, online banking, leader election, one-time passwords and delegated computation. While quantum key distribution (QKD) has already enabled secure key exchange over hundreds of kilometers, a myriad of other quantum-cryptographic primitives are being developed to secure future applications against quantum adversaries. This review surveys the theoretical and experimental developments in quantum cryptography beyond QKD over the decades, along with advances in secure quantum computation. It provides an intuitive classification of the main quantum primitives and their security levels, summarizes their possibilities and limits, and discusses their implementation with current photonic technology.
\end{abstract}


\maketitle

\tableofcontents{}

\section{Introduction}
\label{intro}

\subsection{General motivation}

 Cryptography is the art of communicating or running a computation while providing certain security guarantees against unauthorized parties. By \textit{security guarantees}, we refer to some form of restriction against such parties. Confidentiality, for instance, hides the contents of a message from unwanted parties, while integrity ensures that a message cannot be altered without detection by other parties and authenticity ascertains that a message originates from a specific party. The question of how to achieve these security guarantees, along with others such as anonymity, unforgeability and non-repudiation, will form the core of this review article.

Today, the security of communication networks relies on a handful of fundamental building blocks, known as \textit{cryptographic primitives}, which may be combined with one another to build secure applications. As an example, consider the primitive of \textit{bit commitment}, which takes a bit $a$ from Alice, encrypts it through some publicly-known algorithm, and sends it to Bob. The desired security guarantee is two-fold: Alice must not be able to change $a$ once she has committed to it and Bob must not learn the value of $a$ until Alice reveals it. Provided that both parties have access to secure bit commitment as a resource, they may then engage in secure \textit{coin flipping}, which generates a random bit such that none of them can bias its value. Intuitively, this is achieved by running a bit commitment algorithm on Alice's input $a$: Bob publicly declares a random bit $b$ before the reveal phase, and the outcome of the coin flip is set to $a\oplus b$ once Alice reveals $a$. Treating this new primitive as a resource in turn allows the construction of randomized leader election schemes, which are important primitives in distributed computation and applications such as Blockchain.

In symmetric cryptography, the communicating parties share a uniformly random secret, and there typically exist security guarantees that are achievable regardless of the computational power of an adversary. We refer to this type of security as \textit{information-theoretic security} (i.t.-security). For instance, a message may be encrypted using a key of the same length via the one-time pad method, and later decrypted by the other party holding the key. As long as the key is only used once, the encryption is i.t.-secure \cite{M:CM82,5061224}. 

Many applications do not allow for the prior distribution of symmetric keys however, as this may be impractical or cumbersome when the parties are separated by large distances. This calls for asymmetric cryptography, typically involving a pair of private and public keys. In such scenarios, the most widely used protocols rely on \textit{computational assumptions}: a scheme is considered secure when the time it takes for an adversary to invert the cryptographic function is much larger than the relevant timescale of the protocol. Two widely used asymmetric primitives are Diffie-Hellman key exchange, which relies on the hardness of discrete logarithm \cite{Diffie}, and RSA encryption, which relies on the hardness of factoring large numbers into primes \cite{RSA:ACM78}. 

Crucially, the rise of quantum algorithms providing super-polynomial speed-up for solving these tasks has triggered a response to search for stronger security levels \cite{S:SIAM97,GS:PRL21,Gidney2025factor}. Two main approaches are being actively developed: the first approach uses classical resources to design primitives that are secure against quantum attacks. Labeled \textit{post-quantum cryptography}, this field is evolving very rapidly, both in terms of standardization and cryptanalysis, in which security conjectures are broken and new schemes appear. The second approach, which is the main focus of this review, employs quantum resources to protect against quantum attacks, and is typically referred to as \textit{quantum cryptography}. 

\bigskip   

\centerline { $ * $ }

\bigskip 

In quantum cryptography, information is typically encoded onto quantum properties of light such as polarization, photon number, and spatial or temporal modes. Intuitively, classical information is mapped onto quantum states such that an adversary attempting to access the information and ignoring the encoding basis will alter the states in a way that is detectable by the honest parties \cite{Wie:acm83,BB84}. This encoding method provides i.t.-security for a number of primitives that require computational assumptions in the classical setting. 

So far, the most mature and widely implemented quantum primitive is quantum key distribution (QKD), which allows users to establish secret keys over $1000$ km of optical fiber \cite{LZJ:PRL23} and free space links using satellite technology \cite{Pan:Nat20}. The fact that QKD has achieved this readiness level sheds hope for the investigation and development of other quantum primitives, which can in turn secure network applications for which QKD is insufficient or inadequate \cite{BS:dcc16}.

Over the past decades, developments in quantum cryptography beyond QKD and secure quantum computation have opened many research directions, identifying where quantum mechanics can bring a security advantage over classical cryptography, how these primitives reduce to one another in the quantum world, and how to tune them to current photonic technology. Experimentally, some require standard prepare-and-measure QKD infrastructure, while others require challenging technology such as quantum storage devices with high efficiency or stringent no-signalling constraints. Generally, even primitives that do not require this technology exhibit stricter limits on the tolerated noise and channel losses than QKD.

\subsection{Aims, scope and structure}

In this review, we seek to provide an intuitive guide to cryptography in the quantum world for a broad physics audience. We discuss primitives, protocols and experiments that use \textit{quantum resources} to protect classical and quantum information against quantum attacks, along with secure quantum computation. In that sense, we do not cover the field of post-quantum cryptography, which makes use of classical resources to provide security against quantum adversaries. 

We follow the vision of the initial review on \textit{Quantum cryptography beyond quantum key distribution} \cite{BS:dcc16} and the preliminary steps taken by the \textit{Quantum Protocol Zoo} \cite{qzoo} in covering literature outside the widely studied field of QKD and quantum random number generation. Building on the work of \cite{BS:dcc16}, we survey the rapidly expanding theoretical progress of the past decade, together with practical security proofs, experimental demonstrations and hardware challenges for the first time. These are particularly crucial to the development of secure quantum networks and large-scale deployment strategies.

For the Reader interested in QKD, we reference some existing reviews discussing its theoretical aspects \cite{PR:RevMod22} and practical challenges \cite{Pan:RevMod20,GRT:RevMod02,PAB:AOP20,DL+:npjQI16}. For quantum random number generation, we refer the Reader to \cite{MYZ:npjQI16,HG:RevMod17}.

\bigskip   

\centerline { $ * $ }

\bigskip 

Our review is structured as follows: In Section \ref{sec:principles}, we recall the fundamental principles of quantum cryptography and some mathematical tools that will be useful throughout the review. In particular, we provide some intuition about the search for optimal cheating strategies, which is a core component of many quantum-cryptographic security proofs.

In Sections \ref{sec:trustful} and \ref{sec:mistrustful}, we discuss two classes of quantum primitives in the i.t.-secure setting: the first contains \textit{trustful} primitives in which at least one of the two parties follows the protocol honestly. A textbook example is Wiesner's quantum money, which harnesses the no-cloning theorem of quantum mechanics to generate inherently unforgeable banknotes, that can later be verified by a trusted authority. The second class tackles \textit{mistrustful} primitives in which the two parties have competing interests, and may both act dishonestly. This includes primitives such as digital signatures and weak coin flipping for which quantum theory can provide i.t.-security, but also primitives such as bit commitment and oblivious transfer for which strong no-go theorems have been established. In Section \ref{sec:assumptions}, we explain how such no-go results can be circumvented by making other physical assumptions based on special relativity, the quantum storage capabilities of an adversary and physical unclonable functions.

In Section \ref{sec:computational}, we present primitives for which quantum resources cannot provide i.t.-security, but instead protect against computationally-bounded quantum adversaries. Such guarantees may be achieved for public-key quantum money, one-shot signatures, public-key encryption and zero-knowledge proofs. In this context, we also include a brief introduction to cryptanalysis in the quantum setting.

In Section \ref{sec:multipartite}, we delve into primitives involving more than two parties, either sharing a multipartite entangled state or performing parallel repetitions of two-party quantum protocols. These include secret sharing, data hiding and electronic voting, as well as consensus primitives which find their use in cryptocurrencies and distributed computation.

Alongside the development of quantum cryptography protecting classical information comes the challenge of securing quantum data and quantum computation. In Section \ref{sec:quantumdata}, we focus on primitives which protect quantum information, such as the authentication and anonymous transmission of quantum messages, while in Section \ref{sec:computing} we tackle the delegation of quantum computations to powerful -- yet untrusted -- servers. Securing these building blocks has important implications for future quantum networks, in which not all parties can perform quantum computations themselves. 

In Section \ref{sec:outlook}, we conclude the review by providing an outlook and overview of current theoretical and experimental challenges.

\subsection{How to use this review}

This article may be treated as a look-up or reference book for quantum-cryptographic primitives. Each section is written in a self-contained, independent manner, explaining the security guarantees of a given primitive, what it can achieve using classical and quantum resources, and its experimental implementation when available in the literature. When concepts from previous sections are required, the relevant links are stated, referring mostly to technical tools presented in Section \ref{sec:principles} and other primitives.

Despite the stand-alone nature of most sections, we recommend the Reader to study the introduction to a given chapter before tackling its contents. For instance, if one wishes to learn about quantum bit commitment, it is useful to understand the general concept of mistrustful quantum cryptography beforehand. 

\section{Tools for quantum cryptography}\label{sec:principles}

We introduce mathematical tools and fundamental methods in quantum cryptography. These will be useful throughout the review, typically in clarifying how each primitive achieves its security guarantees.

The first three sections are dedicated to conjugate coding, no-cloning and quantum teleportation, which lie at the core of many security analyses, often underlying the claim of ''quantum advantage'' over classical counterparts. In a fourth section, we discuss the derivation of optimal bounds on quantum cheating strategies. Finally, we cover tools that allow for computational security against quantum attacks, such as equal superpositions of inputs and minimal assumptions in the quantum world.

\subsection{Conjugate coding} \label{sec:conjugate}

The method of conjugate coding was introduced by Stephen Wiesner, opening the entire field of quantum cryptography \cite{Wie:acm83}. It harnesses the uncertainty principle to provide security guarantees that only assume the validity of quantum theory. As a reminder, the uncertainty principle applies to pairs of conjugate variables $(x_1,x_2)$ describing two measurable properties of a quantum system. It states that decreasing the uncertainty on the value of $x_1$ increases the uncertainty on the value of $x_2$, and vice-versa. This can be captured by the bound 
\begin{equation}
    \Delta x_1  \Delta x_2 \geqslant C,
\label{eq:uncertainty}
\end{equation}
where $\Delta x_i$ is the uncertainty (standard deviation) of observable $x_i$ and $C$ is a positive constant, whose value depends on the specific uncertainty relation considered. Examples of this law apply to position and momentum, time and energy, and phase and photon number in quantum mechanics.

In quantum cryptography with discrete variables, the uncertainty of interest relates to measurements performed in conjugate bases described by the Pauli $X$, $Y$ and $Z$ operators. These describe observables such as photonic polarization, particle spin, spatial and temporal modes, etc. Labeling the eigenstates of the Pauli $Z$ operator as $\ket{0}$ and  $\ket{1}$, we may express the eigenstates of the $X$ operator as
\begin{equation}
\begin{aligned}
\ket{+} = \frac{1}{\sqrt{2}}\left(\ket{0}+\ket{1}\right),\\
\ket{-} = \frac{1}{\sqrt{2}}\left(\ket{0}-\ket{1}\right),
\end{aligned}
\end{equation}
and those of the $Y$ operator as
\begin{equation}
\begin{aligned}
\ket{+i} = \frac{1}{\sqrt{2}}\left(\ket{0}+i\ket{1}\right),\\
\ket{-i} = \frac{1}{\sqrt{2}}\left(\ket{0}-i\ket{1}\right).
\end{aligned}
\end{equation}
The recipe for conjugate coding using $X$ and $Z$ eigenstates then proceeds as follows: 

\begin{itemize}
\item Encode bit $0$ as a quantum state uniformly drawn from the set $\{\ket{0},\ket{+}\}$.
\item Encode bit $1$ as a quantum state uniformly drawn from the set $\{\ket{1},\ket{-}\}$.
\end{itemize}

Since vectors from a given set bear a non-zero overlap with one vector from the other set, it is impossible for an external party to retrieve the value of the encoded bit with unit probability. This may be inferred from Eq. (\ref{eq:uncertainty}): performing a projective measurement in one basis will destroy the information encoded in the conjugate basis. This provides a type of security that solely relies on the validity of quantum mechanics, rather than on the computational power of an adversary.  
 \begin{figure*}
    \begin{center}
        \includegraphics[width=0.79\linewidth]{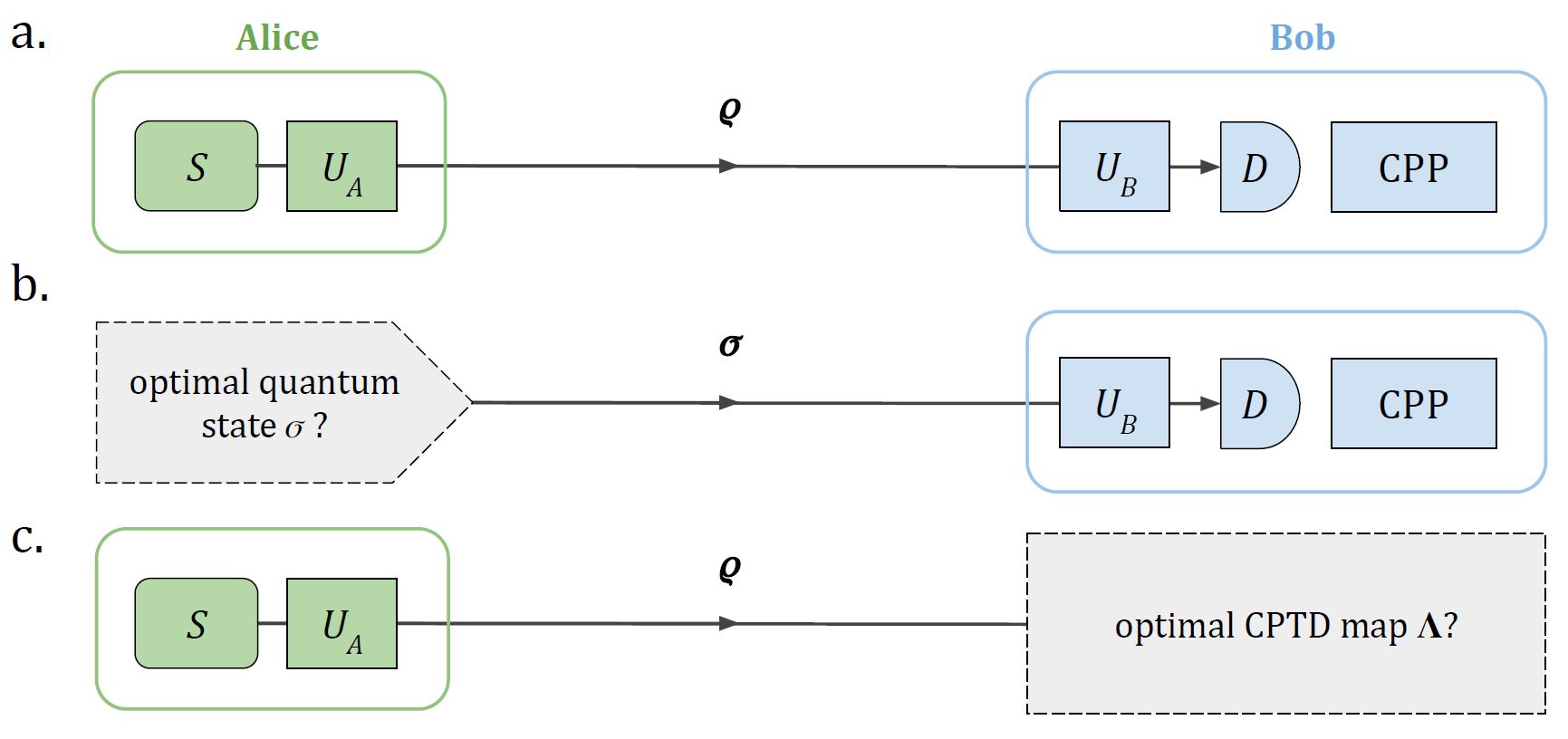}
        \caption{\textbf{Adversarial settings for two-party quantum primitives.} a) Honest protocol: Alice generates a quantum state $\rho$ using her source $S$ and an encoding unitary $U_A$. Bob applies a unitary $U_B$ and performs a measurement with a detection device $D$, possibly followed by some classical post-processing $CPP$ \:  b) Protocol with a dishonest Alice: she can replace $\rho$ with another state $\sigma$, possibly living in a larger Hilbert space \: c) Protocol with a dishonest Bob: he can perform any CPTD map on the state $\rho$ he receives.}
        \label{fig:optimal}
    \end{center}
\end{figure*}

\subsection{No-cloning}\label{sec:cloning}
Initially stated in \cite{WZ:Nat.82}, the no-cloning theorem expresses a fundamental difference between classical and quantum information, underlying the security of many quantum primitives. Informally, the theorem states that an unknown quantum system cannot be copied without introducing noise to the system (i.e. producing two or more perfect copies of that system is impossible). This is in contrast to the classical world, in which a bit of information may be read or measured without destroying its properties, and hence reproduced an arbitrary amount of times.

Formally, suppose we wish to clone an unknown initial state living in $\mathcal{H}_i$ of the form $\ket{\psi}=\alpha\ket{0}+\beta\ket{1}$ where $\left(\alpha,\beta\right)\in\mathbb{C}^2$ and $|\alpha|^2+|\beta|^2 = 1$ onto an initialized copy register living in $\mathcal{H}_c$. There exists no unitary $U$ acting on $\mathcal{H}_i\otimes\mathcal{H}_c$ such that
\begin{equation}
    U\ket{\psi}_i\otimes\ket{0}_c \propto \ket{\psi}_i\otimes\ket{\psi}_c
    \label{clone}
\end{equation}
for all $\left(\alpha,\beta\right)\in\mathbb{C}^2$.

Naturally, this has tight links to the principles introduced in Section \ref{sec:conjugate}: if producing perfect copies of an unknown quantum system were possible, then simultaneously measuring the system's conjugate properties $x_1$ and $x_2$ would also be possible. Conversely, if simultaneously measuring all of the system's conjugate properties were possible, then perfect quantum cloning would become possible.

\subsection{Quantum teleportation}\label{sec:teleportation}
The concept of quantum teleportation allows for the transfer of an unknown quantum state $\ket \psi$ using classical communication and pre-shared entanglement \cite{PhysRevLett.70.1895}. This concept is present in several cryptographic situations throughout the following sections. For reference, we recall the steps involved: 
\begin{enumerate}[label=(\roman*)]
\item Sender and receiver pre-share two halves of a singlet state for every qubit they wish to transfer.
\item The sender makes a Bell-basis measurement of the unknown state $\ket\psi$ together with their half of a singlet.
\item The sender publicly announces the two classical bits $(a,b)$ identifying the outcome of their measurement (out of four possibilities) to the receiver.
\item The receiver transforms the other half singlet into $\ket\psi$ using one of four unitary operators $I, X, Z, XZ$ as prescribed by $(a,b)$.
\end{enumerate}

Note that among other features, quantum teleportation allows for the reversal of a quantum channel from sender to receiver using a classical channel. If Alice wants to send a qubit to Bob, he can send her a half singlet through a reverse quantum channel, after which Alice can teleport an unknown state $\ket\psi$ in the forward direction using the classical channel.

\subsection{Optimal cheating strategies}\label{sec:optimal}

Proving the security of quantum primitives in the i.t.-secure setting often translates to the search for the attacker's optimal cheating strategy. If no explicit strategy can be found, one should at least derive an upper bound on its success probability, which quantifies either how likely it is that the protocol fails to prevent the attack in the first place or how likely it is that a successful attack remains undetected by the honest part(ies). To achieve this goal, the high-level guidelines in this section are based on the following assumptions:

\begin{enumerate}[label=(\roman*)]
        \item All channels, devices and parties are constrained by the laws of quantum mechanics.
        \item No party can access or control another party's classical storage and quantum devices (such as source, encoding unit, and detection apparatus).
\end{enumerate}

In the honest setting (FIG. \ref{fig:optimal}.a.), quantum protocols involve at least one party, Alice, generating a quantum state $\rho$ using a source $S$ and an encoding unitary $U_A$, and a second party, Bob, applying a unitary $U_B$ before performing a measurement with his detection device $D$, possibly followed by some classical post-processing $CPP$. 

When Alice is malicious (FIG. \ref{fig:optimal}.b.), her optimal cheating strategy will involve replacing the expected quantum state $\rho$ by some other quantum state $\sigma$, potentially living in a larger Hilbert space, such that her chances of succeeding according to a protocol-specific figure-of-merit are increased. When Bob is malicious (FIG. \ref{fig:optimal}.c.), his optimal strategy may include any physical operation allowed by the laws of quantum theory, replacing his honest measurement setup by a completely positive trace-decreasing (CPTD) map $\Lambda$ acting on $\rho$. In many cases, $\Lambda$ consists of a carefully-chosen POVM that gives Bob some additional information on which quantum state was sent by Alice. For instance, if $\rho$ is honestly encoded according to Section \ref{sec:conjugate}, Bob will attempt to discriminate which of the states from the set $\{\ket{0},\ket{+},\ket{1},\ket{-}\}$ was sent by Alice with the highest success probability.

\bigskip   

\centerline { $ * $ }

\bigskip 

Deriving optimal bounds on the success probability of a dishonest party can be challenging, especially when quantum state generation and measurements are vulnerable to practical imperfections such as bit flips, phase flips, lossy channels and multi-photon noise. As discussed throughout the review, these may all be exploited by dishonest parties to enhance their cheating probability. 

In some cases, analytical upper bounds may be derived, while in other cases, numerical tools such as semidefinite programming (SDP) may be used. The latter is a branch of convex optimization which allows to optimize over semidefinite positive variables, given linear constraints \cite{VB:SIAM96}. In quantum cryptography, these variables are density matrices, measurement operators, or more general CPTD maps, while the constraints are imposed by the honest part of the protocol \cite{MVW:TQC13,W:LN11}. Once a problem is correctly set up, the dual structure of SDP may be used: a dual minimization problem can be constructed from each primal maximization problem, such that the optimal value $s_d$ of the dual upper bounds the optimal value $s_p$ of the primal, i.e. $s_d\geqslant s_p$. When strong duality holds, we have $s_d= s_p$, and tight bounds on the adversarial cheating probability can be derived \cite{MVW:TQC13,W:LN11}.

\subsection{Equal superposition of inputs}\label{sec:preimages}

Many quantum schemes providing computational security against quantum attacks require the preparation of an equal superposition of all pre-images of a classical function $f$.

To achieve this, we first intialize two registers on an input/output space $\mathcal{H}_i\otimes\mathcal{H}_o$, and apply a Hadamard on the first register to produce an equal superposition\footnote{Note that here and in other places, the normalization constant in front of the equal superposition has been omitted.}:
\begin{equation}
\ket{0}_i\otimes\ket{0}_o\rightarrow \sum_x\ket{x}_i\otimes\ket{0}_o.
\label{eq:equal}
\end{equation}
We then construct a unitary $U_f$ that coherently applies the function $f$ as: 
\begin{equation}
U_f\ket{x}_i\otimes\ket{z}_o=\ket{x}_i\otimes\ket{z\oplus f(x)}_o.
\end{equation}
and apply it to the superposition state from Eq. (\ref{eq:equal}). By measuring the second register, we get a classical value (image) $y$. In doing so, we have collapsed the first register into an equal superposition of the pre-images that match $y$:
\begin{equation}
 \sum_x\ket{x}_i\otimes\ket{0}_o\rightarrow\sum_{x\vert f(x)=y}\ket{x}_i\otimes\ket{y}_o.
\end{equation}
It is straightforward to notice that anyone holding the first register can directly recover an $x$ that matches $y$, in the sense that $f(x)=y$, by simply measuring their register in the computational basis.
Depending on the properties of the function $f$, this superposition provides various security guarantees. For instance, if $f$ is a one-way function, it is hard for any party that does not have the first register to find an $x$ that is a pre-image of $y$. If $f$ is collision-resistant, even holding the first register makes it hard to obtain more than one $x$ that matches $y$. 

This method recovers some notion of (computational) unclonability, where the properties of the function $f$ determine what can and what cannot be recovered or copied. For this reason it plays crucial role in public-key quantum money and one-shot signatures (Section \ref{sec:computational}), as well as in many types of classical-client secure quantum computation (Section \ref{sec:computing}). 
 
\subsection{Minimal assumptions}\label{sec:minimal}

As summarized in Table \ref{tab:summary}, some primitives exploiting quantum resources can achieve i.t.-security under various physical assumptions, while others may only achieve computational security against quantum attacks. In the latter setting, it is important to identify the minimal or weakest assumptions required to achieve a given primitive, some of which are introduced in this section.

 In the classical world, many primitives can be achieved securely under the assumption that one-way functions (OWFs), namely functions that can be evaluated easily but are computationally hard to invert, exist \cite{IL:FOCS89}. In the quantum world however, it turns out that many
such primitives can be securely implemented under even weaker assumptions. 

{\em Pseudorandom quantum states} are efficiently preparable quantum states that are
computationally indistinguishable from Haar-random states, and for which quantum-safe OWFs are sufficient but not strictly necessary (in relativized settings) \cite{kretschmer:LIPIcs.TQC.2021.2,10.1145/3564246.3585225,10.1145/3618260.3649650,AQY:Crypto22}. As an example, let $F: K\times M \rightarrow \{0,1\}$ be a classical pseudorandom function acting on a key space $K$ and message space $M$. We may construct a pseudorandom quantum state associated with key $k\in K$, consisting of a uniform superposition over all strings $m\in M$, as \cite{BS:TOC19,JLS:CRYPTO18}:
\begin{equation}
\ket{\psi_k} = \frac{1}{\sqrt{2^{|M|}}} \sum_{m\in M}(-1)^{F(k,m)}\ket{m}.
\end{equation}
{\em One-way state generators} are quantum generalizations of OWFs in which the output $\phi(x)$
is a quantum state and the function is difficult to invert even when multiple copies of the output state, that is 
$\phi(x)^{\otimes n}$, are provided \cite{MY:CRYPTO22}.

{\em EFI pairs} are composed of two quantum states that are
Eﬃciently generated, statistically Far, and computationally Indistinguishable \cite{brakerski_et_al:LIPIcs.ITCS.2023.24}, and are an analogous counterpart to the classical EFI pairs of distributions \cite{GOLDREICH1990277}.
It could be the weakest assumption that is sufficient for several cryptographic primitives.

\begin{table*}
\begin{ruledtabular}
\begin{tabular}{ccc}
  \thead[l]{\textbf{Security}} &  \thead[l]{\textbf{Assumption}}  &  \thead[l]{\textbf{Primitive}} \\
\hline
\thead[l]{information\\-theoretic}  & \thead[l]{quantum mechanics (QM)} & \thead[l]{unforgeable tokens, unclonable encryption, data locking, \\digital signatures, weak coin flipping, cheat-sensitive imperfect \\ mistrustful primitives (Section \ref{sec:mistrustful}), multipartite primitives \\(Section \ref{sec:multipartite}), private quantum channels, authentication of \\ quantum messages, anonymous communication,\\blind computation with quantum clients} \\
 & \thead[l]{QM + noisy channel} & \thead[l]{covert communication}  \\
 & \thead[l]{QM + special relativity} & \thead[l]{bit commitment, oblivious transfer, strong coin flipping,\\ multiparty computation }\\ 
 & \thead[l]{QM + bounded/noisy quantum storage} &  \thead[l]{data-locked QKD, bit commitment, oblivious transfer,\\ strong coin flipping, multiparty computation\\  } \\
   & \thead[l]{QM + special relativity + bounded entanglement}  & \thead[l]{position verification}  \\
\hline
\thead[l]{computational} & \thead[l]{pseudorandom quantum states, \\ one-way state
generators, \\ EFI pairs } & \thead[l]{bit commitment, oblivious transfer, strong coin flipping, \\zero-knowledge proofs}\\
& \thead[l]{quantum-safe one-way functions } & \thead[l]{encryption with quantum public keys}\\
 & \thead[l]{trapdoor $2$-regular collision-resistant, \\ trapdoor claw-free functions } & \thead[l]{blind computation with classical clients, \\ fully homomorphic encryption}\\
 & \thead[l]{ indistinguishability obfuscation}  & \thead[l]{public-key money }\\
 & \thead[l]{ indistinguishability obfuscation \\+ learning with errors}  & \thead[l]{one-shot signatures}\\
 \hline
\thead[l]{hardware-based} & \thead[l]{one-time memories } & \thead[l]{one-time programs}\\
& \thead[l]{quantum physical unclonable functions} & \thead[l]{device authentication, identification,\\ software licensing} 
\end{tabular}
\end{ruledtabular}
\caption{\label{tab:summary} Classification of quantum primitives according to their achievable security guarantees --either \textit{information-theoretic}, \textit{computational} or \textit{hardware-based}-- and the assumptions required for these to hold. For detailed discussions, please refer to each primitive's dedicated section.}
\end{table*}

\section{Trustful quantum cryptography}\label{sec:trustful}

This section surveys quantum primitives in which at least one of the two parties behaves honestly, i.e. follows the protocol's intended steps. 

The first type of trustful primitives involves both parties, usually referred to as Alice and Bob, acting honestly and collaborating against an external adversary. QKD is the most well-known example \cite{BB84,Pan:RevMod20}, and this section will cover other examples making use of quantum states such as unclonable encryption, covert communication and fingerprinting. The second type involves one party interacting with an untrusted party in an attempt to test or verify the latter's behavior. In unforgeable quantum tokens for instance, a trusted authority verifies that a quantum token has not been duplicated or counterfeited. In quantum position verification, an honest verifier attempts to authenticate a potentially dishonest prover's location by exchanging a sequence of classical and quantum messages and monitoring the timing of the interactions. 

So-called \textit{mistrustful} quantum primitives, in which both parties can deviate from the honest protocol, will be discussed in Section \ref{sec:mistrustful}.

\subsection{Unforgeable tokens}\label{sec:tokens}

The most intuitive cryptographic application of quantum no-cloning was arguably formulated in 1973, and published ten years later \cite{Wie:acm83}. In this pioneering proposal, conjugate coding was introduced as a means to generate quantum objects that are inherently impossible to counterfeit, such as quantum banknotes. From one valid banknote associated with a given serial number, we wish that the probability of producing $k \geqslant 2$ valid copies with the same serial number can be made arbitrarily small in some security parameter. Note that, in the classical world, there exists no such notion of security for banknotes. Instead, the difficulty of producing counterfeits relies on their intricate coloring and watermark design \cite{Chia:09}.

\bigskip   

\centerline { $ * $ }

\bigskip 

In essence, quantum schemes involve a central bank encoding a random secret key onto a sequence of qubits, following the methods of Section \ref{sec:conjugate}. The piece of money, ideally consisting of all qubits stored in a perfect quantum memory, is handed to the user, along with its unique public serial number. Once the user wishes to spend, they forward the quantum money with its serial number to a legitimate verifier in possession of the secret key. The verifier measures each qubit in the basis matching the preparation basis dictated by the key, and checks that the outcomes are consistent with the initial encoding. If no inconsistencies occur, the money is accepted by the verifier.

In a dishonest setting, a malicious user might attempt to counterfeit the quantum money, in order to spend it at least twice. Since the qubits are prepared in a sequence of unknown conjugate bases, no-cloning will prohibit the creation of perfect copies (Section \ref{sec:cloning}). In fact, it was rigorously proven using convex optimization techniques that forging two copies of quantum money will introduce an error rate of $25\%$ upon verification \cite{MVW:TQC13}. Following the techniques introduced in Section \ref{sec:optimal}, the majority of modern security proofs recast the search for optimal cloning strategies as a semidefinite program \cite{BDG:PRA19,AA:pra17,SKS:NatComm23}. This in turn applies to the more general primitive of unforgeable quantum tokens, in which a trusted central authority distributes tokens comprised of quantum states, such that they cannot be successfully verified by multiple spacelike-separated verifiers. 

We note that the presence of a trusted authority for verification is not required in public-key schemes, where the quantum money or token is publicly verifiable. In that sense, private-key schemes are somewhat analogous to credit card schemes (which require verification from a central authority), while public-key schemes are closer to physical coins and banknotes (which can in principle be verified by \textit{any} party). Since the latters may only achieve computational security against quantum adversaries, we postpone their discussion to Section \ref{sec:publicmoney}.

\bigskip   

\centerline { $ *\: * $ }

\bigskip 

 In photonic networks, direct transmission of quantum tokens is impractical, as noisy channels will randomly alter the states, leading to authentic tokens being rejected. Furthermore, adaptive attacks may take place in the case where authentic tokens are passed on and erroneous ones are destroyed  \cite{NSB:QIC16}. As a means to circumvent these issues, classical verification schemes were introduced, enabling remote verification of quantum tokens through the use of challenge questions \cite{Gav:12}. These were later
extended to practical, noise-tolerant schemes  \cite{PY+:pnas12,GK:tqc15,AA:pra17,Kumar2019}, some of which allow for various forms of device independence \cite{BDG:PRA19,HS:NJP20}.

Interestingly, it was only after the first proof-of-principle demonstrations of memoryless quantum tokens were published \cite{BC+:npjQI17,BOV:npj18,GAA:pra18,JBC:SciRep19} that the drastic impact of lossy transmission and detection on security was noticed. To understand this bottleneck better, consider the following attack: when challenged for classical verification under a loss tolerance of $50\%$, a malicious user splits the initial token into two halves. To succeed at the first verifier, they measure the first half honestly and declare the second as lost. To succeed at the second verifier, they measure the second half honestly and declare the first as lost. This perfect strategy causes both verifiers to accept with probability $1$, implying that such quantum
token schemes are insecure beyond $50\%$ losses. In the presence of noise, this threshold decreases further, due to attacks exploiting both noise and loss tolerance \cite{BDG:PRA19,SKS:NatComm23}.

\bigskip   

\centerline { $ *\:*\:* $ }

\bigskip 

Perhaps the most significant challenge for practical quantum tokens resides in the need for efficient, high fidelity quantum storage as a means to ensure flexible spending. Quantum storage times of a few hours or days are currently beyond state-of-the-art, which ranges from a few microseconds to a few minutes \cite{MMZ:natcomms21,VHC:NC18,HE+:jmo16}. Nevertheless, two experimental demonstrations of secure quantum tokens with genuine quantum storage were realized, the first using a cold-atom-based quantum memory  \cite{MNH:SciAdv25} and the second employing color centers in diamond \cite{SBS:Arx25}. 

Memoryless alternatives were also proposed, achieving spending flexibility through a network of trusted agents and authenticated channels, positioned at precise spatial locations with respect to the verification points \cite{Kent:PRA20,K:Proc19}. To some extent, these schemes allow the user to choose the time and place of verification without storing the quantum states. From a security perspective, monitoring the spatial and temporal coordinates of verifiers requires either a trusted Global Positioning System (GPS) or secure position verification schemes discussed in Section \ref{sec:position}. A first partial implementation using weak coherent states was performed in \cite{Kent:npj22}, followed by an experiment demonstrating transaction time advantage over classical schemes \cite{JKP:Arxv24}. Note that the work from \cite{Kent:npj22} also stated three properties that should be simultaneously achieved to claim a quantum advantage: token unforgeability, instant validation and user privacy.

Focusing on other applications, recent experimental works have targeted use-cases that do not require flexible spending, such as the generation of quantum cryptograms to authenticate and secure online transactions \cite{SKS:NatComm23} and the signing of contracts in quantum e-commerce \cite{CLW:SciAdv24}. Since both these schemes require some form of authentication and non-repudiation, they have tight links to the quantum digital signatures presented in Section \ref{sec:signatures}.

\subsection{Unclonable encryption}

Two decades or so after Wiesner's proposal for private-key quantum money, the no-cloning property of quantum mechanics was applied to other forms of cryptography, some of which are nicely summarized in \cite{S:ACM23}. 

In quantum tamper-evident encryption, a classical plaintext is encrypted using quantum states, resulting in a quantum ciphertext that is inherently unclonable \cite{G:QIC03}. Contrary to the classical world, this implies that an eavesdropper attempting to copy the ciphertext in order to store it for later decryption will be detected by the honest parties. If the eavesdropper chooses to store the original ciphertext in a quantum memory instead of producing copies, the honest recipient will not receive the ciphertext, thus also triggering eavesdropping detection \cite{G:QIC03}.

\bigskip   

\centerline { $ * $ }

\bigskip 

Although this first quantum proposal allowed for tamper detection, the question of how much information can be extracted from the original message by splitting the ciphertext into several (imperfect) quantum copies was left open. This was addressed by introducing the somewhat stronger notion of unclonable quantum encryption, whose security guarantees may be understood as a three-partite game between an honest sender Alice and two dishonest receivers Bob and Charlie \cite{BL:TQC20}. Essentially, Alice chooses a message $m$ of length $l$ and an encryption key $k$ of length polynomial in $\lambda$ (a security parameter), producing a quantum ciphertext $\sigma\left(m,k\right)$ on $\mathcal{H}_A$. Following the intuition from Section \ref{sec:optimal}, Bob and Charlie apply some optimal CPTD map $\Lambda:\mathcal{H}_A\rightarrow\mathcal{H}_B\otimes\mathcal{H}_C$, each keeping register $B$ and $C$, respectively, before being isolated from one another. In a later phase, Alice reveals $k$ to
both of them, allowing them to recover two messages $m_B$ and $m_C$ from their respective quantum systems. Bob and Charlie win the game if and only if their messages agree with the original one (i.e. $m_B = m_C = m$). Following \cite{BL:TQC20}, the scheme is $t$-unclonable if their winning probability is upper bounded by $2^{-l+t}+negl(\lambda)$, where $negl(\lambda)$ indicates a negligible parameter in $\lambda$. 

\begin{figure} 
    \begin{center}
        \includegraphics[width=1\linewidth]{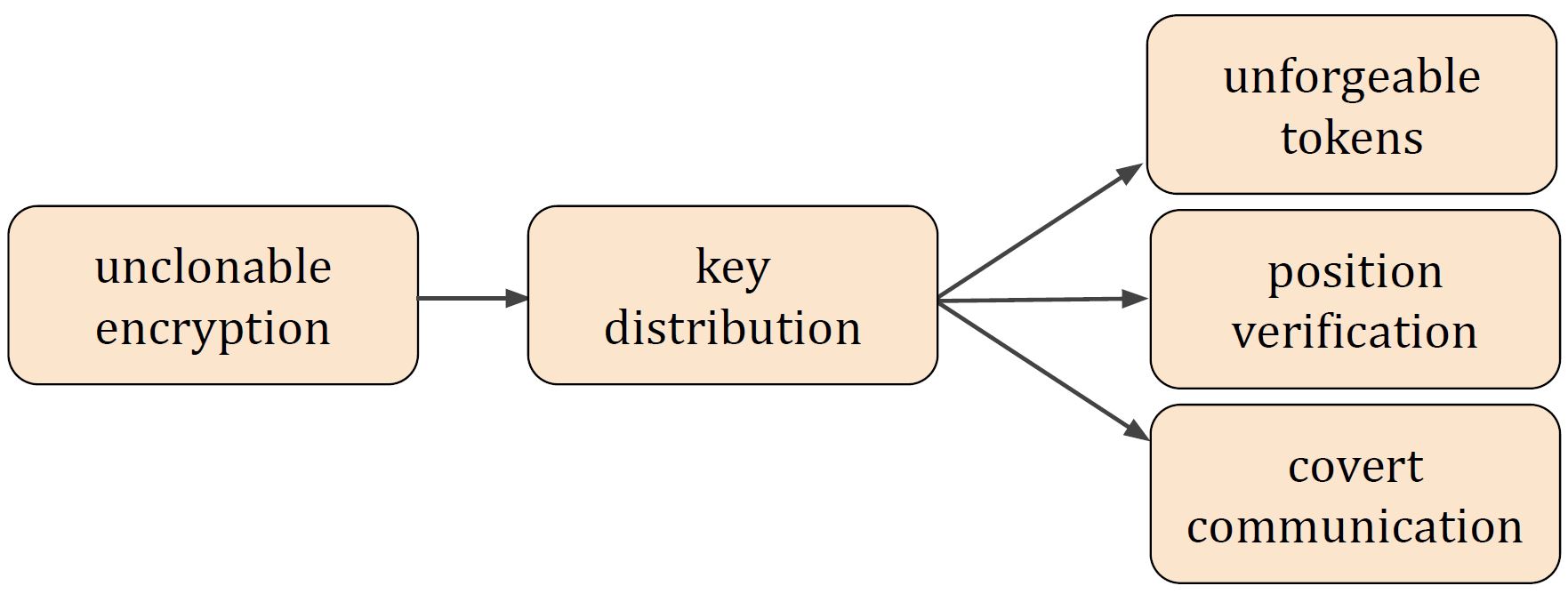}
        \caption{\textbf{Constructions for trustful quantum cryptography.} In the i.t.-secure setting, quantum key distribution may be constructed from unclonable quantum encryption, or from its weaker variant of tamper-evident quantum encryption \cite{G:QIC03}. Once a uniformly distributed secret key has been established between two (or more) parties, the three primitives of unforgeable quantum tokens (Section \ref{sec:tokens}), quantum position verification (Section \ref{sec:position}) and covert communication (Section \ref{sec:covert}) may be securely constructed.    }
        \label{fig:trustful}
    \end{center}
\end{figure}

As shown in FIG. \ref{fig:trustful}, unclonable encryption is in fact the strongest trustful primitive discussed in this chapter, as, given a secure quantum unclonable encryption scheme, Alice and Bob may perform a secure QKD scheme \cite{G:QIC03}, which in turn allows for unforgeable tokens, position verification and covert communication. Later works studied the question of the reusability of the encryption key without the random oracle model, and proposed some private and public-key constructions \cite{AK:TC21}. Note that the guarantees provided by unclonable encryption with reusable keys are closely related to those studied in key recycling \cite{BBB:Springer14,DPS:CRYPTO05,FS:EUROCRYPT17}. 

\bigskip   

\centerline { $ *\:* $ }

\bigskip 

Focussing on symmetric encryption, schemes based on the simple conjugate coding method of Section \ref{sec:conjugate} were proposed \cite{BL:TQC20}, both with and without the assumption of a quantum-secure pseudorandom function. From a practical standpoint, this implies that experimental requirements should be similar to QKD, unless storage of the quantum ciphertext is needed, in which case the technological challenges are similar to those for flexible quantum tokens (Section \ref{sec:tokens}). Since no security analysis incorporating experimental imperfections exists to our knowledge however, no statement can be made regarding the noise and loss tolerance of the scheme, and this should be left for future work.

\subsection{Position verification (tagging) }\label{sec:position}

 \begin{figure*} 
    \begin{center}
        \includegraphics[width=0.65\linewidth]{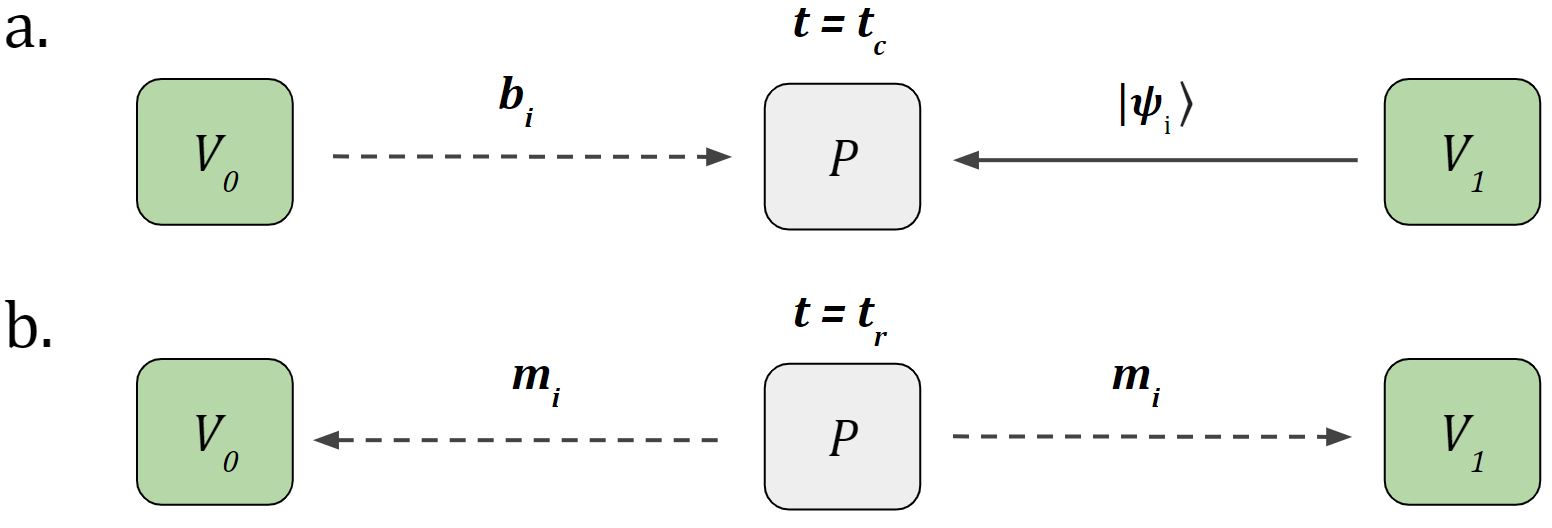}
        \caption{\textbf{Simple quantum position verification in 1D.} We illustrate a variation of Scheme III from \cite{Kent:PRA11}, secure against unentangled dishonest provers. Two verifiers $V_0$ and $V_1$ are placed equidistantly from the honest prover's location $P$. These verifiers can communicate through secure authenticated classical channels and share a trusted GPS reference, and all signals are sent at light speed.
    a) In the challenge phase, $V_0$ sends $N$ uniformly chosen basis bits $b_i\in\{0,1\}$ and $V_1$ sends $N$ uniformly chosen quantum states $\ket{\psi_i}\in \{\ket{0},\ket{1},\ket{+},\ket{-}\}$ such that both signals reach $P$ at time $t=t_c$ --or at matching time-tags for each $i$ if the $N$ states are sent one after the other--. Note that $b_i=0,1$ challenges a measurement of $\ket{\psi_i}$ in the $X$ and $Z$ basis, respectively. b) In the response phase, an honest prover located at $P$ will immediately measure each $\ket{\psi_i}$ in the basis dictated by $b_i$ and broadcast the corresponding measurement outcome $m_i\in\{0,1\}$ such that both verifiers receive their signal at time $t=t_r$. Using their secure channel, $V_0$ and $V_1$ will check that the time-tags of each $m_i$ are equal, and that each $m_i$ is consistent with a measurement of $\ket{\psi_i}$ in basis $b_i$. If these conditions are satisfied, the location of $P$ is successfully authenticated.}
        \label{fig:position}
    \end{center}
\end{figure*}

Many primitives in this review ensure that a given party holds some form of confidential, authenticated, or integrous digital information. However, there exist applications which require other forms of credentials. In military settings for instance, one might wish to verify that a given message originates from a specific geographical position, eg: from a trusted ally military base. In the classical literature, a verifier can convince themself of an upper-bound on the distance of a prover by sending classical radio wave signals and measuring the time they take to travel back, assuming no-signalling constraints \cite{BC:LNCS93}. Note that the exact method may involve pre-shared keys or challenge-response rounds to avoid a dishonest prover sending signals before the verifier sends theirs \cite{SP:IEEE05}. 

Since classical information may be copied and distributed at light speed however, classical protocols are insecure against collusion attacks, in which a dishonest prover splits into several coordinated agents to convince the verifiers that they are at position $P$ even though none of the agents are located at $P$. I.t.-secure protocols are however possible in the Bounded-Storage Model (consult Section \ref{BNS}), which assumes an upper-bound on the amount of information that dishonest parties can store \cite{C:LNCS09}.

\bigskip   

\centerline { $ * $ }

\bigskip 

The first proposal for authenticating a tagging device's classical location through the exchange of quantum states was formulated in a patent entitled \textit{Tagging systems}, whose contents were later published in \cite{Kent:PRA11}. In this work, along with independent results from \cite{M:PRA10,M:IEEE10}, several schemes were proposed for i.t.-secure quantum position verification assuming the validity of quantum theory and no-signalling constraints. In FIG. \ref{fig:position}, we describe a simple scheme that is secure under the assumption that dishonest provers may not share any entanglement \cite{Kent:PRA11}. This assumption is crucial since all previously mentioned schemes are in fact insecure against teleportation-based attacks, in which dishonest provers $P_0$ (resp. $P_1$) located to the right (resp. left) of $V_0$ (resp. $V_1$) pre-share a singlet state for each round $i$. As soon as $P_1$ receives $\ket{\psi_i}$, they perform a teleportation measurement on their part of the singlet and $\ket{\psi_i}$, forwarding all relevant classical data at light speed to $P_0$. As soon as $P_0$ receives $b_i$, they measure their part of the singlet in basis $b_i$, forwarding the measurement basis and outcome to $P_1$, who may then retrieve the correct measurement outcome corresponding to the challenged basis \cite{Kent:PRA11,LL:PRA11}.

Variations of these schemes were later broken by general attacks in which the adversaries share sufficient amounts of entanglement \cite{LL:PRA11,B:SIAM14}, with lower bounds derived in \cite{BK:NJP11,RG:Arx15,TFK:NJP13,CL:PRA15,BCS:NatPhys22}. Secure quantum position verification is thus possible assuming adversaries with bounded entanglement resources, but also in the random oracle model \cite{U:Crypto14,DS:NJP21} or in cases where the tagging device securely stores a pre-shared secret -- i.e. cannot be inspected or modified by the adversary \cite{K:PRA11second}. 

\bigskip   

\centerline { $ *\: * $ }

\bigskip 

Facing unentangled dishonest provers, the scheme from FIG. \ref{fig:position} cannot tolerate losses beyond $50\%$ / $3$ dB \cite{QS:PRA15}. The reason is similar to that for unforgeable quantum tokens (Section \ref{sec:tokens}), in which dishonest parties can replace lossy channels by ideal ones: a dishonest $P_1$ measures each $\ket{\psi_i}$ in a random basis chosen from $\{Z,X\}$ and waits for dishonest $P_0$ to forward the basis bit $b_i$. If their random basis choice agrees with $b_i$, $P_1$ forwards the correct measurement outcome to $V_1$, while if their choice disagrees with $b_i$, they declare the state as lost. Interestingly, this challenge may be overcome by using more than two encoding bases, uniformly distributed over the Bloch sphere: using a perfect single-photon source and in the limit of many encoding bases, an implementation with $1\%$ error rate may then tolerate up to $14$ dB losses \cite{QS:PRA15}. Other practical schemes were proposed to increase the loss tolerance under various assumptions \cite{ABB:Arx23,ABS:Arx22,EL:PRL23,CXS:PRA16,CKP:Arx23}. 

Recently, the potential of quantum-dot single photon sources for quantum position verification was experimentally investigated \cite{KSP:Quantum24}: the interference visibility of two photons emitted by the same quantum dot was measured in the context of the SWAP protocol from \cite{ABS:Arx22b}. In such interference-based protocols, one main challenge lies in improving the quantum interference from photons emitted by different single-photon sources, each owned by one of the provers (see Section \ref{sec:single} for a discussion of single-photon sources).

\subsection{Covert communication}\label{sec:covert}

While message encryption protects against third-party eavesdropping, communicating in hostile environments may call for even stronger security guarantees. For instance, some settings might require that the communication \textit{itself} is hidden from unwanted parties, in which case it is qualified as covert. The quest for such a property dates back to ancient civilizations, where messages were hidden on the scalp of commuting slaves and retrieved upon shaving. Steganography has also been widely investigated, providing methods to embed classical information into seemingly innocent images, videos and physical objects \cite{HL:CRYPTO02,F:2009}, as well as into quantum noise and quantum data \cite{STL:PRA16,G:JMP02,N:Spring06}.

\bigskip   

\centerline { $ * $ }

\bigskip 

Using quantum resources, covertness may be achieved and quantified in the context of measurement theory: security is guaranteed if a quantum adversary cannot reliably distinguish the quantum state they intercept when the parties communicate from the quantum state they intercept when the parties do not communicate. In 2015, i.t-secure bounds were established for covert communication over lossy thermal-noise bosonic channels \cite{BGP:NatComm15}, which quantum-mechanically describe the propagation of a single spatio-temporal mode of the electromagnetic field through a lossy channel with thermal noise. Assuming that Alice and Bob pre-share a long enough secret, and that there exists \textit{some} excess noise that cannot be controlled by the quantum adversary --eg: blackbody radiation--, then Alice can covertly transmit $O\left(\sqrt{n}\right)$ bits to Bob using $n$ photonic modes \cite{BGG:IEEE20}. 

A proof-of-principle experiment was performed using coherent state encoding and homodyne detection on Bob's side, under some assumptions on the adversary's detectors \cite{BGP:NatComm15}. Techniques for covert communication in the microwave regime were also experimentally proposed \cite{DY:PRXQuant21}. Recently, a covert scheme secure against unbounded quantum adversaries was demonstrated over metropolitan distances, exploiting the cross-talk noise from co-propagating classical channels \cite{LAL:IEEE24}.

\bigskip   

\centerline { $ *\:* $ }

\bigskip 

The question of how to covertly transmit quantum information in the form of qubits was also addressed \cite{AS:PRL16}, this time making no assumption on the adversary's ability to control channel noise. Instead, if the excess noise stems from Alice's lab (which the adversary cannot access), then no such assumption is required to derive i.t.-secure bounds. Across $N$ possible time modes, Alice essentially transmits $qN$ qubit signals, whose positions are exactly dictated by the pre-shared secret. For $q<<1$, the $N$-mode state transmitted when communication takes place appears indistinguishable from the thermal state expected when no communication takes place. Schemes exploiting relativistic quantum effects also allow for covert quantum communication \cite{BSW:Arx17,BKS:Arx16}.

Possibilities of performing covert QKD for the re-generation of pre-shared secrets was investigated, first deemed impossible with i.t.-security \cite{AS:PRL16}, but later revisited using more sophisticated coding schemes for information reconciliation and privacy amplification \cite{TB:PRA19}. \hl{Notably, schemes allowing for randomness (or secret key) expansion were designed, based on the observation that a secure covert protocol implies a secure message transmission protocol. In other words, starting from an input secret, a new secret can be communicated to the other party \cite{AA:PRA18}}. For completeness, we note that security bounds on steganography with quantum information were also derived \cite{SB:PRA11,SB:PRA19}.  
  
\hl{Finally, using quantum entanglement, a special class of covert protocols enabling the undetectable sensing of a target phase was demonstrated \cite{HSG:PRL22}, inspired by the covert sensing schemes from \cite{GBD:PRA19,BGD:IEEE17}. A little later, covert sensing designed for the Doppler effect \cite{HZ:PRR23} and target detection \cite{TNG:PRL24} were developed.}
  
\subsection{Fingerprinting}

One important communication task involves two parties comparing their respective bit strings with as little communication and resources as possible, in order to determine whether they are equal. More formally, Alice and Bob each hold an $n$-bit string, labelled $x$ and $y$ respectively, and they wish to check whether $x=y$ or not. The protocol occurs in the setting of simultaneous message passing\footnote{The simultaneous message passing model has many applications in complexity theory, cryptography, distributed systems, IoT, machine learning, and multi-agent systems, where it is essential to reduce direct communication rounds among parties while still achieving computational goals.}, where (a) each party sends their message to a common referee who will determine whether they are identical and (b) the parties do not share any common randomness source (or entanglement). 

\hl{Developing fingerprinting schemes that require as little communication as possible is relevant in many areas of cryptography, including secure identification protocols (eg: password-based identification in which a party wishes to authenticate themselves to a device without revealing their actual password) and secure two-party computation in the context of equality testing \cite{C:Proc18} and the socialist millionaire problem \cite{JY:CRYPTO96}.} Classically, it was shown that fingerprinting requires each player to send $\Omega(\sqrt{n})$-bit messages \cite{NS1996STOC}. 

\bigskip   

\centerline { $ *$ }

\bigskip 

In contrast, by sending quantum bit strings and having the referee perform quantum operations and measurements, the same task can be achieved with $O(\log n)$-qubit messages, thus providing an exponential improvement. Introduced in \cite{BCW:PRL01}, the key idea involves finding $2^n$ quantum states that are pairwise almost orthogonal, and thus distinguishable with high probability, while using $O(\log n)$-qubits. This is achieved by first applying an error-correcting code $E(x)$ that expands the string to $m=cn$ bits, ensuring that each distinct code word differs by at least $\delta m$ bits; then each bit string $x$ is mapped onto the quantum state 
\begin{equation}
\ket{h_x}:=\frac{1}{\sqrt{m}}\sum_{i=1}^m\ket{i}\ket{E_i(x)},
\label{eq:finger}
\end{equation} 
where $E_i(x)$ denotes the $i$th bit of $E(x)$. It is easy to see that $\bra{h_x}h_y\rangle\leq \delta$ where $0<\delta <1$ (as given by the distance of codewords in the code) for $x\neq y$, making the $2^n$ different quantum states almost orthogonal. The referee in fingerprinting protocols then uses an ``equality test'' such as the SWAP-test \cite{BBD+1997SIAM} 
to determine whether the two quantum states received are the same (and thus whether the initial bit strings are equal) or different. Since the two states are almost orthogonal, this test succeeds with high probability, and by repeating a constant number of times, the referee fails only with negligible probability to correctly assess two strings' equality.

The idea of quantum fingerprinting was later extended to other functions $f(x,y)$ of the two inputs with minimal communication complexity \cite{Yao2003STOC}, beyond the equality of two strings that was initially considered. An extension to multiple parties with applications to quantum networks was recently presented \cite{QWYW2021PRR}. Interestingly, it was also shown that one can maintain quantum advantage in another setting, in which the parties are allowed to pre-share randomness and/or entanglement \cite{Sanders2005ISMVL}. In this case, the classical solution is already efficient (i.e. logarithmic to the size of the bit string), but the quantum solution is even more efficient (i.e. constant).  
 
\bigskip   

\centerline { $ *\:* $ }

\bigskip  
  
As a step towards practical schemes, a quantum fingerprinting protocol based on coherent states was proposed in \cite{AL2014PhysRevA}, considerably easier to implement than previous proposals and robust to noise. The way to measure the communication cost, in this case, is slightly different: the protocol requires $O(n)$ optical modes (thus having running time quadratically slower than the classical solutions), but the mean photon number is constant, thus providing an exponential saving in energy. Moreover, the protocol leaks only $O(\log n)$ bits of the initial strings due to the Holevo bound, again an exponential advantage compared to classical solutions and an important figure of merit for certain applications. 

Following the proposal from \cite{AL2014PhysRevA}, proof-of-principle quantum fingerprinting schemes were experimentally demonstrated, using an altered commercial QKD system at telecom wavelengths \cite{xu2015experimentalNatComn} and achieving quantum advantage over $20$km of standard optical fiber \cite{GXY:PRL16}. Based on the hidden matching problem introduced in \cite{BJK:SIAM08}, other implementations of quantum advantage for communication complexity were performed, requiring a simple setup consisting of only one beamsplitter and two single-photon detectors \cite{KKD:NatComm19}.

Overall, two main challenges in quantum fingerprinting are (i) the experimental viability of quantum implementations and (ii) finding realistic use-cases where quantum advantage is beneficial. Classical solutions, in the most commonly used scenario of shared randomness, are already efficient, requiring only logarithmic communication. At the same time, quantum solutions require non-trivial technology, except the coherent state approach of \cite{AL2014PhysRevA}: as can be seen from Eq. (\ref{eq:finger}), the quantum states/fingerprints are generally highly entangled, and thus challenging to prepare for clients and difficult to send to the referee since they are susceptible to noise and losses. Moreover, the referee is required to perform the SWAP-test between these highly entangled states, again a non-trivial operation for standard quantum communication/quantum networks. Finally, although more practical, the coherent state protocols result in worse computation time than classical solutions despite offering other advantages. The main open challenge is therefore to construct experimentally viable and robust quantum fingerprinting protocols that maintain the computation time advantage over classical solutions.

\bigskip   

\centerline { $ *\:* \:*$ }

\bigskip  

Finally, we note the existence of other communication complexity tasks that significantly benefit from quantum advantage \cite{R:ACM99,BCW:ACM98,DM:Schloss20,BCG:Proc16,GSX:PRL23}, and refer the Reader to \cite{BCM:RMP10} for a more detailed overview.

\subsection{Data locking}

In a classical setting, consider Alice encrypting an $m$-bit message into an $m$-bit ciphertext using a $k$-bit key. If Alice sends the ciphertext to Bob without the key, the parties share a mutual information of $m$ bits. Once Alice communicates the key to Bob, their mutual information increases to $(m+k)$ bits. Generally, classical encryption can made i.t-secure if and only if the secret has the same size as the message, i.e. when $k=m$ \cite{M:CM82,5061224}.

\bigskip   

\centerline { $ * $ }

\bigskip 

Remarkably, quantum mechanics allows communication of a $k$-bit key to increase the mutual information between Alice and Bob by an amount strictly larger than $k$ bits \cite{DHL:PRL04}, which is in stark contrast to classical one-time pad encryption. Thus, it is possible to lock a uniformly random $m$-bit message into a quantum state with a key of length $k<<m$ \cite{DFH:Proc13}. Only once the key is revealed can the full $m$-bit data be unlocked. In fact, there exist quantum locking schemes in which the key size is independent of the message length: $n$ bits can be encoded into $n$ qubits using a secret of constant size \cite{FHS:ACM13}. On the other hand, this suggests that any small leakage of the secret key may unlock a large amount of information to an eavesdropper \cite{LWL:PRA14}, which calls for protocols that are fault-tolerant and robust against such leakage \cite{HKL:PRA21}. 

More generally, like many primitives discussed in Section \ref{sec:composability}, the security definition of data locking should be carefully formulated to ensure its composability \cite{KRB:PRL07}. In the context of quantum-locked key distribution for example, the resulting secret key can be securely used for one-time pad encryption if the eavesdropper's quantum storage is limited in time \cite{LL:PRL14}. A more thorough discussion of i.t. security based on bounded and noisy storage assumptions is provided in Section \ref{sec:storage}. 

\bigskip   

\centerline { $ *\: * $ }

\bigskip

In the presence of noise and losses \cite{FHS:ACM13}, the performance of locking protocols was analyzed by formally defining the locking capacity of a quantum channel \cite{GHK:PRX14}. Photonic demonstrations were provided for the scheme of \cite{DHL:PRL04}, in which the eavesdropper’s accessible information is limited to half that of the legitimate receiver, and for the loss-tolerant protocol from \cite{FHS:ACM13}. The first scheme requires the combined channel transmission and detection efficiency to exceed $50\%$, similarly to some trustful primitives in Sections \ref{sec:tokens} and \ref{sec:position}. This was achieved by employing heralded spontaneous parametric down-conversion emitting orthogonally-polarized photon pairs at telecommunication wavelengths, where Alice's heralding employs a superconducting nanowire detector and Bob's detection employs highly-efficient transition edge sensors \cite{LCW:PRA16}. The second, loss-tolerant scheme was implemented using the same setup with active polarization encoding and decoding through Pockels cells, surpassing the performance of classical one-time pad encryption for various values of channel loss \cite{LCW:PRA16}.   

\bigskip   

\centerline { $ *\: *\: * $ }

\bigskip

Generally, data locking has tight links to quantum secure direct communication (QSDC), in which messages are efficiently encoded onto quantum states without prior sharing of a secret key, enabling either the detection of an eavesdropper before information is leaked \cite{LL:PRA02,DL:PRA04,WDL:PRA05} or denying access to the information through the use of entangled pairs \cite{DLL:PRA03,SW:PRA06,ZL:China14}. Along with the first QSDC experiment using frequency coding \cite{HYJ:LSA16}, a quantum data locking implementation targeted to QSDC was provided, where $6$ bits of information were encoded onto a single photon using a key of less than $6$ bits \cite{LHA:PRA16}. While this work also generates photon pairs through parametric down conversion, Alice's phase-scrambling unitaries are implemented through spatial light modulation, and the states are detected on Bob's single-photon detection nanowire array.

A little later, QSDC employing an atomic quantum memory was reported \cite{ZDS:PRL17} and long-distance experiments were provided \cite{ZZS:Bulletin17,ZSQ:LSA22}, along with continuous-variable implementations \cite{ZYG:Research23}. For a more complete review of QSDC and its experimental demonstrations, we refer the Reader to \cite{PLY:IEEE24}.

\section{Mistrustful quantum cryptography}\label{sec:mistrustful}

Contrary to Section \ref{sec:trustful}, we now discuss primitives in which the two parties involved hold competing interests. In that sense, any protocol participant may act dishonestly, deviating from the original instructions. This setting is well-suited to the primitive of digital signatures, along with the two-party building blocks of secure computation: oblivious transfer, bit commitment and coin flipping. We also extend the discussion to one-time programs whose security relies on specific mistrustful primitives.

Following the early success of QKD and unforgeable quantum money (Section \ref{sec:tokens}), researchers started investigating the potential of quantum resources for i.t.-secure mistrustful tasks \cite{BC:CRYPTO90,BCJ:IEEE93,C:JMO94,BBC:Adv91}. As early as 1997, the possibility of i.t.-secure quantum bit commitment was unfortunately ruled out \cite{M:PRL97,LC:PRL97}. Intuitively, suppose that Alice commits to a bit $b$ by generating an entangled state $\rho_{AB}^{(b)}$ on $\mathcal{H}_A \otimes \mathcal{H}_B$ and sending register $B$ to Bob (note that the set of vectors involved in the superposition must be orthonormal on $\mathcal{H}_A$, but not necessarily on $\mathcal{H}_B$). For the protocol to be perfectly hiding, Bob's reduced density matrix must look identical regardless of whether Alice commits to $b=0$ or $b=1$, giving the constraint: 
\begin{equation}
Tr_A\left(\rho_{AB}^{(0)}\right)=Tr_A\left(\rho_{AB}^{(1)}\right).
\end{equation}
From the Schmidt decomposition of $\rho_{AB}^{(b)}$, this condition implies the existence of a unitary that Alice can apply on her register $A$ to transform $\rho_{AB}^{(0)}$ into $\rho_{AB}^{(1)}$, thus violating the binding property of the protocol \cite{M:PRL97,LC:PRL97}. This no-go result was later extended to quantum oblivious transfer \cite{Lo:PRA97} and quantum strong coin flipping \cite{LC:Phys98}.  

Despite such impossibilities, there do exist noteworthy mistrustful primitives capable of achieving i.t.-security with quantum resources, of which two examples are quantum digital signatures and quantum weak coin flipping. Furthermore, cheat-sensitive versions of all quantum primitives affected by the no-go results are possible at the cost of accepting a finite cheating probability. Finally, alternative approaches to i.t.-secure mistrustful quantum cryptography, circumventing the no-go results by relying on additional physical assumptions, will be discussed in Section \ref{sec:assumptions}.   

\subsection{Digital signatures} \label{sec:signatures}

The concept of digital signatures was first introduced by Diffie and Hellman as a mathematical method to verify the authenticity of digital messages and documents \cite{Diffie}. It is the most widely used cryptographic primitive along with encryption. The standardized and most common schemes are the digital signature algorithm (DSA), the elliptic curves DSA and the ElGamal signatures \cite{nist1994dss,johnson2001elliptic,Elgamal1985signature}, all of which rely on the hardness of discrete logarithm and factoring problems. Similarly to encryption schemes, these signature schemes will be broken when large scale fault-tolerant quantum computers become available \cite{S:SIAM97}. While classical solutions offering security against polynomially-bounded or even unbounded quantum adversaries exist \cite{NIST}, an alternative approach is the use of quantum communication to offer some notion of i.t.-secure ``quantum digital signatures'' (QDS).

Generally, digital signatures involve a ``preferred'' party, the \emph{sender}, who signs the message, and one or more parties that act as \emph{receivers}. In any protocol, the following guarantees should be provided:

\begin{enumerate}[label=(\roman*)]
        \item  No party other than the sender can produce a valid signature (unforgeability).
        \item The sender should not be able to deny having signed a message they did sign (non-repudiation).
        \item All receivers allowed by the protocol can verify the correctness of a signature, and will accept a correctly formed signature (correctness).
        \item A receiver should be confident that the message they accepted as valid would also be accepted as valid by other receivers (transferability).
\end{enumerate}
As one can see, the security guarantees involve the possibility of any party being malicious and we thus classify this primitive as \textit{mistrustful}. 

Digital signatures are usually viewed as part of public-key cryptography: one assumes the existence of a Public-Key Infrastructure (PKI) that ensures the identity of each person's public keys. This public key may then be used by anybody to verify the signature formed by the corresponding person/signer. In contrast, a sender wishing to sign a message must use their secret key. Variants exist where PKI is replaced by pre-shared secret keys among the participants, in which case the security of the scheme is guaranteed among those pre-defined parties and special care is required to ensure non-repudiation and transferability of the messages. By exploiting quantum resources, we may achieve this latter type of digital signatures.

\bigskip   

\centerline { $ * $ }

\bigskip 
The first QDS protocol was proposed in 
\cite{GC:Arx01}. Based on the classical scheme of 
\cite{Lamport:1979}, the main idea was to replace the need for a classical one-way function $f(x)$ (Section \ref{sec:minimal}) with a quantum state $\ket{f(x)}$. Lamport's signatures offer only computational security: with unbounded computational resources, it becomes possible to recover the secret key that contains the pre-image $x$ required to sign a message from the public information $f(x)$ by using brute-force inversion. In contrast, for non-orthogonal quantum states, it is impossible to perfectly recover their classical description (Section \ref{sec:conjugate}). To ensure all security properties listed above (i-iv), the protocol requires a \emph{setting-up} phase, where parties ensure that they receive the intended quantum states through point-to-point quantum communication and computation. In the \emph{signing} phase however, each party acts locally, allowing the signing and verification steps to happen without further communication. 

The protocol from \cite{GC:Arx01} presented a number of practical limitations that were gradually lifted: the requirement for parties to perform quantum computation in the setting-up phase \cite{ACJ:PRA2006}, the requirement to store the quantum ``public'' key between the setting-up and signing phases \cite{DWA:PRL14}, the hardware requirements, simplified to those needed for QKD \cite{WDK:PRA15} and the requirement for an authenticated quantum channel  \cite{AWK:PRA16,Yin:PRA2016}. A measurement-device independent version was also developed \cite{PAW:PRA2016}, along with an extension to include multiple parties in \cite{AWA2016}. Inspired by those quantum works, an i.t.-secure classical scheme offering an efficient solution for the signing of larger messages was proposed \cite{AAWA:ACNS2018}, requiring point-to-point secret keys that can be achieved with QKD.

Current state-of-the-art protocols have the following (high level) structure. In the setting up phase, the sender quantumly communicates with each of the participants, ending up sharing an imperfectly correlated bit string for each of the possible message values (similarly to QKD before information reconciliation). The participants then ``symmetrize'' their bit-strings with each other privately (in an ideal honest run, their bit-strings may already be identical). In the signing phase, the sender chooses the message they wish to sign and sends the classical description of the quantum states used to produce the bit-strings they share with all participants. The receivers accept the message if the number of errors lies below a certain threshold, which depends on several factors including the protocol's robustness, related to property $(iii)$, and its transferability $(iv)$. Only the sender has information about the bit-strings of all parties, guaranteeing the unforgeability property from $(i)$, while each participant only knows their respective bit-string symmetrizations, implying that the sender cannot make one party accept and others reject, providing non-repudiation $(ii)$.

\bigskip   

\centerline { $ *\: *$ }

\bigskip 

In parallel to these theoretical developments, we discuss the key experimental progress related to QDS. The first proof-of-principle experiment used phase-encoded coherent states \cite{CCD:NatComm12}, allowing each participant to split their state in two and compare each half with another participants' half through interference on a shared beamsplitter, along with an unambiguous state discrimination (USD) measurement. The setting-up and signing phases were performed one after the other to avoid the use of a quantum memory. A protocol that did not require any quantum storage was later implemented \cite{CDD:PRL2014}, removing the need to sign immediately after the setting-up phase. An unambiguous state elimination measurement (generalization of the USD measurement used earlier) was implemented. The first demonstration of QDS in the continuous variable setting was performed in \cite{CPH:PRL16}, using free-space channels and heterodyne detection for the measurements. Experiments exceeding the kilometer range were then achieved \cite{DCK:PRA2016}, followed by measurement-device independent schemes \cite{YWT:PRA2017,RLY:NatCommn2017}. More recently, an i.t.-secure implementation over an eight-user quantum network was provided \cite{PPV:NJP2022}, \hl{along with efficient QDS based on one-time universal hashing \cite{HYC:nwac23,LXC:PRAppl23} and their application to e-commerce scenarios \cite{CLW:SciAdv24}. }

Overall, QDS are in the same status as QKD from a technological and experimental standpoint. While many implementation challenges of QDS have been lifted, either through theoretical developments (and simpler protocols) or experimental solutions, there remain a few outstanding challenges. Perhaps the most significant is the requirement that every party should communicate with every other party during the setting-up phase, combined with the fact that signatures are ``one-time'' and fresh bit-strings should be shared to sign new messages. Practical imperfections affecting the achievable distance and key rate in QKD therefore play a similar role in QDS, but are magnified by the requirement that multiple parties need to communicate. Furthermore, the theoretical limitations (fixed number of parties, setting-up phase, etc.) imply that QDS can only replace classical signature schemes in restricted settings: for each application, one must ensure that QDS outperform the classical protocols offering similar guarantees (e.g. \cite{AAWA:ACNS2018}), especially when considering larger messages. 

Signatures that offer extra or different guarantees, such as blind signatures, have also been considered. Of special interest are signatures that have a one-time property with various applications. Since these rely on quantum-computational security assumptions, we postpone their discussion to Section \ref{subsec:one-shot signatures}.

\subsection{Bit commitment }\label{sec:qbitcomm}

A bit commitment (BC) scheme is a mechanism to disclose a certain binary value displaced in time. Instead of immediately revealing the value, it is stored temporarily in a {\em commitment} where it is unreadable yet uniquely determined. More formally, one of two mutually distrustful parties transfers a bit to the other party, postponing the moment when the receiver will discover its actual value. In analogy, it is like storing the bit in a closed safe box, in front of everybody, so that the content cannot be read until the combination to the safe is disclosed. Once the information is stored inside the safe, it cannot be modified by the committing party. 

BC is broken down to two phases: the {\em commit} phase and the {\em unveil} phase. In the former phase, a message $com(b)$ is sent from the sender to the receiver. You should think of this message as a strong encryption of $b$ which does not yet disclose the value of $b$ itself. However $com(b)$ should uniquely determine $b$. In the latter phase, a second message $b,unveil(b)$ is sent to the receiver to disclose the value $b$. The receiver owns an efficient public  algorithm $\operatorname{Vrfy}(c,b,u)$ that outputs True if and only if $u$ constitutes a valid unveiling of bit $b$ committed as $c$. BC is designed to satisfy two properties:

\begin{enumerate}[label=(\roman*)]
        \item The message $com(b)$ uniquely determines the value of $b$ (binding).
        \item The message $com(b)$ perfectly hides the value of $b$ (hiding).
\end{enumerate}


It is a folklore argument in the classical realm that BC is impossible in an i.t.-secure sense: if $com(b)$ uniquely determines $b$ then it must be possible to identify the value of $b$ and thus defeat hiding. Similarly, if $com(b)$ hides $b$, then it must be possible to unveil either value and thus defeat binding.

In a computational setting, where at least one party is time-bounded in its computations, BC exists since the 1980's. The terminology of commitments, influenced by the legal vocabulary, first appeared in Shimon Even's contract signing protocols \cite{crypto-1981-1276}, although it seems fair to attribute the concept to Manuel Blum who implicitly used it for coin flipping around the same time \cite{crypto-1981-917}. In his work, Even refers to Blum's contribution saying: \textit{“In the summer of 1980, in a conversation, M. Blum suggested the use of randomization for such protocols”} \cite{crypto-1981-1276}. So apparently Blum introduced the idea of using random hard problems to commit to something, eg: coins, contracts, etc. However, one can also argue that earlier work on ``mental poker" implicitly used commitments as well \cite{shamir1981mental}, since in order to generate a fair deal of cards, Alice encrypts the card names under her own encryption key, which is the basic idea for implementing commitments. 

Under such computational assumptions, BC comes in two dual flavors: binding but computationally hiding and hiding but computationally binding. Commitments of the first type may be achieved from any one-way function \cite{10.1007/BF00196774,doi:10.1137/S0097539793244708}. The construction of computationally binding and statistically hiding BC from any one-way function was solved more recently in \cite{doi:10.1137/080725404}.

\bigskip   

\centerline { $ * $ }

\bigskip 

The first ever proposal for quantum BC (QBC) appeared in \cite{BB84}. Actually, the protocol they described was designed for coin tossing, but it is obvious how to interpret it as a BC. The authors however also presented an attack on their own scheme in the same paper. In the quantum realm, it was first believed that i.t.-secure QBC was actually possible as exemplified by \cite{BCJ:IEEE93}. However, it was later realized that their security claim was incorrect because in the quantum setting obtaining a chosen value out of two possibilities is easier than obtaining both values (these notions are equivalent classically). As discussed in our introduction to Section \ref{sec:mistrustful}, it was finally proven that QBC is impossible in the i.t.-secure setting \cite{M:PRL97,LC:PRL97,LC:Phys98}. 

Under the assumption that computationally hard-to-invert post-quantum one-way functions exist, the same duality is possible in the quantum world: binding but computationally hiding BCs and hiding but computationally binding BCs. The definition of binding is considerably more subtle in the quantum case. An excellent survey of the situation may be found in \cite{cryptoeprint:2015/361}, where several definitions are exhibited, compared and analyzed. It is a general result however that QBC of both types can be obtained from each other as shown in \cite{10.1007/3-540-44987-6_5,hhan2023hardnessdetectingsuperpositionscryptography,10.1145/3564246.3585198}.

There is nevertheless the possibility that the seemingly weaker quantum assumptions introduced in Section \ref{sec:minimal} may be enough to obtain secure BC, namely pseudorandom quantum states \cite{JLS:CRYPTO18,AQY:Crypto22}, one-way state generators \cite{khurana2024commitmentsquantumonewayness,cryptoeprint:2022/1336}  or EFI pairs of states \cite{brakerski_et_al:LIPIcs.ITCS.2023.24}.



\bigskip   

\centerline { $ *\: * $ }

\bigskip

Despite the no-go result for perfect QBC, i.t.-secure schemes are possible in a setting where both parties can bias the binding and hiding conditions with a non-negligible probability, strictly bounded away from $1$. Some of these schemes have the interesting benefit that any cheating attempt can be detected by the honest parties with a non-zero probability, causing the protocol to abort \cite{HK:PRL04,H:SciRep15,BCH:PRA08,LWL:QIP13}. As discussed later in Section \ref{sec:SCF}, these protocols are commonly used as a subroutine for cheat-sensitive quantum strong coin flipping \cite{BBB:PRA09,PJL:NC14,BBB:NC11}.

Without cheat-sensitivity, optimal bounds of $P_s=P_r\approx 0.739$ were derived using quantum weak coin flipping with arbitrarily small bias \cite{IEEE:CK11}. Here, $P_r$ denotes the probability that a dishonest receiver successfully guesses the committed bit before the reveal phase, while $P_s$ is upper-bounded by $1/2 \left(P_0+P_1\right)$, where $P_i$ denotes the probability that a dishonest sender successfully unveils bit $i$ during the reveal phase. Device-independent versions with finite cheating probabilities were considered in \cite{SCA:PRL11,AMP:NJP16}.

\hl{Remarkably, values of $P_s$ and  $P_r$ arbitrarily close to $1/2$ may be achieved using additional physical assumptions such as no-signalling constraints or the fact that an adversary's quantum storage is imperfect or bounded. These schemes, along with their experimental demonstrations, are covered in Section \ref{SR}.}

\subsection{Oblivious transfer }\label{sec:qot}

In its original form, oblivious transfer (OT) involves one of two mutually distrustful parties sending a message such that it is only delivered with probability $1/2$ to the other party \cite{Rabin81}. The desired security properties are as follows:

\begin{enumerate}[label=(\roman*)]
        \item The sender does not find out if the receiver gets the message or not.
        \item The receiver does not get any information about the message when it is not delivered.
\end{enumerate}
In a second form named $1$-out-of-$2$ OT, the sender transfers two messages and the receiver obtains only one message of their choosing \cite{EvenGoLe82}. In that case, security requires that:

\begin{enumerate}[label=(\roman*)]
        \item The sender does not find out which of the two messages is received.
        \item The receiver is completely ignorant about one of the two messages.
\end{enumerate}

The two versions of OT presented above were proven equivalent in \cite{C:CRYPTO87}. In the context of this review, it is particularly interesting to point out that Stephen Wiesner, who introduced the fundamental idea of conjugate coding in the 1970s (Section \ref{sec:conjugate}), also proposed an early application of quantum cryptography named ``multiplexing''  which is essentially identical to $1$-out-of-$2$ OT \cite{Wie:acm83}. Unfortunately, he failed to see the cryptographic significance of his discovery. While early papers discussed specific cryptographic applications of OT, it is Kilian who first generalized its use to any secure two-party computation \cite{K:ACM88}.
Noticeable simplifications and efficiency improvements can be found in \cite{CVT:CRYPTO95,IPS:CRYPTO08}. 

In classical settings, it was shown that OT reduces to trapdoor one-way permutations \cite{GMW:ACM87}, but it is widely believed that simple one-way functions are not enough to implement a secure OT \cite{IR:ACM89}. The picture is quite different in the quantum realm.

\bigskip   

\centerline { $ * $ }

\bigskip 

The first quantum-based protocol for OT (QOT) was introduced in \cite{Wie:acm83}, but the author also explained how to defeat it. Intuitively, one message is encoded onto a string of quantum states prepared in the $Z$ basis, while the second message is encoded onto a string prepared in the $X$ basis. Note that, for each quantum state generation, the sender randomly chooses whether they encode the next bit of the first message or that of the second message. Unaware of which quantum state encodes which message, the honest receiver chooses one of two possible measurement bases in order to fully retrieve one message (while the second message, measured in the wrong basis, is in principle destroyed). 

The first ever connection between OT and BC (Section \ref{sec:qbitcomm}) can be traced back to \cite{CK:IEEE88}. The authors introduced the idea that BC may be used to force someone to make measurements on a quantum system. Since then, it is widely believed that the security of a QOT can rest on that of QBC. Recent work
showed however that this belief requires considerably more sophisticated techniques to formally reduce QOT to one-way functions when a modern notion of security is used \cite{GLS:LNCS21,BCK:Adv21}. OT has also been reduced to the apparently weaker assumptions introduced in Section \ref{sec:minimal}: the existence of pseudorandom quantum states \cite{AQY:Crypto22,JLS:CRYPTO18}, one-way state
generators \cite{morimae2024onewaynessquantumcryptography} or EFI pairs of states \cite{brakerski_et_al:LIPIcs.ITCS.2023.24}.


\bigskip   

\centerline { $ *\: *$ }

\bigskip 

The quantum technology necessary to implement QOT is essentially the same as QKD. The initial prototype of \cite{jofc-1992-14098} had its software adapted by \cite{BBC:Adv91} to implement QOT. The main difference from QKD was the use of non-interactive error-correction. Interactive error-correction, performed in the original QKD, was not appropriate to the distrustful two-party scenario.
\hl{Along with experimental demonstrations of QOT assuming only symmetric primitives to implement commitments \cite{LSG:PRXQuantum25},} modern composable security definitions have encouraged the development of more advanced QOT implementations as proposed in
\noindent\cite{DGI:Arx24}. 

Similarly to its BC counterpart, there exist variations of quantum OT that are i.t.-secure with non-negligible cheating probabilities, with some bounds and experimental demonstrations provided in \cite{ASR:PRXQu21,SHS:PRXQuantum23,CGS:Chicago16}. Note that practical schemes and implementations relying on bounded or noisy storage assumptions are covered in Section \ref{sec:assumptions}. For a more complete and technical review of oblivious transfer, we refer the Reader to \noindent\cite{LRY:AppSci20}.\bigskip   

\centerline { $ *\: *\: *$ }

\bigskip 

\hl{For completeness, we note the existence of a related primitive known as quantum private queries \cite{GLM:PRL08,JSG:PRA11}, in which a user wishes to access an entry from a database while keeping their choice of entry private from the database provider. While secure $1$-out-of-$n$ OT could provide such a service, security can be achieved in a weaker cheat-sensitive model in which the provider does not wish to risk being detected cheating. A fault-tolerant protocol along with its proof-of-principle demonstration were provided in \cite{CLM:SciRep14}. }

\subsection{Strong coin flipping }\label{sec:SCF}

In strong coin flipping (SCF) or \textit{coin flipping by telephone} \cite{Blum:ACM1983}, two remote parties agree on a random bit such that its value cannot be biased. This primitive is a fundamental building block of distributed computation and finds its relevance in consensus primitives such as randomized leader election (Section \ref{sec:leader}). An ideal SCF protocol ensures the following properties:
\begin{enumerate}[label=(\roman*)]
        \item  In the honest setting, the output is a uniformly random bit (fairness).
        \item A dishonest party $D\in\{A,B\}$ cannot force their opponent to declare outcome $i\in\{0,1\}$ with probability higher than $P_D^{(i)}=1/2+\epsilon_D^{(i)}$, where $\epsilon_D^{(i)}$ is one of the four protocol biases (security).
        \item The biases $\epsilon_A^{(0)}=\epsilon_A^{(1)}=\epsilon_B^{(0)}=\epsilon_B^{(1)}$ (balance). For protocols that are not balanced, the bias $\epsilon$ is defined as the largest of all four biases.
\end{enumerate}
 A naive attempt at performing i.t.-secure SCF with zero bias involves Alice (Bob) generating a private random bit $x_A$ ($x_B$). At a later pre-agreed time, both parties publicly broadcast their bits and agree to the following flip outcome: $x_A\oplus x_B$. At first glance, this appears secure since none of the parties can learn or influence the other party's bit before they communicate. However, simultaneous broadcasting must be enforced, which effectively requires a trusted third party \cite{BBB:PRA09}. Secure classical SCF is nevertheless possible when allowing for a probability $0< \Delta < 1/2$ that the flip does not provide an outcome \cite{HW:TCC11}, yielding an optimal bias of:
\begin{equation}
\epsilon_c = 1/2-\sqrt{\Delta/2}.
\label{eq:classicalbound}
\end{equation}
When $\Delta=0$, classical SCF requires computational assumptions such as completely-secure one-way functions \cite{Blum:ACM1983}. 

\bigskip   

\centerline { $ * $ }

\bigskip 

As the first quantum scheme for i.t.-secure coin flipping were proposed  \cite{BB84}, the authors were already aware of their vulnerability to entanglement-based attacks and coherent quantum measurements. An optimal bound on the achievable bias of $\epsilon\geqslant 1/\sqrt{2}-1/2 \approx 0.207$ was later proven using semidefinite programming \cite{K:PC03,ABD:IEEE04} and quantum weak coin flipping (see Section \ref{sec:WCF}) as a subroutine \cite{IEEE:CK09}. In fact, many explicit SCF schemes relying on quantum bit commitment see their impossibility rooted in the no-go theorems previously established \cite{LC:Phys98,NS:PRA03}. After the original quantum SCF proposal from \cite{BB84}, several works nevertheless studied explicit schemes with finite biases, from $\epsilon\approx 0.41$ \cite{ATA:ACM00} to $\epsilon = 0.25$ \cite{A:STOC01,SR:PRA01,NS:PRA03}.

\bigskip

\centerline { $ *\: *  $ }

\bigskip

 \begin{figure} 
    \begin{center}
        \includegraphics[width=0.70\linewidth]{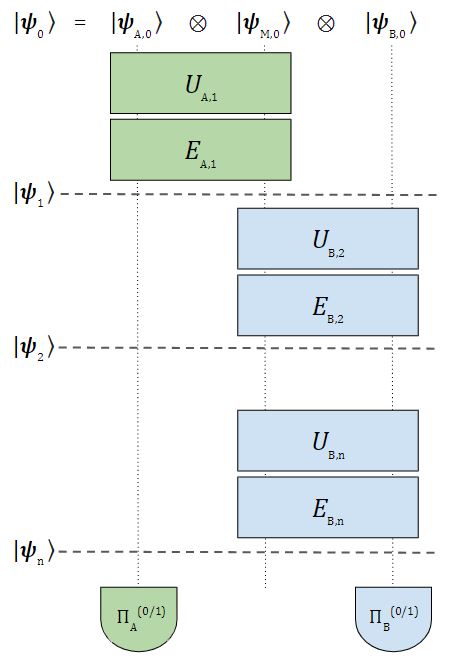}
        \caption{\textbf{General representation of a quantum coin flipping scheme.} Adapted from \cite{ACG:SIAM16}, the protocol starts with a separable state $\ket{\psi_0}$ living in Hilbert space $\mathcal{A}\otimes\mathcal{M}\otimes\mathcal{B}$, where $\mathcal{A}$ and $\mathcal{B}$ are Alice and Bob's spaces, respectively, and the message register living on $\mathcal{M}$ is sent back and forth between the two parties at each round. At each odd round $i$, Alice applies a unitary $U_{A,i}$ and projection $E_{A,i}$ on the space $\mathcal{A}\otimes\mathcal{M}$, while at every even round $i$, Bob applies a unitary $U_{B,i}$ and projection $E_{B,i}$ on the space $\mathcal{M}\otimes\mathcal{B}$. After the final round $n$, each party measures their private register to obtain the outcome of the coin flip.}
        \label{fig:coinflip}
    \end{center}
\end{figure}
 
Despite these limitations, quantum resources can provide SCF with biases that outperform the best classical bound from Eq. (\ref{eq:classicalbound}). Combining this simple classical benchmark with loss-tolerant quantum SCF proposals \cite{BBB:PRA09,PCD:PRA11}, experimental works have demonstrated quantum advantage over optical fiber spools of a few tens of kilometers \cite{PJL:NC14,BBB:NC11}. These implementations all rely on some form of (imperfect) quantum bit commitment: Alice commits to a bit $x_A$ through a single quantum state, while Bob uniformly generates and declares a bit $x_B$. During the reveal phase, measurement bases and outcomes are checked for consistency, and the output of the flip is set to $x_A\oplus x_B$. Although the biases achieved remain high for cryptographic purposes ($\epsilon\approx 0.30-0.40$), these experiments stand as elegant examples of mistrustful quantum schemes outperforming classical schemes \textit{despite} the established no-go theorems \cite{M:PRL97,LC:PRL97}.

It must be noted that in most of these experiments, loss tolerance is improved by allowing the measuring party to back-report detection time-tags to the sending party. As in many implementations of mistrustful quantum cryptography, this subroutine can leak some information about the receiver's measurement basis choice \cite{BCD:PRX21}. It remains an open question how to perform such a sub-routine in a secure manner.

\subsection{Weak coin flipping (and arbitrarily small biases) }\label{sec:WCF}

Weak coin flipping (WCF) is a version of coin flipping in which the parties a priori desire opposite outcomes: it effectively designates a winner and a loser. This implies that only two of the four biases from the set $\{\epsilon_A^{(0)},\epsilon_A^{(1)},\epsilon_B^{(0)},\epsilon_B^{(1)}\}$ in Section \ref{sec:SCF} should be constrained. Classically, i.t.-secure WCF with bias $\epsilon$ can therefore be constructed from i.t.-secure SCF with bias $\epsilon$.

\bigskip   

\centerline { $ * $ }

\bigskip 
In the quantum world, WCF holds a special status: unlike its other two-party counterparts, it can achieve biases arbitrarily close to zero \cite{M:arx07,ACG:SIAM16}. Following the general structure of quantum coin flipping shown in FIG. \ref{fig:coinflip}, the core of the security proof models the protocol, consisting of a sequence of unitaries and projections applied by Alice and Bob in turn, as a semidefinite program. Recalling the brief intuition from Section \ref{sec:optimal}, the dual problem associated with this program then provides a lower bound on the achievable bias. Notably, this milestone result was used as a subroutine in the construction of optimal quantum bit commitment \cite{IEEE:CK11} and optimal quantum SCF \cite{IEEE:CK09}, as well as in quantum leader election schemes \cite{G:QIP17,AS:NJP10}. 

Although Mochon's work proved the existence of quantum WCF with arbitrarily low bias, no explicit protocol was provided until \cite{ARV:SODA21}. This was the result of a line of research investigating explicit quantum constructions from biases of $\epsilon\approx0.239$ \cite{KN:IPL04} to $\epsilon=1/\sqrt{2}-1/2\approx0.207$ \cite{SR:PRL02}. Note that the latter bias incidentally coincides with the quantum SCF lower bound, but was shown to in fact belong to a larger family of WCF protocols with bias $\epsilon=1/6\approx0.167$ \cite{M:FOCS04,M:PRA05}. This was followed by a new explicit family of protocols achieving $\epsilon=1/10$ \cite{ARW:STOC19}, and finally arbitrarily low biases \cite{ARV:SODA21}.  

\bigskip

\centerline { $ *\: *  $ }

\bigskip
While quantum SCF was experimentally demonstrated several times (Section \ref{sec:SCF}), no implementation was proposed for quantum WCF until very recently. This may be explained by two reasons. First, it is challenging to find an encoding and implementation which is robust to losses: a dishonest party may always abort the protocol by declaring a loss when they are not satisfied with the predicted flip outcome. For this reason, the detection probability cannot be amplified by performing several repetitions of the protocol like in QKD or quantum SCF. Second, none of the explicit constructions translate trivially into a simple experiment: they require either performing single-shot generalized measurements or generating beyond-qubit states. 

A connection between quantum WCF unitaries and linear optics was established in \cite{BCK:PRA20}, where a two-round photonic implementation was proposed exploiting photon-number encoding, a large single-photon interferometer and an optical switch for verification. This was later realized in laboratory conditions, showing that quantum WCF is possible over a few kilometers of optical fiber spool \cite{NYC:NatComm23}. Despite these demonstrations, the question of how to lower the bias by increasing the number of rounds remains open, especially with recent results showing that the bias scales inefficiently with the number of rounds \cite{M:STOC20}. 

\subsection{One-time programs}

One-time programs (OTPs) were introduced as a computational paradigm for security applications, ranging from software protection and program obfuscation to the temporary transfer of cryptographic abilities \cite{GKR:CRYPTO08}. Ideally, an OTP acts as a black box function which can only be evaluated on a single input before self-destructing. This implies that a user cannot access the inner workings of the program besides a single input-output relation. \hl{More formally, the OTP cryptographic primitive for a function $f$ ensures that an efficient receiver can evaluate $f(x)$ at run time, but cannot learn anything about $f(y)$ for $y\neq x$ beyond what they can learn from $f(x)$.}

OTPs can be constructed from hypothetical devices called one-time memories (OTMs) acting on a single bit, modeled from non-interactive 1-out-of-2 oblivious transfer \cite{GKR:CRYPTO08}: at the end of the query, the user should only be able to access one of the two bits stored in the OTM. Classically, it remains unclear how to build such hardware in a secure manner.
\bigskip

\centerline { $ *$ }

\bigskip
 Using quantum resources, i.t.-secure OTPs with a deterministic output do not exist \cite{BGS:Crypto13}, although allowing for a non-zero error probability in the output circumvents this no-go result \cite{RKB:NatComm18}. In fact, such noisy schemes were demonstrated experimentally, with the four single-bit gates encoded onto the states summarized in Table \ref{tab:OTP}. The input to each program is realized through a projective measurement in the $Z$ basis for $0$ and $X$ basis for $1$: this provides a measurement outcome interpreted as the output of the gate, which succeeds with probability $\approx 85.36\%$. The gate rate and loss tolerance of this protocol were later improved using remote state preparation \cite{RKF:npjQI21}. 

Finally, it is worth noting that i.t.-secure OTPs are possible assuming quantum mechanics and a linear number of queries to stateless tokens \cite{B:Quantum21}, which can be re-run but keep no record of previous interactions \cite{CGS:EUROCRYPT08}.

\begin{table}
\begin{ruledtabular}
\begin{tabular}{ccc}
Gate & Truth table & Encoding
 \\
\hline
 Constant $0$ & 0 \:\:\: 0&  \\
& 1 \:\:\: 0& $\frac{1}{\sqrt{2+\sqrt{2}}}\left(\ket{0} +\ket{+}\right)$  \\
 Constant $1$ & 0 \:\:\: 1&   \\
& 1 \:\:\: 1& $\frac{1}{\sqrt{2+\sqrt{2}}}\left(\ket{1} -\ket{-}\right)$    \\
Identity & 0 \:\:\: 0&   \\
& 1 \:\:\: 1&  $\frac{1}{\sqrt{2+\sqrt{2}}}\left(\ket{0} +\ket{-}\right)$\\
Not & 0 \:\:\: 1&    \\
& 1 \:\:\: 0&    $\frac{1}{\sqrt{2+\sqrt{2}}}\left(\ket{1} +\ket{+}\right)$ 

\end{tabular}
\end{ruledtabular}
\caption{Summary of the four single bit gates and their corresponding truth tables, along with their quantum encodings proposed in \cite{RKB:NatComm18}. In the middle column, each line indicates a desired input/output relation for the considered gate. Measuring each gate in the $Z$ basis for input $0$ and $X$ basis for input $1$ yields a measurement outcome satisfying the truth table with probability $\approx 85.36\%$.}\label{tab:OTP}
\end{table}

\section{Security based on other physical assumptions}\label{sec:assumptions}

As a brief reminder, Section \ref{sec:mistrustful} established the impossibility of i.t.-secure mistrustful cryptography using quantum resources (with the noteworthy exceptions of digital signatures and weak coin flipping). The no-go result is a consequence of entanglement-based attacks in which a cheating party can always store their half of a shared pair and delay their measurement in time.

This section discusses how the additional physical assumptions summarized in Table \ref{tab:summary} may be combined with conjugate coding to provide i.t.-secure implementations of mistrustful primitives, thus circumventing this no-go result. In a nutshell, the first two approaches enforce that a dishonest party measures their quantum states at a given point in time, either by exploiting no-signalling constraints or by making assumptions on the size/noise of the adversary's quantum storage device. The third approach relies on physical unclonable hardware assumptions.

\subsection{Special relativity and trusted agents}
\label{SR}

Mistrustful quantum cryptography may be combined with no-signalling constraints to achieve new security guarantees \cite{BGK:ACM88,K:DIMACS90,K:PRL12,K:PRL99}, evading the no-go theorems of Section \ref{sec:mistrustful} for most primitives. In this scenario, Alice typically splits into several agents $\{A_i\}$ who can message each other using secure classical channels, and likewise for Bob splitting into $\{B_i\}$. Before making any security claim, the following physical assumptions must hold within the context of special relativity \cite{K:PRL99}: 

\begin{enumerate}[label=(\roman*)]
        \item  The protocol's spacetime geometry is approximately flat.
        \item Both parties have access to correct and accurate descriptions of their local light cones.
        \item Signalling between spacelike-separated points is impossible.
\end{enumerate} 

With these at hand, one can enforce that a cheating party measures their quantum states in a given region of spacetime before pursuing the remaining protocol steps. Note that such constraints may also prevent other cheating actions, depending on the protocol and scenario considered \cite{KTH:IEEE13}. 

\begin{figure} 
    \begin{center}
        \includegraphics[width=0.92\linewidth]{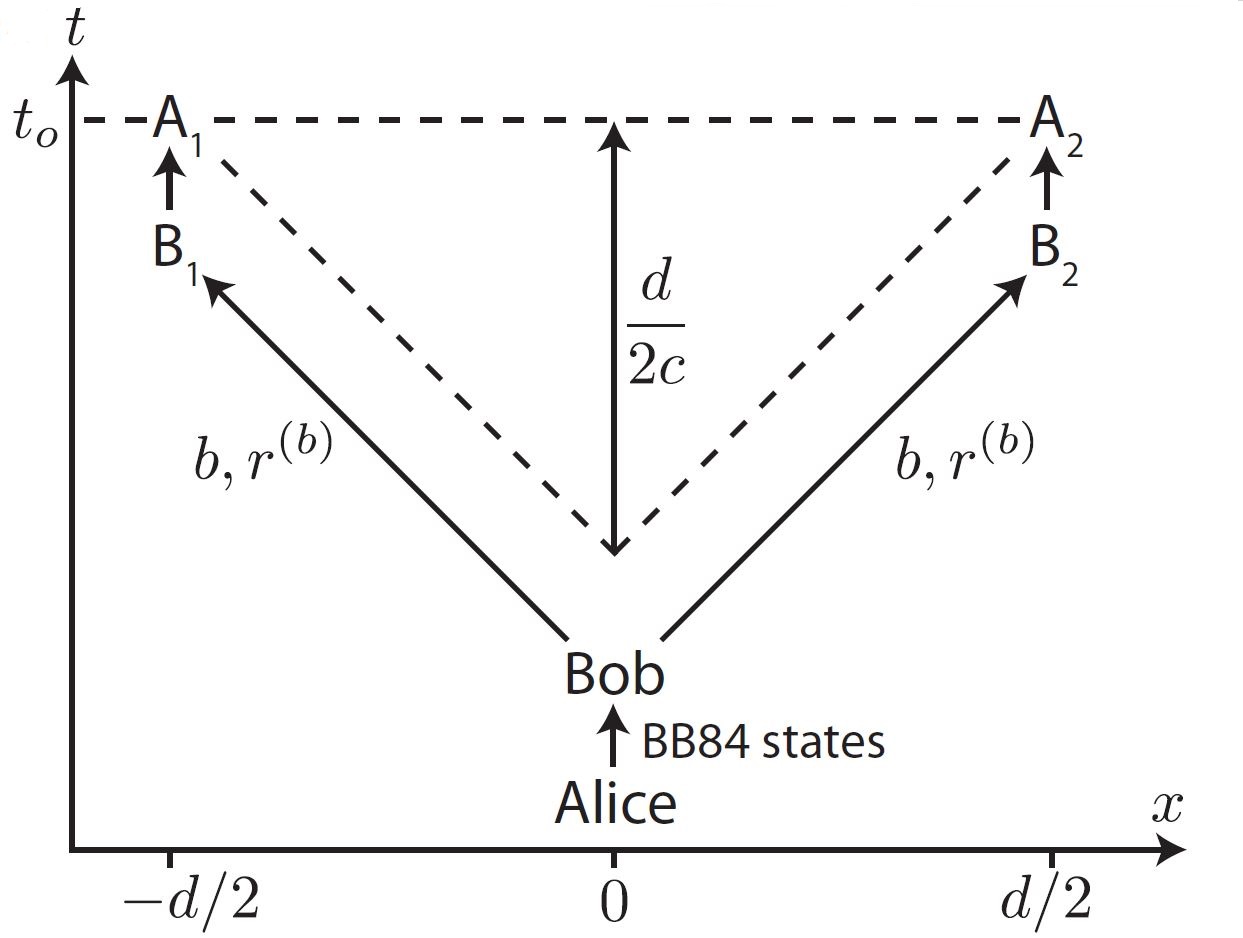}
        \caption{\textbf{Quantum bit commitment based on special relativity.} First proposed in \cite{K:PRL12}, this space-time diagram $(x,t)$ was taken from \cite{Zbind:PRL13}. In the commit phase, Alice sends Bob a string of BB84 states using conjugate coding (Section \ref{sec:conjugate}). Bob commits to a bit $b$ by measuring all states in the same basis $b$, and sending his measurement outcomes $r^{(b)}$ at light speed to agents $B_1$ and $B_2$. In the reveal phase, $B_1$ (resp. $B_2$) communicates these strings to Alice's agent $A_1$ (resp. $A_2$). In the verification phase, $A_1$ and $A_2$ cross-check that both messages from $B_1$ and $B_2$ were received at time $t_0$, and that $b$ and $r^{(b)}$ are consistent with the initial description of the BB84 states. A successful verification implies that Bob must have measured and committed no later than $t_0-d/2c$, where $d$ is the distance separating $\{A_1,B_1\}$ from $\{A_2,B_2\}$ and $c$ is the speed of light. }
        \label{fig:relativity}
    \end{center}
\end{figure}

As an example, consider the quantum bit commitment protocol described in FIG. \ref{fig:relativity}, in which Bob commits to a bit $b$ by measuring all of Alice's states in basis $b$ (labeling either the $Z$ or $X$ basis) and sending the measurement outcome string $r^{(b)}$ at light speed to agents $B_1$ and $B_2$. By allowing Alice's spacelike-separated agents $A_1$ and $A_2$ to cross-check the timing of messages received from $B_1$ and $B_2$ at $t_0$, no-signalling implies that Bob could not have measured his states any later than $t_0-d/2c$, where $d$ is the distance separating $\{A_1,B_1\}$ from $\{A_2,B_2\}$ and $c$ is the speed of light.  

\bigskip

\centerline { $ *$ }

\bigskip

This protocol was later refined and experimentally demonstrated in \cite{Zbind:PRL13,Pan:PRL14}. From a practical standpoint, it is crucial to note that the commitment time is a function of distance $d$, thus imposing large distances between a party's trusted agents. Locating the agents in Singapore and Geneva, for example, provides a commitment time of $15$ ms \cite{Zbind:PRL13}. \hl{By designing multi-round classical protocols, arbitrary long commitment times secure against classical attacks are in principle possible \cite{CCL:PRA15,LKB:PRL15}, and were experimentally demonstrated up to $24$ hours \cite{VMH:PRL16}}. Generally, one must consider that monitoring the spatial and temporal coordinates of verifiers requires a trusted Global Positioning System (GPS), which might open the door to spoofing attacks and time-stamp manipulation \cite{spoof11}. This may however be mitigated if secure position verification schemes are used (Section \ref{sec:position}). Furthermore, the composability of such relativistic protocols must be carefully assessed \cite{VPL:NJP19}.

For completeness, we note the existence of other mistrustful protocols basing their security on spacetime constraints, such as quantum oblivious transfer \cite{P:PRA16,P:PRA19}, device-independent bit commitment \cite{AK:PRA15}, multiparty coin flipping \cite{P:Proc21} and zero-knowledge proofs in the multi-prover setting \cite{BGK:ACM88,CL:EUROCRYPT17,GSY:IEEE19,ABC:Nat21}. 

\subsection{Bounded and noisy storage }\label{sec:storage}
\label{BNS}
At any point in time there is so much technology one can acquire to attack cryptographic protocols, and the same goes with the quality of the equipment one can achieve. Such limitations are particularly true regarding quantum technologies. At the time of writing these lines, quantum computers can be constructed for sizes up to a few thousand qubits with $99.8\%$ reliability. Assuming no adversary can do one million times better is not a serious security threat. Based on similar considerations, security in the bounded or noisy storage model was first introduced classically for key distribution \cite{256484} and extended to mistrustful cryptography in \cite{CCM:FOCS98}. 

\bigskip

\centerline { $ *$ }

\bigskip

In the quantum realm, i.t.-secure oblivious transfer and bit commitment can be obtained under the assumption that dishonest participants are at some point in time limited in the total amount (or reliability) of quantum storage available during their attack. In such a scenario, if an adversary eventually acquires more or better storage at a later time, it will not impact the security of prior executions of the protocol. This is called \textit{forward security}.

The work of \cite{DFS:SIAM08} introduced protocols for OT and BC in the bounded quantum storage model, while other papers did the same for the noisy quantum storage model \cite{PhysRevLett.100.220502,STW:QIC09}, with later generalizations including \cite{KWW:IEEE12}. The latter model is the following:
the adversary has unlimited classical storage and computational resources and
whenever the protocol requires the adversary to wait for a time $\Delta t$, they must measure/discard all their quantum information except what they can encode into a fixed size memory. At that time, this information undergoes noise. All the previous (bounded or noisy) models are special cases of the latter. 

More recently, i.t.-secure constructions under the bounded storage model were proposed for asymmetric key encryption and signature schemes \cite{BS:Arx25}. 

\bigskip

\centerline { $ *\:*$ }

\bigskip

 Since secure demonstrations of these schemes do not require any quantum storage on the honest parties' side, experimental requirements are in fact very close to standard prepare-and-measure or entanglement-based QKD. Various practical security proofs were proposed \cite{WCS:PRA10,LPA:PRR23}, and most implementations have used polarization encoding of weak coherent states \cite{ASP:JO15} or  entangled photon pairs generated by spontaneous parametric down-conversion \cite{NS:NatComm12,ENG:NatComm14}. As in many implementations of mistrustful quantum cryptography however, it remains unclear how to back-report the detection time-tags in such a way that no information is leaked about the receiver's measurement basis choice \cite{BCD:PRX21}. We also note implementations of noisy storage in the continuous-variable setting \cite{FGS:NC18}.

 \bigskip

\centerline { $ *\:*\: *$ }

\bigskip

As an alternative or addition to quantum storage assumptions, we note the existence of BC schemes in which the emission time of each quantum state is randomized by the sender, which yields secure protocols under the assumption that a dishonest receiver cannot perform quantum non-demolition measurements \cite{DV:QIP11,LAA:PRA14,LAD:PRA16}. 

\subsection{Physical unclonable functions}


In cryptography, an alternative to relying on computational assumptions and trusted parties/infrastructures is to adopt hardware assumptions. Physical unclonable functions (PUFs) are one of such hardware assumptions, exploiting the practical “unclonability” of physical systems to obtain a notion of device \textit{fingerprint}. A PUF can be modeled as a set of challenge-response pairs (CRPs), extracted by physically querying the PUF and measuring its responses. The inherent unpredictability of such devices implies that even the manufacturer of the PUF, with access to many CRPs, cannot predict the response to a new challenge. As an example, it was shown that highly-scattering optical media act as good PUFs with a large CRP space \cite{PRT:Science02}: challenges are materialized by the angle of incidence of a laser beam on the medium, while responses are given by the resulting speckle patterns. Further works studied the rigorous entropic aspects of optical PUFs, for instance in \cite{TS:book07}.

PUFs have multiple applications, including identification schemes, device authentication, software licensing, random number generator, anonymous computation, etc. However, classical PUFs are typically susceptible to side-channel and machine-learning attacks \cite{RSS:ACM10}. On the other hand, considering quantum PUFs (qPUFs) can offer provable unclonability.

\bigskip

\centerline { $ *$ }

\bigskip

The first quantum approach was introduced in \cite{K:IJQI12}, where the concept of quantum readout PUF (QR-PUF) was defined. The idea is that a quantum challenger prepares a superposition of challenges and receives back the corresponding superposition of responses. Any party intercepting the quantum states would be essentially detected, thus ensuring that a challenger is indeed probing a specific QR-PUF, even when it is under adversarial control. The security here relies on the principles of quantum theory. A variant involving coherent states with low mean photon number was later implemented \cite{GHM:Optica14}, while another protocol, based on continuous variable with homodyne detection, was demonstrated in \cite{ND:SciRep17}. Furthermore, a theoretical framework for QR-PUFs was developed in \cite{gianfelici2020theoretical}. While offering interesting prospects, these protocols all present two main disadvantages. First, their security is proven against specific types of attacks (e.g. \cite{Nikolopoulos2018PhysRevA,FNA:Cryptography19}), but not in full generality. Second, the verifier requires some information on the quantum channel implemented by the PUF, typically in order to perform the quantum readout. This differs from the initial classical notion of PUFs.

A more general notion of qPUF was provided in \cite{ADD:Quantum21}, where the qPUF is treated as a general quantum channel, and the verifier does not require any further knowledge about the device. Three different types of security one might expect from such a functionality were defined, and the weakest one (quantum selective unforgeability) was shown to be possible. Identification protocols using qPUFs were later given \cite{DKD:ACM21}, and their security proven against any quantum polynomial-time adversary, providing the first scheme with a complete security proof.  Another interesting dimension was also explored in \cite{DKKK2022QuantScieTech}, where the existence of secure qPUFs was demonstrated to be equivalent to the existence of pseudorandom quantum states (Section \ref{sec:minimal}). 

\bigskip

\centerline { $ *\:*$ }

\bigskip

Overall, qPUFs appear easy to implement, while the theoretical connection with other concepts such as pseudorandom quantum states and $t$-designs may offer many alternative options. Crucial challenges remain for qPUFs to become practical, such as the security of different schemes within specific applications and the need for concrete, experimentally simple and robust, constructions.



\section{Computational security against quantum attacks}\label{sec:computational}

In the review so far, we have covered possibilities and no-go results for quantum cryptography in the i.t.-secure setting, and discussed additional physical assumptions that open possibilities for mistrustful quantum cryptography. However, since the field's growth was initially triggered by the threat of quantum computers to computationally-secure classical primitives, it is equally relevant to design and analyze primitives that are secure against computationally-bounded quantum adversaries. 

Currently, there are two main approaches: the first approach, which is not covered in this review, uses classical resources to design primitives that are secure against quantum attacks. Labelled \textit{post-quantum cryptography}, this field is evolving very rapidly, both in terms of standardization (through the NIST competition) and cryptanalysis, in which security conjectures are broken and new schemes appear. The second approach, which is the main focus of this section, uses quantum resources to protect against quantum attacks, unlocking new protocols summarized in Table \ref{tab:summary} such as public-key quantum money, one-shot signatures, public-key encryption and zero-knowledge proofs. For completeness, we provide a very brief discussion of cryptanalysis in Section \ref{sec:cryptanalysis}. 

\subsection{Public-key quantum money}\label{sec:publicmoney}

In Section \ref{sec:tokens}, we described one of the first cryptographic applications for which quantum resources can provide i.t.-security: private-key quantum money. While achieving a useful task that was impossible classically, most private-key schemes suffer from three shortcomings \cite{AC:stoc12}:

\begin{enumerate}[label=(\roman*)]
        \item The ``big database'' problem: the bank needs to store a database with details about every banknote in circulation.
        \item The ``verifiability'' problem: only the bank can verify a banknote.
        \item The ``online attack'' problem: an attacker that can submit multiple banknotes for verification and get them back can compromise the scheme's security.
\end{enumerate} 
Interestingly, all of these may be addressed by transitioning from i.t.-security to computational security against polynomial-time quantum adversaries, while even stronger properties can potentially be achieved.

\bigskip

\centerline { $ *$ }

\bigskip

Problem $(i)$ was already considered in \cite{BBB:AdvCrypto83}, where the quantum token $\ket{\psi_s}$ depends on a public serial number $s$. Instead of storing a random sequence of quantum states for each serial number, the authors proposed the use of a quantum-secure, pseudorandom function $f_k$, where the key $k$ is only known to the bank and the output $f_k(s)$ determines the states to be prepared and the information required for verification. From a security perspective, an adversary with unlimited computational power can recover $k$ and break the scheme, but any computationally-bounded adversary cannot, since this would imply breaking the pseudorandomness property of $f_k(s)$.

Problems $(ii)$ and $(iii)$ may be addressed by designing a protocol in which \textit{all} parties, not only the bank, are able to verify the quantum money. First introduced in \cite{A:IEEE09,FG+:itcs12,AC:stoc12}, this concept is known as \textit{public-key quantum money}. The basic idea is to produce a state $\ket{\psi}$, such that the bank can publish a description of a quantum circuit that lets other parties recognize $\ket{\psi}$, but does not allow them to prepare or even copy $\ket{\psi}$. This may not be achieved in an i.t.-secure way, but becomes possible with computational assumptions. Whenever this is the case, it is crucial to find assumptions that we are confident cannot be broken (such as the minimal assumptions introduced in Section \ref{sec:minimal}). A lot of research effort is dedicated to weakening assumptions and eventually resorting to so-called ``standard cryptographic assumptions'', such as the existence of a quantum random oracle, the hardness of Learning With Errors (LWE) or the existence of post-quantum secure one-way functions. For public-key quantum money, a scheme whose security relies on the hardness of a knot theory problem -- a non-standard assumption --, was provided in \cite{FG+:itcs12}, and its security is yet to be proven (or disproven). In other works, the quantum money state takes the form
\begin{equation}
\ket{A}:=\frac{1}{\sqrt{\vert A\vert}}\sum_{x\in A}\ket{x},
\label{eq:publicmoney}
\end{equation}
i.e. an equal superposition of all bit strings of a subset $A$ that consists of half the possible bit strings \cite{AC:stoc12}. The classical description of the subset $A$ remains hidden from the users and is only known to the bank. The bank also publicizes the details of how to check that the note is correct using calls to membership oracles, with essentially the property that $V_A\ket{A}=\ket{A}$. While verification requires a single call to the verification oracle $V_A$, an adversary could cheat with an exponential number of queries to that oracle. Note that giving access to a program without leaking information about its inner workings can be achieved with a cryptographic notion called indistinguishability obfuscation (IO).


\bigskip

\centerline { $ *\: *$ }

\bigskip

Another major development appeared in \cite{Z:JC21}, where a stronger notion of public-key quantum money became possible, subsequently termed ``public-key semi-quantum money''. In addition to enabling public verification, the bank becomes fully classical, simply instructing the untrusted participants to generate the quantum money themselves. Based on the idea of quantum lightning  \cite{Z:JC21}, the trick is to prepare an equal superposition of inputs as shown in Section \ref{sec:preimages}, using a collision-resistant hash function $H$. Specifically, one can generate a state $\ket{\psi_y}=\sum_{x\vert H(x)=y}\ket{x}\ket{y}$ for a random $y$. It is then easy to see that nobody can prepare the same state twice without breaking the collision resistance of $H$. For its use in quantum money, the crucial property is to be able to verify it, which is possible if one can distinguish between the states $\ket{x}$ and $\ket{\psi_y}$, a property known as non-collapsible collision-resistant hash functions\footnote{It is an open question whether such functions exist.}. A construction for public-key semi-quantum money based on weaker assumptions, namely IO and LWE, was later developed \cite{S:STOC22}.

This functionality is powerful and has many potential applications, opening a research direction in which the constructions should be improved, developing weaker and stronger versions based on standard assumptions. A notable application was discussed in \cite{CS:Quantum20}, where a classical blockchain enhanced with quantum resources offers new possibilities such as smart contracts and the property that lost/damaged quantum banknotes can be retrieved by the legitimate owner without a central authority. 

\bigskip

\centerline { $ *\: *\: *$ }

\bigskip

It is worth stressing that all protocols discussed in this section require quantum computation that is well within the fault-tolerant era, since only such computations could be hard to invert. Furthermore, most of the schemes do not rely on standard cryptographic assumptions, and their security is essentially conjectured: attacks questioning their security have for instance been explored in \cite{roberts2021security,BDG2023cryptanalysis,AHY2023plausibility}.

\hl{Recently, a first step towards noise-tolerant public-key quantum money was proposed, along with its security challenges \cite{Y:Quantum25}. Essentially, the subspace state from Eq. (\ref{eq:publicmoney}) may either be treated as a codeword in a Calderbank-Shor-Steane code, or encoded and decoded using an additional layer of quantum error correction.}

\subsection{Tokenized and one-shot signatures}
\label{subsec:one-shot signatures}

Building on the techniques from public-key quantum money (Section \ref{sec:publicmoney}), it is possible to construct signature schemes that offer extra security properties. The first relates to settings in which a central signing authority, Alice, wants to delegate the ability to sign a document on their behalf to another entity, Bob. In this context, Alice does not want to give Bob the ability to sign more than one message (or more than a fixed number of messages). This notion of \textit{tokenized} signatures can be achieved by using quantum states as tokens, which Bob needs to measure, and therefore destroy, in order to sign. Ultimately, this security guarantee relies on computational quantum unclonability: the message signed by Bob can then be verified by everyone, using public information. The second property strengthens this notion of tokenized signatures, by adding the requirement that even the signing authority, using all her secret information, cannot generate a second valid signature corresponding to the same public key. Schemes that satisfy this property are known as \textit{one-shot} signatures.

\bigskip

\centerline { $ *$ }

\bigskip

Quantum tokens as digital signatures were introduced in \cite{BS:Quantum23}. To get a little intuition, consider the simple case of signing a one-bit message. Alice selects a subspace $A$ and prepares the corresponding quantum token $\ket{A}$ exactly as in Eq. (\ref{eq:publicmoney}) from Section \ref{sec:publicmoney}. Bob can sign the message $m=0$ by returning an $x$ that belongs to the subspace $A$, which he obtains by measuring the quantum token in the computational basis. To sign the message $m=1$, Bob must return an $x$ that does not belong to $A$, which he obtains by first applying the Hadamard $H$ to all qubits to obtain $\ket{A^{\bot}}$ and then measuring the register in the computational basis. Bob then returns the pair $(m,x)$ as a signature and Alice checks if $x$ is compatible with the message. To turn this into a publicly verifiable signature scheme, Alice can use the same methods as in Section \ref{sec:publicmoney}, and give obfuscated access to the membership test, i.e. can allow other parties to check if a given $x$ belongs to $A$ or $A^\bot$ without revealing the sets. 

\bigskip

\centerline { $ *\:*$ }

\bigskip

One-shot signatures were introduced in \cite{AGK:STOC20}. In this setting, we require that even Alice, the signing authority, cannot sign a second message corresponding to the same public key. Interestingly, by achieving this, one can now delegate the generation of the quantum token to Bob, making Alice completely classical. Alice picks a collision-resistant hash function $H$ and instructs Bob to generate an equal superposition of a random image using the trick of Section \ref{sec:preimages}. The first register $\ket{\psi_y}=\sum_{x\vert H(x)=y}\ket{x}$ is the token. To sign a message, Bob needs to output a pre-image such that the first bit of the string is equal to the message he wants to sign. To construct an algorithm that can ``select'' a pre-image matching the message that Bob wants to sign without any secret information, the hash function requires extra properties, such as being non-collapsible. The triplet $(m,x,y)$ materializes the signature, and anyone can verify its validity by checking the first bit of $x$ and computing $H(x)$ to check that it is equal to $y$. 

From a security perspective, the state of the token cannot be copied, and if one could sign two different messages, this would directly imply a collision for the hash function, yielding a contradiction.

\bigskip

\centerline { $ *\:*\:*$ }

\bigskip

Signatures with these strong properties have many applications. For example, one might delegate budget signatures, in which a party can sign as many messages as they want provided they do not exceed a fixed ``budget'' amount. Another application is quantum money with classical communication, in which the quantum token can be delegated to the signing party and the bank does not need to prepare or communicate any quantum state. Further applications include decentralized blockchain-less cryptocurrency, ordered signatures, key-evolving signatures and proof-of-stake blockchains. The latter solves one of the major shortcomings of proof-of-stake blockchains, making them as secure as proof-of-work blockchains but with the key advantage of being much more energy-efficient.

\bigskip

\centerline {$ *\: *\: *\:\: *$ }

\bigskip

Despite these exciting applications, two main challenges stand in the way of one-shot signatures being practically realized. The first one, that is shared with public-key quantum money, is the hardware requirements, as it requires a large scale (to have sufficient qubits), fault-tolerant (to have sufficient depth), quantum computation in order to resist brute-force attacks. The second issue relates to practicality. \hl{Recently, a concrete scheme was proposed to realize the functionality, relying solely on quantum secure indistinguishability obfuscation (iO) and the hardness of Learning with Errors, thus resolving the theoretical question of whether OSS are possible  \cite{SZ:eprint2025}. However, the said construction would lead to very high overheads if one wanted to implement it, meaning that further research on reducing the complexity and cost is essential before being realized in practice.} 

\subsection{Public-key cryptography}

The notion of public-key cryptography was introduced in \cite{Diffie} and first implemented by RSA \cite{RSA:ACM78}. \hl{Besides the latter, public-key encryption algorithms include ElGamal \cite{Elgamal1985signature} based on discrete logarithm or elliptic curves \cite{K:MOC87}, Cramer-Shoup \cite{CS:CRYPTO98} and lattice-based key encapsulation recently standardized by NIST \cite{NIST2024}.} 

A public-key encryption (PKE) scheme comprises three efficient algorithms: $\operatorname{Gen, Enc, Dec}$. The first one, $\operatorname{Gen}$, is used to generate a pair of secret and public keys $sk,pk$. The public key is published for everyone to use while the secret key is kept secret by the creator of the system.
The second algorithm, $\operatorname{Enc}$, maps the public key $pk$ and message $m$ to a ciphertext $c$. Finally, the third algoritm $\operatorname{Dec}$ maps the secret key $sk$ and ciphertext $c$ back to message $m$.  

The great advantage of PKE is that no secret exchange between sender and receiver is necessary to send encrypted messages. It suffices for the sender to get hold of the public key associated with the receiver. In the classical setting, PKE can be obtained from trapdoor one-way functions \cite{Diffie}, and it is simple to understand why i.t.-secure PKE is impossible: if $pk$ determines $sk$ completely, then any receiver can compute $sk$ and decrypt every possible ciphertext $c$.

\bigskip

\centerline { $*$ }

\bigskip

Interestingly, it was shown that i.t.-secure quantum public-key encryption (qPKE) is also impossible \cite{N:PRA08,NI:PRA09,SNA:PRA12,BGH:TOC23}. Due to the public key being a quantum state, this impossibility is much less obvious than for the classical case. In fact, some of the existing constructions of qPKE had been conjectured to be unconditionally secure \cite{Gottesmann2005}, a conjecture that is invalidated by the work from \cite{BGH:TOC23}.

In the classical world, it is widely believed that one-way functions are insufficient to obtain  public-key encryption \cite{IR:ACM89}. It is also believed that quantum public-key encryption with entirely classical keys from quantum-safe one-way functions
is impossible \cite{bouazizermann2023impossibilityquantumpublickey}.

Recently, a couple of  works showed that post-quantum one-way functions are sufficient to obtain public-key encryption schemes where the public keys are quantum states \cite{C:arx23,BGH:TOC23}. New definitions of security have to be introduced since the obvious extension of the classical definition doesn't work in the quantum realm. Classically, when the encryption keys are public, anyone can efficiently encrypt messages at will and thus no encryption oracle is necessary for Chosen-Plaintext Attack (CPA) security. With quantum public keys however, in order to encrypt messages, a polynomial number of public-key states are provided to the adversary. Both approaches in \cite{C:arx23,BGH:TOC23} explore this direction to define CPA security and construct qPKE in this model under the sole assumption that post-quantum one-way functions exist. The work of \cite{BGH:TOC23} also offers a new definition of (quantum) Chocen-Ciphertext Attack (CCA)  security and a construction to achieve it.

When quantum keys are involved, we need an extra generation procedure $\operatorname{Genp}$ to produce public keys $\ket{pk}$ from secret keys $sk$. It is generally assumed that $\ket{pk}$ is a pure quantum state. The encryption process $\operatorname{Enc}$ maps the public key
$\ket{pk}$ and classical message $m$ to a ciphertext $\rho_c$ (generally a mixed state) and $\operatorname{Dec}$ maps the private key
$sk$ and ciphertext $\rho_c$ back to the classical message $m$.

An example scheme from \cite{BGH:TOC23} assumes the existence of a quantum-safe pseudorandom function (PRF),
which can be obtained from an arbitrary quantum-safe one-way function \cite{6375347}. Given a PRF ensemble $\{f_k\}_k$,
the public key consists of a pair of pure quantum states $(\ket{qpk_0},\ket{qpk_1})$ and the secret key
consists of a pair of strings $(dk_0,dk_1)$ such that, for all $b\in\{0,1\}$,
\begin{equation}
\ket{qpk_b} =
\frac{1}{\sqrt{2^n}}
\sum_{x\in\{0,1\}^n}
\ket{x,f_{dk_b}(x)}.
\end{equation}
 
To encrypt a bit $b$, one measures $\ket{qpk_b}$ in the computational basis. The result is $(x,f_{dk_b}(x))$ for some uniformly random $x\in\{0,1\}^n$ and this is the ciphertext, i.e., $(qc_0,qc_1) = (x,f_{dk_b}(x))$.

To decrypt a ciphertext $(qc_0,qc_1)$, one computes the value
of $b\in\{0,1\}$ such that $f_{dk_b}(qc_0) = qc_1$. In case this occurs for neither or both keys, the decryption simply aborts.

\subsection{Zero-knowledge proofs}\label{sec:ZKP}

Zero-knowledge proofs constitute an important tool of modern cryptography invented by Goldwasser, Micali and Rackoff \cite{GMR:SIAM89}. Through them it is possible to demonstrate the validity of certain mathematical statements in a way that does not disclose anything more than their sole validity. Zero-knowledge does not only quantify the (zero) amount of information disclosed but also the (zero) amount of computational advantage communicated. The scenario considered is that of a {\em prover} of unbounded computing power trying to convince a polynomial-time {\em verifier} of the membership of a certain string $x$ to a language $L$. It is required that:

\begin{enumerate}[label=(\roman*)]
        \item Strings $x\in L$ lead the verifier to acceptance with overwhelming probability (completeness).
        \item Strings $x\not\in L$ lead the verifier to rejection with overwhelming probability (soundness), regardless of the prover's strategy\footnote{When the soundness is only guaranteed with respect to polynomial-time provers, the word {\em argument} is used instead of proof as introduced by \cite{BCC88,BC89}.}. 
        \item The honest prover will not provide any computational advantage to any polynomial-time verifier beyond what such a verifier can compute by oneself (zero-knowledge), given that $x\in L$.
\end{enumerate} 
The zero-knowledge property of proofs or arguments is demonstrated by exhibiting a polynomial-time {\em simulator} capable of efficiently producing transcripts of would-be conversations between the prover $P$ and an arbitrary polynomial-time verifier $\widetilde{V}$ for every $x\in L$. These are generated without any interaction with $P$.
The distribution of these transcripts is compared to that of the real conversations between $P$ and the same verifier $\widetilde{V}$. If they are identical, we qualify the zero-knowledge property as {\em perfect}. If they are statistically indistinguishable, we qualify the zero-knowledge as {\em statistical} and if they are computationally indistinguishable, we qualify the zero-knowledge as {\em computational}. Zero-knowledge is often achieved using bit commitment (BC) schemes (Section \ref{sec:qbitcomm}): perfectly (resp. statistical or computational) hiding BCs lead to perfect (resp. statistical or computational) zero-knowledge.

The main tool for constructing a simulator is {\em rewinding}, a concept by which simulated transcripts can be generated via a trial-and-error process whereas real ones must be produced in a straightforward fashion. Rewinding allows the simulator to create the transcript by first guessing a prover's answer and then hoping the verifier will ask the corresponding question. If this is the case, the simulation may go on and otherwise the last attempt is forgotten and the simulator tries again.

\bigskip

\centerline { $ *$ }

\bigskip

For a while, it was believed that quantum simulators could not be implemented because ``forgetting the last attempt'' was incompatible with $\widetilde V$'s unitary evolution. Consult the work from \cite{V:Thesis98} for a discussion. However, it was later proven that rewinding of quantum verifiers is actually possible if the simulator measures as little as possible \cite{W:SIAM09}. Instead of generating an answer followed by a question hoping that they match, the simulator can generate a superposition of all answers and all questions and only measure whether they match. If they do, one may proceed to measuring the answer and the question (that necessarily match now). If they don't, it is possible (under certain extra conditions) to return to the previous superposition state with (near) certainty.

In the classical world, the class of languages that admits a statistical zero-knowledge proof system is $\operatorname{SZK}$.
Examples of languages in this class are graph isomorphism, quadratic residuosity and discrete log.
When the prover and verifier are quantum devices, the class $\operatorname{QSZK}$ was defined by \cite{1181970}. At first it is not obvious what the relation between $\operatorname{SZK}$ and $\operatorname{QSZK}$ should be. Having quantum provers who can send quantum messages to the verifier may make more languages in $\operatorname{QSZK}$ and having quantum verifiers that are harder to simulate may make fewer
languages in $\operatorname{QSZK}$. But because $\operatorname{SZK} \subseteq \operatorname{QSZK_{HV}}$ (the honest-verifier version of $\operatorname{QSZK}$), all problems in SZK have quantum interactive proof
systems that are statistical zero-knowledge against quantum honest verifiers. 
Subsequently, it was shown that honest-verifier and general-verifier quantum statistical zero-knowledge are equivalent \cite{W:SIAM09}, and therefore $\operatorname{SZK} \subseteq \operatorname{QSZK}$.
The question
of whether every problem in SZK has a classical interactive proof system that is
zero-knowledge against quantum attacks is not answered yet.

Putting zero-knowledge aside for a moment, it is a well known result of interactive proof theory that $\operatorname{QIP}=\operatorname{IP}=\operatorname{PSPACE}$ \cite{S:ACM92,JJU:ACM11}.
Under a computational assumption, it was demonstrated that {\em everything provable is provable in zero-knowledge} \cite{IY:LNCS87,BGG:CRYPTO88}, i.e.
$\operatorname{IP} = \operatorname{PSPACE} = \operatorname{ZKIP}$.
Since it is also the case that honest-verifier and general-verifier quantum computational zero-knowledge are equivalent \noindent\cite{K:TOC08}, it follows that $\operatorname{QIP} = \operatorname{PSPACE} = \operatorname{QZK}$.
For a thorough review of this topic, consult \cite{TCS-068}.

Note however that this last result does not automatically yield the same implications concerning classes within $\operatorname{QIP}$. In particular, it took several more years before zero-knowledge proof systems were provided for $\operatorname{QMA}$ \cite{BJS:FOCS16}. It is also rather recent that there are classical zero-knowledge {\em arguments} for $\operatorname{QMA}$ \cite{Vidick2020}.

In the context of multiprover interactive proofs (Section \ref{SR}), it was established that $\operatorname{MIP} = \operatorname{NEXP}$ for classical provers \cite{89520} and that $\operatorname{MIP^*} = \operatorname{RE}$ (the class of all recursively enumerable languages) for quantum provers \cite{JNV:ACM21}.
Their zero-knowledge counterparts were later demonstrated in the classical case \cite{BGK:ACM88,FK:ACM94} and in the quantum case
\cite{GSY:IEEE19}.

In cryptographic applications, it is often necessary that the zero-knowledge proof not only be sound but also be a {\em proof of knowledge} \cite{BG:CRYPTO92}. This notion was defined in the quantum case in \cite{U:EUROCRYPT12} with demonstration that several existing protocols are also proofs of knowledge.





\bigskip

\centerline { $ *\: *$ }

\bigskip

\hl{Two experimental demonstrations of zero-knowledge proofs have been proposed for the problem of graph $3$-coloring. The first aims for i.t.-security by relying on multiple provers that cannot communicate with each other \cite{ABC:Nat21}, harnessing the relativistic assumptions of Section \ref{SR}. The second uses device-independent quantum randomness and a hash-based signature scheme to construct a non-interactive zero-knowledge proof without resorting to the random oracle model \cite{LZZ:PNAS23}.}

\subsection{Cryptanalysis}\label{sec:cryptanalysis}

The security of a cryptosystem ultimately depends on how secure it remains if an adversary performs the best possible attack on the system. Cryptanalysis is the field of cryptology that deals exactly with this aspect: finding the best possible attacks and determining the security level\footnote{In cryptography we say that a system offers $n$-bit security if an attacker needs $2^n$ operations/resources to break it.} offered by a given cryptosystem. Generally, cryptanalysis is mostly relevant for protocols that offer security with computational guarantees. By definition, protocols that offer information-theoretic security cannot be broken, so cryptanalysis for such protocols is relevant only (a) to determine the best practical attack that saturates the optimal theoretical attacks (e.g. \cite{CDK:PRA22}) and (b) to find potential gaps/errors in security proofs. 

Cryptanalysis is crucial for the adoption of \emph{any} cryptographic functionality. 
The belief that a system based on the (assumed) hardness of a computational task is secure relies on the confidence that the best attacks have been found. Specifically, the hardness of the problem along with the specific choice of security parameters and key lengths that are used in standardized cryptosystems should be well defined and stable against new attacks. Any further algorithmic improvements should not be expected to change this evaluation significantly.  

While we have chosen not to include post-quantum cryptography in this review, the cryptanalysis of classical and post-quantum cryptosystems requires quantum computing and quantum algorithms expertise. We therefore provide a brief discussion here since it is relevant to the audience of the review.  

\bigskip

\centerline { $ *$ }

\bigskip

Due to Shor's algorithm \cite{S:SIAM97}, we know that the most commonly used cryptosystems will break when a sufficiently large fault-tolerant quantum computer is available. Cryptanalysis, however, deals with specific numbers that arise from the parameters of each cryptosystem, and in this case, we need to determine what resources would be needed to actually break standard cryptosystems such as RSA \cite{RSA:ACM78}. This of course depends on many parameters. Indicatively: the most optimised quantum factoring algorithm, the noise level and speed of gates of the quantum hardware and the quantum error-correcting code used. A very thorough analysis was provided in \cite{Gidney2021}, where the authors show that RSA 2048 can be broken in eight hours using a superconducting quantum computer with 20 million qubits of the same quality (fidelity, coherence, gate-speed) as those from concurrent papers. \hl{However, such estimates can significantly change, purely from theoretical developments in quantum error correction and quantum algorithms optimization. In a very recent follow-up work, and making the same assumptions on the quality of qubits,  \cite{Gidney2025factor} used three novel methods to reduce the required physical qubits to break RSA 2048 to under one million -- giving a twenty-fold improvement over prior state-of-the-art.}


In symmetric cryptography, the situation is very different to that of public-key cryptography. Quantum attacks that do not use the structure of the cipher can offer a moderate quadratic speed-up due to Grover's algorithm, implying that by just doubling the key size, the same security level can be maintained. A thorough analysis for AES is given in \cite{BNS2019_AES}. 
Cryptographic hash functions such as SHA2 or SHA3 admit similar unstructured-search Grover-type attacks. This becomes relevant in applications such as proof-of-work blockchains, with the most famous example being the bitcoin. It was shown for instance that a quantum adversary cannot compromise the proof-of-work part of bitcoin as long as suitable modifications are made to what  constitutes the computational majority \cite{CGK:Quantum23}.

\bigskip

\centerline { $ *\: *$ }

\bigskip

NIST (the US National Institute of Standards and Technology) runs a multiple-round competition to establish the new, post-quantum secure, cryptographic standards. The winning cryptosystems \cite{NIST} are expected to replace existing pre-quantum cryptography in the near future. The (hard) problems on which most of these systems rely include lattice-based problems, with the main one being the Shortest-Vector-Problem (SVP), as well as other closely related problems such as Learning with Errors (LWE). A lattice may be defined  given a set of basis vectors, and by considering all the integer linear combinations of these basis vectors. If one is given a random (far-from-orthogonal) basis, the task of finding the lattice vector with the smallest length (except the zero vector) is believed to be hard for both classical and quantum computers. This applies to both the exact version and the approximate\footnote{Finding a small vector that is at most a factor $\alpha>1$ larger than the shortest vector.} versions.

Classically, there exist two approaches: sieving and enumeration. Sieving has better time-complexity -- still exponential -- but requires exponential space too. The classical record for finding SVP is for lattices of dimension $170$, while cryptosystems standardized by NIST use lattices of dimension greater than $400$. Quantum approaches exist that attempt to improve classical methods specifically by performing better enumeration techniques.
However, neither standard quantum algorithms \cite{bindel2023quantum,bai2023concrete,prokop2024grover} nor near-term quantum algorithms \cite{JCLM21_PRA,APSW23_Quantum} have been very successful, since these can barely outperform state-of-the-art classical algorithms, and are far from posing a threat to post-quantum cryptography.

One interesting thing to note, however, is that novel quantum approaches that do not closely follow their classical counterparts could drastically change this picture. There have been a few suggestions, eventually proven wrong, that came close to completely breaking lattice problems such as SVP and LWE. One recent attempt was that from \cite{chen2024quantum}, now withdrawn, that made the world believe that post-quantum cryptography, and potentially all public-key cryptography, could be quantumly broken. This highlights the need for serious quantum algorithms research for post-quantum cryptanalysis, well before we move to these cryptosystems and have quantum computers large enough to run such fault-tolerant algorithms.

\section{Multipartite quantum cryptography}\label{sec:multipartite}

In the context of quantum networks, most applications will involve more than two parties interacting, which justifies a section dedicated to multipartite quantum primitives. Naturally, the protocols we study here either extend two-party quantum primitives to the multipartite setting \cite{P:Proc21,ABD:IEEE04}, or intrinsically require more than two parties, such as secret sharing, data hiding, Byzantine agreement, randomized leader election and electronic voting. Since QKD and its multipartite extension are beyond the scope of this review, we omit the study of quantum conference key agreement and refer the interested Reader to the review from \cite{MGK:AQT20}.

In multipartite quantum cryptography, typical quantum resources involve $n$-partite graph states in which each of the $n$ parties holds one subsystem \cite{MK:Cryptography20}, or local equivalents such as $n$-qubit GHZ states of the form:
\begin{equation}
    \ket{\psi^{(n)}_{shared}} = \frac{1}{\sqrt{2}}\left(\ket{0}^{\otimes n}+\ket{1}^{\otimes n}\right).
\end{equation}
All parties may perform local quantum operations on their subsystem and classically communicate with the other parties. In the case where the distribution of the quantum resource cannot be trusted, additional protocol rounds should be added to verify the entangled state \cite{PCW:PRL12,MP:NatComm16,YDK:PRA21}. 

It is interesting to note that multipartite quantum primitives do not always require large multipartite entangled states. Throughout the following sections, we will discuss some scalable alternatives involving parallel repetitions of two-party quantum primitives or sequential applications of unitary operations by several parties on a single qubit. Most of these alternatives rely on specific assumptions on the reliability of classical broadcast channels.

\subsection{Secret sharing }
\label{sec:sharing}

Secret sharing allows a central authority to distribute a secret amongst $n$ parties, such that it is retrievable only if a subset of at least $k$ parties collaborate \cite{S:ACM79,B:IEEE79}. Any subset of fewer than $k$ parties, or any subset that was not originally designated by the authority, should gain no information about the secret.

Textbook applications include that of nuclear weapon detonation, for which it might be preferrable to split the password access between several parties. This enforces them to convince each other that the crisis is grave enough to actually detonate the weapon. Beyond its military applications \cite{S:EUROCRYPT89}, secret sharing has far-reaching applications in the digital world, from secure and reliable online data storage \cite{PK:IS09,R:ACM89} to distributed Blockchain systems \cite{KR:IEEE19} and secure federated learning \cite{XLL:IEEE20}.

Typically, secret sharing between three parties unfolds as follows: Alice sums (modulo $2$) her plaintext message with a random bit string, forwarding the resulting string to Bob and one copy of the random string to Charlie. Individually, neither Bob nor Charlie can retrieve the original message since both strings appear random. By collaborating and summing their strings, they may recover the original message. 

\bigskip

\centerline { $*$ }

\bigskip

\hl{The motivation behind the first quantum schemes was to render eavesdropping attempts on \textit{both} strings sent by Alice detectable \cite{HB:PRA99,ZZH:Polonica98}.} In the honest scenario, each of the three parties holds one subsystem of a $3$-partite GHZ state and performs a measurement in a basis uniformly chosen from $\{X,Y\}$. This is repeated for many GHZ states. After all three parties publicly declare their measurement bases (but not the outcomes), Bob and Charlie may combine their outcome strings to infer Alice's string in half of the cases. Similarly to QKD, the presence of eavesdropping (whether from an external party or from Bob/Charlie intercepting Charlie/Bob's subsystem) may be tested by public comparison of a subset of their measurements. 

Crucially, it must be noted that the order in which parties broadcast their basis or measurement outcomes may affect the security of quantum secret sharing \cite{HB:PRA99,KKI:PRA99}. Some schemes require the order to be pre-established while others should be chosen uniformly at random (either by a trusted party or using mistrustful primitives such as coin flipping). Additionally, broadcast channels should be authenticated. A detailed cryptanalysis of early quantum secret sharing was provided in \cite{QGW:PRA07}. 
 
Proposals exploiting other quantum resources followed, making use of $2$-particle entanglement \cite{KKI:PRA99}, graph states \cite{MS:PRA08}, weak laser states \cite{GXL:OptExp21}, continuous variable states \cite{KXH:PRA17,AFG:PRA19,WWH:PRA20,LLZ:PRA21,GQ:PRA19} and the fact that parties can apply, one after the other, a unitary on the same single photon \cite{ZLZ:PRA05,THZ:PRA15}. Some attacks related to the latter schemes were discussed in \cite{CTX:QIP18}, as well as in \cite{H:PRL07} regarding the single-qubit experimental demonstration from \cite{STB:PRL05}. Countermeasures were proposed in \cite{STB:PRL07}.

In fact, some of these schemes mentioned above allow for the splitting of \textit{quantum} secrets, which has important applications in the context of quantum error correction \cite{CGL:PRL99,ZLW:CPB11,GSP:QIP24,CGS:EUROCRYPT05,OTZ:PRA17,OGR:PRA23}.

\bigskip

\centerline { $*\: *$ }

\bigskip

Photonic implementations of multipartite protocols such as quantum secret sharing require three key ingredients:

\begin{enumerate}[label=(\roman*)]
        \item High-rate generation of entangled photon pairs: this is typically achieved through non-linear processes in which a single pump photon spontaneously down-converts into two single photons with lower energy exhibiting the required quantum correlations (in polarization or time for instance).
        \item High-visibility multi-photon interference: entangled pairs should be fused with one another to produce larger multipartite entangled states, requiring high photonic indistinguishability. 
        \item Low-loss channels: for $n$ parties, the probability of detecting an $n$-fold coincidence decreases exponentially as $\eta^n$, where $\eta$ is the channel transmission.
\end{enumerate} 
Given the low count rates achievable for $3$- and $4$-photon entanglement at the time \cite{PDG:PRL01,BPD:PRL99}, experimental quantum secret sharing was first demonstrated using \textit{pseudo-GHZ states}, in which $3$-partite correlations were extracted from two down-converted photons and one pump photon instead of three down-converted photons \cite{TZG:PRA01}. A full scheme was later demonstrated using $4$-qubit polarization GHZ states, in which one of the photons is used to herald a successful $3$-partite GHZ state with visibility up to $83\%$ \cite{CZZ:PRL05}, followed by a $4$-partite scheme \cite{GKB:PRL07}. As an alternative to multipartite entangled states, a scalable "single photon" alternative was demonstrated between $6$ parties, in which each party sequentially imprints a phase on a single-photon that they forward to the next party \cite{STB:PRL05}. This was later extended to a travelling single qudit experiment \cite{SET:npjQI16}. 

\hl{Alongside phase-encoded proposals for quantum secret sharing in the measurement device-independent setting \cite{FYC:PRL15,LFL:PRR23}, a weak coherent state scheme was demonstrated \cite{WSC:PRApp24}. Finally, it is worth noting the implementation of continuous-variable schemes \cite{ZYY:PRL18,LLP:npjQI23,QCM:PRR24} and secret sharing protocols involving quantum secrets \cite{LZC:PRL16,BMH:NatComm14}. For a more detailed discussion of quantum primitives aiming to protect quantum information, we refer the Reader to Section \ref{sec:quantumdata}.}

\subsection{Data hiding}
Introduced in \cite{TD:PRL01,DL:IEEE02}, quantum data hiding is a natural strengthening of quantum secret sharing (Section \ref{sec:sharing}) to the local operations and classical communication  (LOCC) model. By this it is meant that any subset of fewer than $k$ parties, or in general any subset that was not originally designated by the authority, should gain no information about the secret despite the fact that the $n$ parties can exchange arbitrary classical information about the secret. Even if the parties publish their classical knowledge to each other, it should remain impossible for any subset of fewer than $k$ parties to correlate with the secret contained in the $n$ quantum shares. First developed for classical secrets in \cite{TD:PRL01,PhysRevLett.89.097905}, it was later extended to quantum data in \cite{DL:IEEE02,10.1023.A.1026013201376}, and shown that perfect data hiding is impossible; the proposed protocol allows the sharing parties to obtain an exponentially small amount of information about the secret.

For two-party hiding, the basic protocol goes as follows: to hide a classical bit $b$, generate a large state involving several random Bell states from the set $\{\ket{\phi^+},\ket{\phi^-},\ket{\psi^+},\ket{\psi^-}\}$, under the only constraint that the number of states $\ket{\phi^-}$ modulo 2 equals $b$. The two parties involved get half of each Bell state. It is a well known fact that the four randomly selected Bell states are not distinguishable in LOCC. The authors showed that from an LOCC point of view, the state $\rho_0$ of a hidden bit $0$ is exponentially close to the state $\rho_1$ of a hidden bit $1$. Therefore, the amount of information available about $b$ in LOCC is exponentially small in the number of Bell states involved. On the other hand, if the parties recombine their Bell states pairwise, then a simple Bell measurement will identify each Bell state accurately, and so the number of states $\ket{\phi^-}$ can be determined perfectly. Note that it suffices to distinguish the state  $\ket{\phi^-}$ from the other three Bell states in order to count them. In practice, this is a simpler task.

An application to bit commitment (Section \ref{sec:qbitcomm}) was offered in \cite{DL:IEEE02}, analogous to that of Section \ref{SR} but only assuming that quantum communication is impossible. But since quantum teleportation reduces quantum communication to classical communication supplemented by entanglement (Section \ref{sec:teleportation}), the latter assumption requires that the parties involved do not pre-share entangled states.

\bigskip

\centerline { $*$ }

\bigskip

All operations required to implement the scheme above are currently available in the laboratory. However, in most applications, the limiting factor will be that each party must store their half of the Bell states for rather long times between the hiding and  disclosing of secrets.

\subsection{Byzantine agreement}\label{sec:byzantine}

The Byzantine agreement problem has widespread applications, especially in distributed computation where the processors involved might be faulty or dishonest. Initially formulated by \cite{LSP:ACM82}, it may be stated as follows: several divisions of the Byzantine army, each led by an independent general, are camped around an enemy city. Conditioned on the enemy's actions, all Byzantine generals should agree on a common plan of action (eg: attack or retreat), even in the presence of potential traitors attempting to scramble the plan. Given the geometry of the situation, generals can only communicate via messenger. In the case of one commanding general sending an order to their $(n-1)$ lieutenant generals, the following conditions must be met for the action plan to succeed \cite{LSP:ACM82}:

\begin{enumerate}[label=(\roman*)]
        \item All loyal lieutenants obey the same order (consistency).
        \item If the commanding general is loyal, then every loyal lieutenant obeys the order they send (validity).
\end{enumerate} 
Note that, if condition $(ii)$ holds, then condition $(i)$ immediately holds. However, no assumption can be made on whether the commanding general is honest or dishonest. When generals communicate through pairwise authenticated channels, there exists no solution to the Byzantine problem unless the total number of generals satisfies $n> 3k$, where $k$ is the number of traitors \cite{PSL:ACM80}. In the simple case of $n=3$, this implies that not a single traitor can be tolerated.

\bigskip

\centerline { $*$ }

\bigskip

A cheat-sensitive version of Byzantine agreement was proposed in the $3$-general setting by adding quantum communication to the pairwise authenticated classical channels \cite{FGM:PRL01}. In the honest protocol, the commanding general distributes $j$ copies of a $3$-partite entangled state $\ket{\psi}$ consisting of qutrits spanned by $\{\ket{0},\ket{1},\ket{2}\}$:
\begin{equation}
\ket{\psi} = \frac{1}{\sqrt{6}} \left(\ket{012} + \ket{120} + \ket{201} - \ket{021} - \ket{102} - \ket{210}\right)
\label{eq:byzantine}
\end{equation}
Intuitively, the agreement can be reached since measuring all qutrits in the same basis should yield three strictly distinct outcomes. After announcing the bit $b$ they wish to broadcast to the other generals through classical authenticated channels, the commanding general measures all $j$ copies of their qutrit in a pre-agreed basis and keeps a record of which measurement outcomes match the value of $b$. They send the corresponding labels to the lieutenants, who can check individually (and with respect to one another) that all three measurement outcomes are strictly different. When all generals agree, the protocol is successful and they pursue action plan $b$. 

In the dishonest setting, at least one of the generals will declare some inconsistency, which can yield a useful outcome in some cases. When both lieutenants announce an inconsistency however, the protocol must abort \cite{FGM:PRL01}. Cheat-sensitive quantum schemes were later extended to tolerate $n>k$ faulty generals under various assumptions, either placing constraints on the adversary's computational power or assuming pairwise channels secured by QKD for instance \cite{FGH:ACM02,IG:PRA04,SKS:Entropy20}. 

\bigskip

\centerline { $*\: *$ }

\bigskip

With the goal of photonic demonstration in mind, cheat-sensitive schemes using more practical resources were designed, replacing the tripartite entangled qutrit state from Eq. (\ref{eq:byzantine}) with $4$-partite entangled qubit states \cite{C:PRA03,GFB:Quantum24}, multipartite entangled states in the continuous variable setting \cite{NRA:PRA08,FSZ:IJTP19}, $3$-partite GHZ states \cite{QZL:IS23} and trusted Bell pair sources \cite{AS:AppSci23}. 

After the first photonic implementation was provided using a $4$-photon polarization-entangled state \cite{GBK:PRL08}, a proof-of-concept version was implemented using a single (i.e. non-entangled) qutrit state, on which each party imprints a phase before forwarding it to the next party \cite{SET:npjQI16}. The qutrit's three levels are encoded onto spatial modes in a $3$-arm Mach-Zehnder-type interferometer. After the last party measures in the Fourier basis, all parties can check for consistency by declaring their respective applied phases. This setup has the interesting benefit that the number of single-photon detectors does not scale with the number of parties $n$. \hl{Recently, an i.t.-secure Byzantine agreement framework breaking the $n > 3k$ classical limit was developed by using a decentralized three-party relationship of quantum digital signatures (Section \ref{sec:signatures}) to eliminate channel independence, along with a recursive structure to extend it to multipartite scenarios \cite{CRY:Research23}. By utilizing an integrated polarization-entangled photon pair source and optical fiber spools of several kilometers, this  framework was experimentally demonstrated \cite{JQW:SciAdv24}, employing the one-time universal hashing signature scheme from \cite{LXC:PRAppl23}.}

\subsection{Randomized leader election}\label{sec:leader}

In Section \ref{sec:byzantine}, we discussed how quantum resources can help achieve consensus in the presence of faulty or dishonest processors. Here, we study a related consensus primitive in which the $n$ parties must agree on a random unbiased bit or leader. Electing a random leader amongst a consortium of dishonest parties naturally extends the primitive of weak coin flipping (Section \ref{sec:WCF}) to the multipartite setting. Using classical resources, secure leader election is possible under computational assumptions \cite{C:STOC86}, and i.t.-secure versions exist assuming a majority of honest parties \cite{BN:SIAM00,F:FOCS99}. For distributed computing, this implies that $n$ processors can agree on an unbiased leader only if $n> 2k$, where $k$ is the number of faulty processors.

In most protocols, the parties can flip a private random coin and communicate in an asynchronous way, broadcasting bits one after the other through authenticated channels. This may proceed according to a simple \textit{pass-the-baton} game \cite{S:SIAM89}: the player initially holding the baton randomly selects the next player to whom they pass the baton. In turn, this player selects another player from the subset of players who have not yet participated. The last player to hold the baton becomes the leader. Evidently, the security analysis must allow the subset of dishonest parties to collude in order to collectively increase their chance of being elected.

\bigskip

\centerline { $*$ }

\bigskip

Based on the milestone result of quantum weak coin flipping with arbitrarily small bias (Section \ref{sec:WCF}), i.t.-secure leader election becomes possible in the quantum realm with no assumption on the number of dishonest parties: in the honest setting, each party gets elected with probability $\frac{1}{n}$, while in the dishonest setting, an honest party will be elected with probability $\frac{1}{n}-\epsilon$, where $\epsilon>0$ is the arbitrary protocol bias \cite{G:QIP17,AS:NJP10}. Intuitively, this is achieved by performing a knockout tournament of two-party weak coin flips with bias $\epsilon$, where the winner of each flip moves on to the next round until a single party wins the tournament. Note that implementing such a tournament may require additional assumptions such as a uniformly random initial labeling of the players and authenticated classical channels.

This idea was in fact proposed earlier than the discovery of quantum WCF with arbitrarily small bias, in the context of multiparty quantum coin flipping \cite{ABD:IEEE04}. Despite the high biases of explicit quantum WCF schemes at the time, optimal bounds were derived by imposing a penalty on parties who get caught cheating: this parameter could be tuned at each round, assuming that dishonest parties are more likely to risk cheating (with a non-zero probability of getting caught) at later rounds.  The less trivial case of performing a tournament with an odd number of parties was studied in \cite{G:QIP17}.

\bigskip

\centerline { $*\:*$ }

\bigskip

Since all theoretical proposals so far rely on the ability to perform secure quantum weak coin flipping, the practical possibilities and challenges are closely linked to those discussed in Section \ref{sec:WCF}.

\subsection{Electronic voting}

Formally defining the security properties of electronic voting schemes is a challenging task: many physical assumptions underlying paper-based schemes, such as isolated voting booths and trusted tallies, must be replaced by rigorous cryptographic assumptions. In a similar way that bit commitment requires somewhat conflicting security properties like binding and hiding (Section \ref{sec:qbitcomm}), ideal e-voting should achieve simultaneous properties that are not necessarily compatible with one another \cite{CFP:TTE10}:

\begin{enumerate}[label=(\roman*)]
        \item All parties, including passive observers, can check the validity of a vote, the computation of the tally, and the list of participating voters (universal verifiability).
        \item The content of a vote must be hidden from all other parties including the tallier (voter privacy) or the voter's identity must be hidden from all other parties including the tallier (voter anonymity).
        \item A voter neither obtains nor is able to construct a receipt proving the content of their vote (receipt-freeness).
    
\end{enumerate} 

Classical e-voting schemes achieve some of these properties by relying on computational assumptions \cite{CCC:IEEE09,A:CSS08,KZZ:EUROCRYPT15}, some conjectured hard against quantum computers \cite{CGG:SIP16,KCR:FC21}. However, i.t.-secure protocols were also proposed in the presence of trusted election authorities \cite{CFS:EUROCRYPT96} or under the assumption of simultaneous broadcast channels with a dishonest majority \cite{BT:ASIACRYPT07}. 

\bigskip

\centerline { $*$ }

\bigskip

Following the classification established in \cite{ALK:ACM21}, i.t.-secure quantum proposals may be divided into several categories, mainly dependent on how the voting ballot is implemented and verified. Even within a given category, one should keep in mind that trust assumptions regarding the election authorities, the talliers, the voters and the communication channels may vary.

The first category consists of dual-basis measurement-based protocols, in which untrusted copies of an entangled state are distributed and shared between the voters \cite{HWL:PRA14,WYG:PRA16}: one subset of these copies is used as the empty ballot while the remaining subset is used for verification. The latter phase consists of one voter challenging all others to measure and simultaneously broadcast their measurement outcomes in either the computational or Fourier basis, and checking for the appropriate correlations.

The second category refers to schemes based on conjugate coding (Section \ref{sec:conjugate}), where each voter's blank ballot is mapped onto a tensor product of BB84 states to which unitaries are applied by the voters \cite{OST:NTT08,ZY:Arx13}. 

Finally, the third and fourth categories apply to referendum-type elections, where the output of the vote is a binary yes/no value. In one case, a single ballot qudit travels from voter to voter, each applying a unitary conditioned on the value of their vote \cite{VSC:PRA07,BBH:PRA11,HZB:PLA06,LZ:OR08}. In the other case, the ballot is instead distributed amongst the voters using a large entangled state, each voter additionally receiving a "yes" and a "no" qudit that they send back to the trusted tallier \cite{BBH:PRA11,HZB:PLA06,VSC:PRA07}. 

In fact, while providing formal definitions of properties $(i)$ and $(ii)$ and attempting to unify security definitions for quantum e-voting, security flaws were discovered for all the aforementioned schemes \cite{ALK:ACM21}. A year later, \cite{CDK:PRApplied22} proposed an i.t.-secure scheme without the need for simultaneous broadcast channels or trusted election authorities, combining the protocol from \cite{WYG:PRA16} with the multipartite entanglement verification schemes from \cite{PCW:PRL12,MP:NatComm16,YDK:PRA21}, thus closing one security loophole.

\section{Quantum information protection}\label{sec:quantumdata}

Following the vision of a future quantum internet \cite{WEH:Sci18,S:NatPhot17} in which all parties exchange quantum messages through direct optical links or quantum teleportation \cite{HPB:Nature22,SMC:NatPhot16}, cryptographic schemes should be tailored to quantum data rather than classical inputs. 

This section discusses the theoretical progress and relevant experiments (when available) covering the protection of quantum information, including private quantum channels, authentication of quantum messages and anonymous quantum communication. Topics related to the privacy and secure delegation of quantum computations will be treated in Section \ref{sec:computing}.

\subsection{Private channels}\label{sec:privchannel}

Since the early days of quantum teleportation \cite{PhysRevLett.70.1895}, it was realized that this protocol in fact constitutes an encryption method for the quantum state that is teleported. In this scheme, the encryption key is the entangled (Bell state) pair shared by the sender and receiver, while the ciphertext is the uniform two-bit classical message sent by the sender in the opening phase. Since these bits are uniform and independent of the teleported state, this scheme is perfectly secret in the sense of Shannon, and known as the one-time quantum pad.

An alternative encryption scheme uses classical keys and produces a quantum ciphertext \cite{AMT:SFCS00,BR:PRA03}. If $\ket{\psi}$ is a qubit to be encrypted, then the encryption scheme uses a pair of random bits $(a,b)$ as the classical key and $\ket{\psi_{ciph}} = X^a Z^b \ket{\psi}$ as the quantum ciphertext. For any party who is completely ignorant about the key, \hl{the security guarantee is that the mixed state associated with $\ket{\psi_{ciph}}$ is indeed the completely mixed state, i.e. appears random to that party}. Since the latter is independent of the input $\ket{\psi}$, this scheme is also perfectly secret in the sense of Shannon, and known as the quantum Vernam cipher.

\subsection{Authentication}
In cryptographic terms, authentication is a process by which two users can make sure that the messages they exchange really come from each other. A message authentication code (MAC) is a combination of two algorithms $\operatorname{Mac, Vrfy}$. The former maps messages and keys to tags and is used to authenticate a message $m$. Using key $k$, the sender computes $t \leftarrow \operatorname{MAC}_k(m)$ and sends $(m,t)$ in the public communication channel. The latter function is used to validate that the sender knew $k$ when they created the tag. Upon receipt of $(m,t)$, the receiver computes $v \leftarrow \operatorname{Vrfy}_k(m,t)$. If $v$ is true, the receiver accepts the message as valid and rejects the message otherwise. In most cases, $\operatorname{Vrfy}_k(m,t) = (t=\operatorname{MAC}_k(m))$ is used as a verification procedure, which is just checking that the received $t$ is indeed the correct tag for $m$. 

\hl{Informally, we consider a MAC to be secure if existential unforgeability against a chosen message attack holds, where:}
\begin{enumerate}[label=(\roman*)]
        \item \hl{\textit{existentially unforgeable} means that the attacker is not able to generate a valid MAC tag on any message, without being in possession of the key $k$.}
        \item \hl{\textit{chosen message attack} means that the attacker is able to obtain MAC tags on any other message before performing the attack.}
\end{enumerate} 

\bigskip

\centerline { $*$ }

\bigskip

In a quantum setting, these notions have to be adapted because the sender does not have several copies of $\ket m$ and thus it cannot both be used to create $\ket t$ and sent alongside with it. In this case, we can think of the authentication process as a quantum mapping
of a message $\ket{m}$ and key $k$ to its authenticated form $A_k\ket{m}$, for some isometry $A_k$. The receiver will use a measurement dependent on $k$ to determine the validity of the message and recover the message itself.

As mentioned in \cite{BCG:IEEE02},
authentication of quantum messages is rather straightforward using quantum secret keys.
If a sender and a receiver share maximally entangled states, they can use them for quantum teleportation (Section \ref{sec:teleportation}).
All that is needed to authenticate the transmission of the quantum state is an extra classical key to authenticate the classical bits transmitted during teleportation.
The problem is more interesting when seeking a solution that solely uses classical keys.

The approach proposed in \cite{BCG:IEEE02} is to encrypt the message $\ket m$ as in Section \ref{sec:privchannel}, mix it with some fixed maximally-mixed check qubits and apply a quantum error-correcting code. The quantum encryption of the message makes it indistinguishable from the check positions. The check positions are present to detect any significant modification of the transmission whereas the quantum error-correcting code is present to recover from any minor modification of the message. 

In the same work, the following statements are proven:

\begin{enumerate}[label=(\roman*)]
        \item Any quantum authentication scheme must encrypt the message it is authenticating.
        \item The minimal number of key bits used for a message of $m$ qubits is $2m$.
        \item Publicly-verifiable authentication (digital signatures) of quantum messages is impossible.
\end{enumerate} 

Note however that a number of reasonable scenarios have been proposed, in which the latter impossibility result may be bypassed \cite{BS:arx23}.

\subsection{Anonymous communication}\label{sec:anonymous}

While some primitives ensure the confidentiality of messages, anonymity ensures the confidentiality of identity. In scenarios such as online auctions, one party may wish to broadcast a message to a set of network users while keeping their own identity secret \cite{SA:IH00}. Using classical resources, i.t.-secure anonymous broadcasting is possible assuming pairwise authentic private channels and reliable broadcast channels, as in the Dining Cryptographers Problem \cite{C:JOC88}, or using trusted servers. Depending on the exact application, some protocols may only require sender or receiver anonymity, while others may require both (full anonymity). \hl{Informally, anonymity for a sender (resp. receiver) is guaranteed if the communication taking place throughout the protocol does not change the a priori uncertainty about the identity of the sender (resp. receiver). }  

In other applications such as electronic voting, a subset of communicating parties may wish to protect both their identities \textit{and} the contents of their message. Such anonymous messaging schemes are possible with i.t.-security under the previously stated assumptions and a majority of honest parties \cite{RB:ACM89} or under an arbitrary number of honest parties with the price that a single cheating party may cause the protocol to abort \cite{BT:ASIACRYPT07}. Computationally-secure schemes were also studied, allowing to drop the requirement for a reliable broadcast channel \cite{WP:EUROCRYPT89} or using public-key cryptography \cite{C:sec03}. 

\bigskip

\centerline { $*$ }

\bigskip

 \begin{figure*}
    \begin{center}
        \includegraphics[width=1\linewidth]{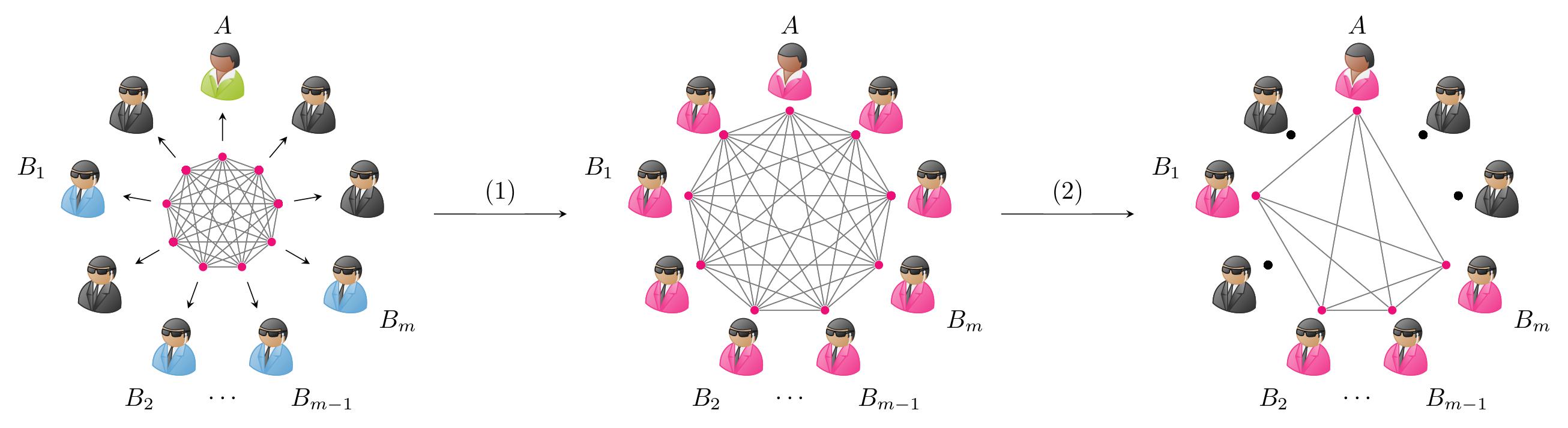}
        \caption{\textbf{Anonymous multipartite entanglement generation.} Initially proposed as a subroutine for quantum conference key agreement and taken from \cite{HJP:PRXQuantum20}, this diagram pictures a sender $A$ distributing an $n$-partite GHZ state to herself and $(n-1)$ other parties. To the left of arrow $(1)$, $A$ uses a classical protocol to anonymously notify a subset of $m<n$ blue receivers $\{B_1,B_2,...,B_m\}$ that she will later establish a secret key with them \cite{BT:ASIACRYPT07}. To the right of arrow $(1)$, all parties are colored in pink since the shared $n$-partite GHZ state is agnostic of the receivers' identities. To reach the right of arrow $(2)$, all unnotified parties measure their subsystem in the $X$ basis and broadcast their measurement outcomes in a random order (or through a simultaneous broadcast channel), while the notified parties declare random bits to preserve their anonymity. $A$ applies a phase flip to her subsystem if the parity of the unnotified parties' outcomes is odd, yielding the desired $(m+1)$-partite GHZ state.}
        \label{fig:anonymous}
    \end{center}
\end{figure*}

Anonymous broadcasting of classical information with quantum states was first proposed assuming Bell pairs shared between neighbouring parties \cite{B:thesis02} or an $n$-partite GHZ state shared between all parties, along with reliable classical broadcast channels \cite{CW:LNCS05}. Note that these in principle allow for i.t.-secure anonymity without any assumption on the number of honest parties. Intuitively, the second scheme relies on the fact that any party may locally flip the phase of the GHZ state conditioned on the bit $b$ they wish to broadcast, leaving the total state invariant regardless of who applied the phase:
\begin{equation}
\frac{1}{\sqrt{2}}\left(\ket{0}^{\otimes n}+\ket{1}^{\otimes n}\right) \xrightarrow[\text{bit $b$}]{\text{broadcast}}\frac{1}{\sqrt{2}}\left(\ket{0}^{\otimes n}+{(-1)}^b\ket{1}^{\otimes n}\right).
\end{equation}

For the broadcast of quantum information, two communicating parties may establish an anonymous entanglement link through local operations on a larger network state \cite{CW:LNCS05}. This allows one of them to anonymously teleport a quantum message, such as a qubit, to the other party, while the classical data needed to complete the teleportation is sent using an anonymous protocol for classical bits. A simple example of how to anonymously turn an $n$-partite GHZ state into a smaller GHZ state is provided in FIG. \ref{fig:anonymous}.

The requirement for a trusted quantum resource was later dropped in the case of sender anonymity \cite{BS:IEEE07} and full anonymity detecting potentially disruptive adversaries \cite{BBF:Asiacrypt07,YHS:SciRep16}. More practical, noise-tolerant schemes were then designed either using W-states \cite{LMW:PRA18} or GHZ states combined with multipartite entanglement verification schemes \cite{UMY:PRL19}. 

\bigskip

\centerline { $*\ *$ }

\bigskip

Recently, the distribution of pairwise secret keys using QKD was performed over an $8$-user deployed quantum network, allowing the implementation of i.t.-secure classical anonymous protocols \cite{HKA:npjQI22}. On the quantum information side, anonymous variations of quantum conference key agreement, in which a sender wishes to anonymously establish a secret key with $m$ other parties from a total of $n$ parties, were designed \cite{HJP:PRXQuantum20,JHE:Quantum23,GMJ:PRXQuantum22}. Some of these protocols were implemented using $4$-photon linear cluster states \cite{RBJ:PRR23} and multipartite polarization-entangled GHZ states ranging from $4$ \cite{THJ:NJP21} to $6$ parties \cite{WHG:Optica24}.

\section{Secure quantum computation}\label{sec:computing}

In the modern digital world, the need to delegate expensive computations either to a single powerful server or to multiple entities/cloud is common. This is usually performed to enable access to more powerful computing resources, or to store and access large data using light devices such as a mobile phone. Clients using such services typically require extra guarantees, the main ones being privacy (of their inputs, outputs, or even of the programs running) and the integrity/correctness of the computation\footnote{In a commercial setting, programs can be proprietary, requiring extra privacy guarantees such as hiding the computation performed. At the same time, it is essential for clients to be able to verify the correctness before paying for services. Privacy is also important in other scenarios, for example in computations using medical or financial data.}. In the classical world, there exist many variations of secure delegated computing, each of them addressing slightly different situations, dependent on the number of parties and their access permissions, the communication allowed between the parties, the efficiency or generality of the way in which computations are performed, etc.

With the anticipated development of fault-tolerant quantum computers that will significantly outperform classical ones, it would be desirable to offer the same privacy and integrity guarantees while performing a \emph{quantum} computation. Moreover, it appears very likely that, at least in the near and medium term, quantum computing hardware will be located at only a few companies or national facilities, and the majority of commercial and scientific activity will need to be performed via delegation. The field of secure quantum computing deals exactly with this type of question. We refer the Reader to earlier reviews on private quantum computing \cite{F:npjQI17} and verifiable quantum computing \cite{GKK2019verification}.

In this section, we will present the main types of secure quantum computing: with clients that can use quantum communication; with clients that can only communicate and compute classically; with clients that communicate only once; with multiple parties involved. In each scenario, we will study versions that are private only, as well as private \textit{and} verifiable.

\subsection{Blind computation}\label{sec:blind}

The first scenario involves one server with a (universal) quantum computer and one client with limited computational capabilities. The client wishes to delegate their computation \hl{in such a way that their input, output and computation remain perfectly hidden from the server. In other words, we have that}

\begin{enumerate}[label=(\roman*)]
        \item \hl{the distribution of classical information obtained by the server during the client's instructions and from the output of the computation}
        \item \hl{the quantum state of the systems received by the server during the protocol}
\end{enumerate}
\hl{must appear independent of the client's input.}

In most cases, we assume that the client can only prepare and send single qubits to the server. This setting was first termed \emph{blind quantum computing} (BQC) in \cite{arrighi2006blind}, where a first, limited, example of delegation of some quantum computation was given. The first protocol that allowed to delegate universal quantum computation with minimal resources was introduced in \cite{BFK2009universal}. 

\bigskip

\centerline { $ *$ }

\bigskip

The blind quantum computing protocol given in \cite{BFK2009universal} forms the basis of many subsequent works and it is therefore instructive to explain it, following FIG. \ref{fig:ubqc}. It starts with a single round of quantum communication happening ``off-line'', i.e. before the client decides which computation they wish to delegate. In this off-line phase, the client prepares and sends multiple qubits to the server, each prepared in one of the states 
\begin{equation}
\ket{+_\theta}=\frac{1}{\sqrt{2}}(\ket{0}+e^{i\theta}\ket{1}),
\label{eq:plusblind}
\end{equation}
where the angle $\theta$ for each position is chosen uniformly at random (and independently) by the client and kept secret from the server. The protocol uses measurement-based quantum computing (MBQC), which is a universal model of quantum computation in which the computation is performed by consecutive single-qubit measurements and the outcome of the previous measurement defines the next measurement setting, on a generic large entangled ``resource'' state that is independent of the computation \cite{MBQCreview2009measurement,MBQCreview2021measurement}). The single-qubit measurement angles $\phi$ determine the unitary implemented by an MBQC ``pattern''. 

The basic trick to make the protocol blind is the following observation. Within MBQC, the effect of measuring a single qubit in the basis $\ket{\pm_\phi}$ on the standard resource state -- where qubits are initialized in the $\ket{+}$ state and then entangled with $\wedge Z$ gates, is identical to measuring a resource state (that is similarly rotated) in the basis $\ket{\pm_{(\phi+\theta)}}$ -- where the qubits are prepared in the $\ket{+_{\theta}}$ state prior to getting entangled. The client, by deciding and hiding the pre-rotation angle $\theta$, can instruct the server to measure in the $\ket{\pm_{(\phi+\theta)}}$ basis, without revealing any information on the angle $\phi$ -- essentially $\theta$ ``one-time-pads'' the angle $\phi$. However, as stated above, the angles $\phi$ fully determine the unitary performed, and thus the server is completely blind as to the computation they are performing. The actual instructions have the client asking the server to measure in an angle with an extra $r \pi $ flip, where $r$ is a random bit that is secret to the server. The purpose of this bit is to hide (one-time-pad) the measurement outcomes.

The full protocol includes further subtleties, the main one being that it requires communication (fully classical) between the client and the server during the computation. In MBQC, the qubits of the resource state are measured one-by-one, and the measurement results of previous rounds affect the measurement angles of the subsequent ones. In our setting however, the server that knows the measurement outcomes does not know the actual measurement angles due to blindness, and thus cannot correct the future measurements on their own. This is why the client needs to communicate the corrections during the computation (hence the need for an interactive protocol), and to do so in a way that the new ``corrected'' measurement angles do not leak further information -- something that becomes possible by introducing the extra randomness $r$ in their instructions.

\begin{figure*}
    \begin{center}
        \includegraphics[width=0.6\linewidth]{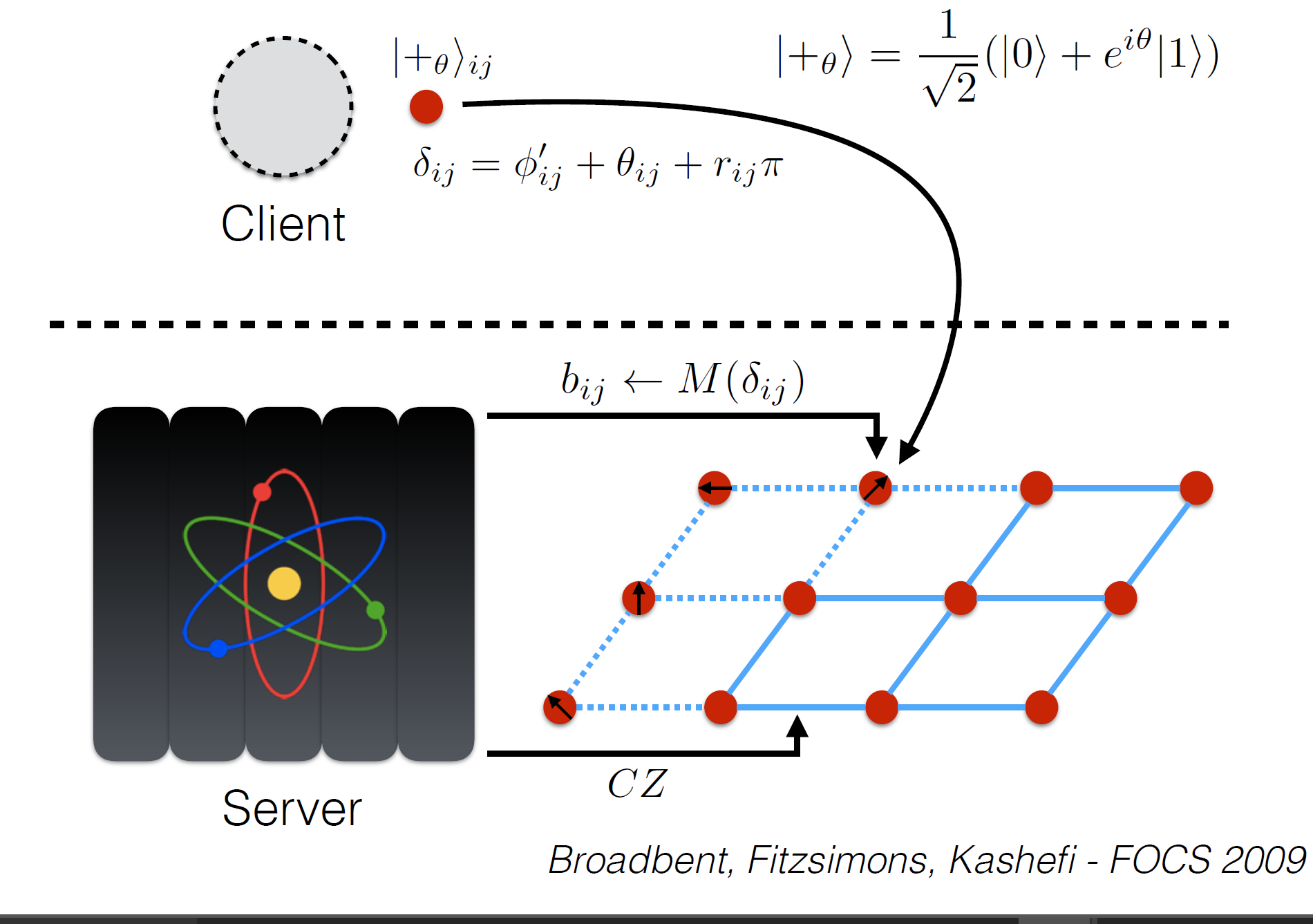}
        \end{center}
        \caption{\textbf{Universal blind quantum computing}. First proposed in \cite{BFK2009universal}, this figure was taken from \cite{GKK2019verification}, and reproduced with permission from the authors. \hl{For each position $(i,j)$ in the rectangular cluster state, the client prepares a random qubit of the form $\ket{+_\theta}_{ij}$ as defined in Eq. (\ref{eq:plusblind})}, sends it to the server, and requires them to prepare the rotated graph state. \hl{The client then instructs the server to perform a measurement $M(\delta_{ij})$ in a basis with angle $\delta_{ij}$, which cancels the pre-rotation $\theta_{ij}$ and performs the desired computation dictated by $\phi_{ij}$ in a blind way. Note that the extra term $r_{ij}\pi$ essentially one-time pads the measurement outcome $b_{ij}$.}}
        \label{fig:ubqc}
\end{figure*}

\bigskip

\centerline { $ *\:*$ }

\bigskip

Small-scale blind implementations of the Deutsch-Josza and Grover search algorithms were demonstrated using four-qubit polarization-entangled cluster states \cite{BKB:Science12}. Over a thousand different $4$-qubit measurements describing the possible branches of each computation were performed by the server, as an equivalent to the active feedforward operations from \cite{PWT:Nature07}. More recently, a photonic demonstration tackling the challenge of multiple clients was performed  \cite{PLL:NatCom23}, with tight links to the notion of multiparty computation presented in Section \ref{sec:MPC}. Based on the linear network architecture from \cite{DHM:Arx23}, this experiment has the practical benefit that each client sequentially applies only a single-qubit rotation on an untrusted travelling qubit. Since each rotation is applied privately, this keeps their inputs secret from other clients, while nevertheless establishing a joint secret between them. 

From a practical security perspective, the effect of multiphoton emission as a potential leakage mechanism of the blind phases, along with the analysis of loss-dependent attacks, remained an open question. A first step towards the study of practical blindness was performed by introducing the notion of $\epsilon$-blindness \cite{DKL:PRL12}, quantifying the maximal probability of successfully distinguishing
between the actual protocol and the ideal protocol from \cite{BFK2009universal}. The work from \cite{DKL:PRL12} already proposed the use of low-cost sources such as attenuated lasers to prepare the blind qubits, and its efficiency was significantly improved by adding a decoy-state subroutine, along with an experimental demonstration at telecom wavelengths \cite{JWH:PRL19}. Other studies discussed the implementation of BQC in the presnce of noisy and lossy channels \cite{TFI:PRA16,SZ:PRA18}. 

In terms of MBQC computation, there is no overhead in running the blind and the non-blind computations. On the other hand, the client is required to send qubits, and those should be included in the quantum computation on the server's side. This is straightforward if both communication and computation are photonic, but challenging if the computation is performed on a different hardware (see experiments discussed in Section \ref{sec:verifiable}). 

\subsection{Verifiable computation}\label{sec:verifiable}

Once blindness has been achieved, one can modify the protocols from Section \ref{sec:blind} to give the client the possibility of testing/verifying whether the computation was performed correctly. The first two schemes achieving this were proposed in the MBQC model in \cite{FK2017unconditionally} and in the circuit model in \cite{broadbent2018verify}. In both protocols, the server needs to prepare and send single qubits. An alternative protocol in which the client needs to \emph{measure}, rather than prepare, single qubits, was also introduced \cite{morimae2016measurement}. One important difference between \cite{morimae2016measurement} and \cite{FK2017unconditionally,broadbent2018verify} is that, in the former, the client requires their quantum ability to be online (i.e. when the computation takes place), while in the latter, the quantum capabilities and actions of the client happen offline (i.e. before the computation is decided). Finally, a verifiable protocol that uses neither trusted preparations nor trusted measurements was introduced \cite{kashefi2024verification}, where the security is based on trusted single-qubit rotations.

 Following FIG. \ref{fig:vbqc}, the main idea in \cite{FK2017unconditionally} was the concept of ``traps''. The client embeds the (blind) quantum computation within a larger computation, where some parts of the larger computation are, in an honest execution, simple to compute and pre-determined. These parts are called traps. Once the client receives the outcomes of the blind computation, they check whether all traps have the correct (pre-determined) outcome, and if this is the case they can then accept the computation as correct. Two elements are crucial to this construction. First, it needs to be performed in such a way that the server cannot distinguish which parts are traps and which parts are the true (delegated) computation. Moreover, any deviation that would cause an error in the computation should be detected by the traps with high probability. In other words, one needs a protocol in which corrupting the computation while avoiding the traps happens with negligible probability. The second important element is the overhead, i.e. how much larger, compared to the blind-only version, does the verifiable computation version need to be.

In \cite{FK2017unconditionally}, the traps were isolated single qubits of the form $\ket{+_\theta}$, where the isolation was achieved by surrounding trap qubits with ``dummy'' qubits prepared in the computational basis $\{\ket{0},\ket{1}\}$. The trap qubits are not entangled with the rest of the resource state because their dummy neighbours are in the computational basis and the $\wedge Z$ gates do not entangle them. Therefore, 
measuring trap qubits in the suitable basis returns a deterministic result. To hide the positions of the traps, and to ensure that the probability of accepting a corrupted computation becomes negligible, quantum error correcting codes and a graph that scales quadratically compared to the non-verifiable version are used.

\begin{figure*}
    \begin{center}
        \includegraphics[width=0.55\linewidth]{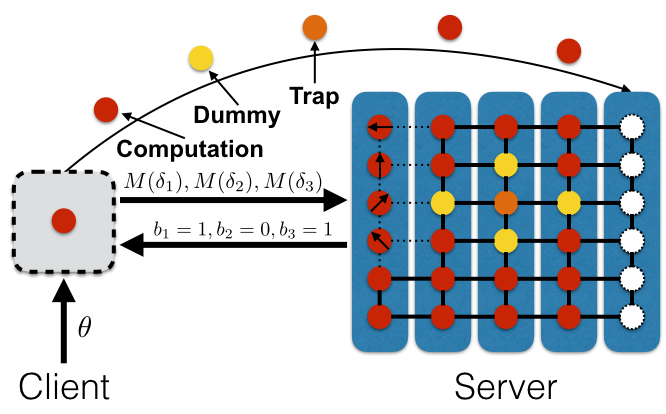}
        \end{center}
        \caption{\textbf{Verifiable blind quantum computing}. First proposed in \cite{FK2017unconditionally} this figure was taken from the talk prepared for presenting \cite{GKK2019verification}, and reproduced with permission from the authors. \hl{As defined in Eq. (\ref{eq:plusblind}), the client generates qubits in a random basis string dictated by $\theta$.} Red qubits denote the computation qubits, yellow qubits are dummy qubits that disentangle their neighbors from the graph and orange qubits are trap qubits. Trap qubits produce determinstic measurement outcomes, so the client can check if the received outcome $b_i$ matches the honest execution. \hl{For each qubit $i$, the client orders the server to perform measurement $M(\delta_i)$ in a basis with angle $\delta_i$, and to send the outcome $b_i$ back to them.} }
        \label{fig:vbqc}
\end{figure*}


\bigskip

\centerline { $ *$ }

\bigskip

There are three major issues that made the original proposal from \cite{FK2017unconditionally} hard to implement on a large scale. The first is the extra overhead in qubits and computation in order to achieve blind verification: in order to hide which part of the graph is ``computation'' and which is ``traps'', a quadratically larger graph state is required. In \cite{KW2017optimised}, a new type of graph state known as the dotted-triple graph was introduced in order to reduce the overhead to a constant factor. In the special case of classical outputs, this was further reduced to a constant number of repetitions of the non-blind MBQC computation in \cite{leichtle2021verifying} .

The second issue is the robustness of the protocol to noise and its compatibility with quantum error correction. Existing devices are noisy and, unless the protocols are robust to noise and can be modified to incorporate fault-tolerant quantum computation, they cannot be used in practice. In \cite{GKW:NJP15}, the robustness of the scheme from \cite{FK2017unconditionally} to imperfect single-qubit preparations was proven. Combined with the rigidity of the CHSH-game proven in \cite{RUV2013classical}, this led to a device-independent VBQC protocol. Finally, it was shown that, with some care, verifiable blind quantum computation can be made compatible with fault-tolerant computation without any further leakage of information during the error-correction process \cite{gheorghiu2018simple}.

The third issue is the compatibility of different quantum hardware technologies. It is clear that all protocols involving quantum communication are essentially based on photonic qubits. On the other hand, quantum computing platforms have been, so-far, mostly dominated by matter qubits, whether these are ion traps, superconducting, neutral atoms, or silicon-based. The protocols presented in this section require both quantum computation (the target computation) and quantum communication between the client and the server. Developing hybrid photon-matter systems might become a crucial challenge to the practical adoption of blind and verifiable quantum computing.


\bigskip

\centerline { $ *\:*$ }

\bigskip

A purely photonic version of VBQC was demonstrated on a four-qubit polarization-encoded cluster state \cite{BFK:NatPhys13}, in which the verifier chooses measurement settings on the cluster such that any qubit can become a trap qubit . This allows to check the correctness of quantum correlations between subsets of measurement outcomes, while verifying the prover's entanglement capabilities using Bell inequalities. A few years later, a scheme inspired by \cite{M:PRA14}, in which the client only performs quantum measurements instead of quantum state generation, was implemented on a similar photonic platform \cite{GRB:NJP16}. 

Remarkably, the first demonstration of VBQC on a hybrid photon-matter platform was recently performed \cite{DNM:PRL24}, leveraging the advantage of photons for the communication part and trapped ions for the computation part. In this experiment, the server possesses an ion trap quantum processor containing one $^{43}$Ca$^+$ ion for the memory qubit (with a long coherence time of around $10$ s) and one $^{88}$Sr$^+$ ion as an interface to the photonic platform. The ion-photon entanglement is generated by inducing a spontaneous decay on one transition of the $^{88}$Sr$^+$ ion, which generates a single photon that is coupled to an optical fiber and sent to the client for measurement and verification. Such a scheme circumvents the need for post-selection in optical quantum computing, which can give significant overheads \cite{LHM:PRX15}, although the fidelity of the combined CZ and SWAP gate should be improved to allow for verifiability. Furthermore, frequency conversion to telecommunication wavelengths would be required to increase the distance and practicality of the photonic part. 


\subsection{Computation with classical clients}\label{sec:classclients}

A limitation of the schemes from Sections \ref{sec:blind} and \ref{sec:verifiable} is their requirement for potentially substantial quantum communication between the client and the server. This not only adds technological challenges to the wide adoption of secure quantum computing, but also limits its generality: compatibility between the quantum computing and quantum communication parts becomes essential, thus excluding clients with light devices and no quantum technology. Therefore, the possibility of schemes with fully classical clients becomes an essential question. Unfortunately, it was shown in \cite{aaronson_et_al:LIPIcs.ICALP.2019.6} that achieving secure blind quantum computing with a fully classical client and i.t.-security is impossible\footnote{The exact statement was subject to a very reasonable complexity-theoretical assumption.}.

Following this result, the research community focused on approaches that ``relax'' the requirement for i.t.-security. In \cite{MDM:PRX17} for instance, the ambiguity in the flow of the computation within MBQC was used to enable classical clients to partially hide the computation from an adversary (server) with polynomial-time quantum resources.  
Inspired by methods first introduced in the context of public-key quantum money \cite{A:IEEE09}, and specifically the idea of equal superposition of pre-images (see Section \ref{sec:preimages}), a new line of research emerged. 

In \cite{M:SIAM18}, a method to achieve classical-client blind quantum computing was given\footnote{A protocol that sits between \cite{M:SIAM18} and \cite{BFK2009universal} was given in \cite{Zhang2021STOC}, where the client had to prepare quantum states but their number was independent of the size of the problem considered. On the other hand, the states were involved (unlike single qubits), and the security relied on the quantum random oracle, a computational assumption that is weaker than LWE.}. Importantly, this protocol requires no interaction between the client and server, which links to Section \ref{sec:homomorphic} on fully homomorphic encryption. The key idea involves a server with a classical encrypted message (ciphertext) to apply a quantum CNOT gate conditioned on the value of the decrypted message (plaintext), but without leaking any information about it. This relies on a certain type of function known as a trapdoor claw-free function, and the security of the implementation relies on the hardness of the Learning-With-Errors (LWE) problem. Obviously, the security is only computational, since an arbitrarily powerful server could search for the set of pre-images and break the security\footnote{The important point is that such an adversary could also use this attack to break the LWE problem, and thus it seems very unlikely for an adversary with polynomial-time quantum resources.}. Further work  gave the first quantum verification protocol with a fully classical client/verifier \cite{M:FOCS18}. The protocol does not hide the computation but uses similar methods, specifically a gadget that enables the server to perform a random single-qubit measurement without knowing what measurement they are performing, in a way that the client/verifier can check the server.

A different, modular, approach was taken in \cite{CCK:LNCS19,cojocaru2021possibility}, where the use of similar mathematical tools (trapdoor, $2$-regular, collision-resistant functions) enabled a fully classical client to instruct a polynomially-bounded quantum server to prepare a BB84 state  in a way that the server has no information about the state beyond what can be extracted by measuring it. This enables a classical client to ``replace'' the need for communicating single qubits via a quantum channel with this primitive called ``classical-client remote state preparation'' (CC-RSP). This could remove the need for a quantum channel in a variety of quantum protocols, including verifiable versions of blind quantum computing, as was done in \cite{GV:IEEE19}. 

It is worth stressing here the limitations of this approach: it requires the quantum side (server) to have a full quantum computer, requires a significant overhead and relies on computational assumptions. This means that CC-RSP and similar methods cannot practically replace the quantum channel in most quantum-cryptographic applications. Specifically, unless the application involves a party that has quantum-computational power anyway (e.g. a powerful server), CC-RSP and related methods are not applicable. 
 
\bigskip

\centerline { $ *$ }

\bigskip

Assuming two quantum servers that share EPR pairs but cannot communicate with one another, a blind implementation of Shor's factoring algorithm was performed for the number $N=15$, instructed by a fully classical client \cite{HZM:PRL17}. This involves the client splitting the computation into two parts, delegating one part to each of the photonic servers. Note that the CNOT gate is implemented differently for each of the two photonic servers, whose roles in the computation are asymmetric: for one server, it suffices to probabilistically post-select on the outputs of a polarizing beamsplitter, while the other server follows an implementation closer to \cite{KSW:PRL05}. By randomly alternating between computation and dummy tasks, the classical client is able to verify the behaviour of the servers and performing a CHSH test based on the protocol from \cite{RUV2013classical}. Following up on the proposal from \cite{HZM:PRL17}, an extension to the blind factorization of $N=21$ was proposed \cite{DS:PRA22}. \hl{More recently, a simplified version of the classical client scheme from \cite{M:FOCS18} was demonstrated using a small trapped-ion quantum processor \cite{SCR:QST24}.}

\subsection{Fully homomorphic encryption}\label{sec:homomorphic}

In many applications, it is desirable to securely delegate a computation without the need for any (classical or quantum) communication between the client and server, except one round at the start to communicate the input and one round at the end to communicate the output. Specifically, assume that the client holds a message $m$ and wants to evaluate a function $f(m)$. The client encrypts the message $m$ as $c(m)=\textrm{Enc}_k(m)$ using key $k$, sends $c(m)$ to the server, and waits for the server to return $c(f(m))$. The client can now decrypt $\textrm{Dec}_k(c(f(m)))$ and recover the output of the function $f(m)$ evaluated on their desired input.

\bigskip

\centerline { $ *$ }

\bigskip

Our task is to construct a quantum version of this scenario, where, instead of evaluating a function on a classical message, we wish to apply a unitary on a quantum state: the client sends $\textrm{Enc}\ket{\psi}$ to the server and expects to receive $\textrm{Enc} (U\ket{\psi})$ back. 
Such a task was achieved for unitaries with a constant (low) $T$-gate depth in \cite{BJ2015quantum}. Since this construction does not handle arbitrary unitaries however, it is not a ``fully'' homomorphic scheme. One first observation is that, for Clifford gates, homomorphic evaluation can be easily performed by applying the gates and modifying the ``corrections'', i.e. modifying the one-time-pad on the quantum state. In \cite{BJ2015quantum}, the authors construct a gadget in which the client hands an (unknown) quantum state to the server, enabling the server to apply a $T$-gate on the encrypted quantum state. While this can be done non-interactively, the cost of the overall method scales exponentially with the $T$-depth of the unitary (using a classical FHE scheme in parallel).

The first fully homomorphic quantum encryption (QFHE) scheme was presented in \cite{DSS:Crypto16}. As suggested by \cite{aaronson_et_al:LIPIcs.ICALP.2019.6}, this demonstrated the impossibility of i.t.-secure solutions, as the scheme is computationally secure and relies on the hardness of the ``garden-hose'' model -- a problem introduced in \cite{BFSS2013}. The solution involves the use of quantum gadgets to enable homomorphic application of $T$-gates, but the number of required gadgets (and their resources) does not scale exponentially with the relevant parameters, in contrast to the work from \cite{BJ2015quantum}. Each of these gadgets includes a number of EPR pairs, chained and shared between the client and server. The scheme, while requiring only polynomial resources, demands considerable amounts of entanglement shared between client and server, thus appearing fairly impractical.

We note here that the protocol from \cite{M:SIAM18}, involving only classical clients, is also a quantum fully homomorphic protocol that relies on computational assumptions. This scheme was discussed in Section \ref{sec:classclients}. 

\bigskip

\centerline { $ *\:*$ }

\bigskip

In the classical world, the ability to verify a fully homomorphic computation enables FHE to be used in numerous applications such as zero-knowledge proofs (Section \ref{sec:ZKP}), secure multiparty computation (Section \ref{sec:MPC}) and obfuscation. This verification happens by providing the client with a copy of the transcript of the server's homomorphic evaluation, such that they can check whether the correct circuit was performed. In the quantum world however, no-cloning forbids this approach, making vQFHE a difficult task to achieve. In \cite{ADS:Asia17}, this issue was evaded by using a different technique. One of the construction's key elements is the trap-based quantum authentication scheme introduced in \cite{BGS:Crypto13}, that uses a similar idea to verifiable quantum computing (Section \ref{sec:verifiable}), mixing the encoded quantum states with ``traps'' that produce deterministic outcomes. The vQFHE protocol uses the $T$-gadget based on the result from \cite{DSS:Crypto16} that relies on the hardness of the garden-hose model, encrypts the state with the trap-based authentication scheme and uses a classical FHE scheme assuming the hardness of LWE. A direct application of this vQFHE scheme, as given in \cite{ADS:Asia17}, is a method that can use any classical one-time-program and lift it to a quantum one-time-program.


\bigskip

\centerline { $ *\:*\:*$ }

\bigskip

On the experimental side, a proof-of-principle computation on encrypted quantum data was performed using polarization-encoded single photons and linear optics \cite{FBS:NatComm14}. In this scheme, the client is only required to prepare and send single qubits, encrypted through the quantum equivalent of a one-time pad (Section \ref{sec:privchannel}). Since client and server agree on the computation beforehand, the overhead in terms of auxiliary qubits is proportional to the number of non-Clifford gates in the computation. Single-qubit Clifford gates are naturally performed using a set of waveplates, while the CNOTs require two-photon interference at a partially polarizing beamsplitter (giving a success probability of $1/9$) and the non-Clifford gates are actively implemented using various Pockels cells to prepare the auxiliary qubit. While a full practical security analysis is not provided, the leakage of information due to imperfect quantum encryption on the client's side and multiphoton emission from the SPDC source is quantified. Two years later, a similar protocol was implemented in the continuous-variable setting at telecom wavelengths, showing that security can be achieved with up to $10$km of lossy channel between the client and the server \cite{MJS:NatComm16}. 

Using a similar polarization-encoded approach to  \cite{FBS:NatComm14}, the non-interactive scheme from \cite{BJ2015quantum} for shallow $T$-depth circuits was demonstrated \cite{TFB:PRX20}. In parallel, the scheme from \cite{RFG:PRL12} was implemented using quantum walks in an integrated interferometer \cite{ZPT:npjQI21}: $m$ photons are input into an interferometer with $n$ spatial modes (such that $m\leqslant n$ and only $1$ photon can enter each spatial mode), while the client's hidden computation is encoded onto the photons' polarization degree of freedom. Such an encoding allows to control which photons will interfere on the chip. Very recently, the approach from \cite{BJ2015quantum} was also implemented on a photonic chip, bringing these schemes a step closer to scalability and compactness \cite{LCL:PRL24}.

\subsection{Multiparty computation}\label{sec:MPC}


In secure multiparty computation (SMPC), a number of participants $P_1,...,P_n$ are involved in a joint calculation on their individual private inputs $x_1,...,x_n$. Their goal is to produce $(y_1,...,y_n) \leftarrow F(x_1,...,x_n)$ for some $F$ they have agreed upon. The function needs to be efficiently computable, i.e. if a single party were to perform the computation, they would produce the output in polynomial time. At the end of the computation, each $P_i$ receives a private output $y_i$. The cryptographic challenge is to perform this calculation in such a way that each party or subset of parties $\{P_{i_1},...,P_{i_k}\}$ learns as little as possible about the inputs $\{x_i\}$ and outputs $\{y_i\}$ of parties outside the subset. Specifically, ``as little as possible'' translates to nothing beyond what can be inferred from the legitimately obtained inputs/outputs of the ``attacking'' coalition. Applications and more restricted use-cases of SMPC are numerous, including e-voting, auctions, data analytics, genetic testing and blockchain.

In proving the security of SMPC, the real protocol should be indistinguishable from a protocol simulating the ideal functionality, in which a trusted third party receives the inputs, computes the joint function and returns the outputs. There exist various types of SMPC that admit different security proofs, depending on the number of parties involved (proofs are very different in the two-party case), the fraction of parties that are allowed to be adversarial, the verifiability (i.e. if correctness is also guaranteed beyond the privacy of inputs/outputs), the number of outputs and the weakness/strength of the adversaries\footnote{A weak notion of adversary is the ``honest-but-curious'' adversary, who does not actively deviate from the ideal protocol but only keeps a record of transcripts and tries to learn from these. While such an adversary is very weak, there exist methods to uplift scheme that are secure in the honest-but-curious setting to schemes that are secure against active adversaries using a compiling method (either using zero-knowledge proofs or the so-called cut-and-choose technique). Quantizing these compiling methods is of interest in its own right but goes beyond the scope of this review.}.

\bigskip

\centerline { $*$ }

\bigskip

In the quantum setting, the function $F$ is evaluated using a quantum computer, meaning that the joint computation corresponds to a quantum circuit. The first work studying this problem was \cite{CGS:STOC02}, proposing a solution based on a verifiable quantum secret sharing (vQSS) scheme. This solution enables the secure computation of any efficient quantum circuit, provided that no more than $n/6$ parties are dishonest. The core idea uses vQSS along with fault-tolerant computation methods that enable the parties to jointly perform the computation such that, if some ``shares'' are corrupted, the computation will still provide a correct output. In terms of technology requirements, each party needs to be able to perform fault-tolerant quantum computation.

In \cite{BCG:IEEE06}, an improved protocol that saturates the number of honest parties to a majority (the maximum achievable for ``fair'' SMPC\footnote{A ``fair'' protocol is a protocol in which it is impossible for an adversary to cause abort of the protocol after they receive their own desired output. As we see below, removing the fairness requirement makes it possible to have SMPC with dishonest majority.}) was proposed, making use of approximate quantum-error correcting codes. Note that the case of two-party computation is special, both in terms of applications and due to the fact that one cannot have an honest majority if one of the two parties is dishonest. In \cite{DNS2010crypto}, the case of a weak honest-but-curious adversary\footnote{There was a subtle difference in the definition of this weak adversary compared to the classical case, which led the authors to use the term ``specious'' adversary.} was considered, and a protocol achieving secure two-party quantum computation was given, based on two secure boxes:  a private SWAP and a harder-to-implement secure AND. Based on the result from  \cite{DNS2010crypto}, a two-party quantum computation scheme secure against fully malicious adversaries was proposed \cite{DNS2012crypto}, through a mechanism that enforces the adversary to be honest-but-curious. The central idea uses a quantum authentication code in a way that allows for ``gates to be performed'' while providing a way to check the authentication code without revealing what was authenticated.

 Following the rapidly evolving field of classical SMPC (one of the most interesting topics in modern cryptography), numerous improvements and versions have been discussed. Recent examples include an asymmetric protocol in which multiple mistrustful clients delegate their computation to a single untrusted server \cite{KP:Crypto17}, a method that uses classical clients with limited capabilities and quantum resources to delegate their universal \emph{classical} computation to an untrusted server \cite{CPE:PRA17} and a scheme in which the quantum resources required for secure multiparty quantum computation are reduced \cite{LRW:PRA20}. An important result was established in \cite{DGJ+2020crypto}, offering a scheme for in which a single party is honest\footnote{As noted earlier, this is only possible by dropping the ``fairness'' property.}. This result was achieved by using a testing protocol for magic-state inputs and a publicly verifiable Clifford authentication code, both relying on classical SMPC. In \cite{KKL+2023asymmetric}, another protocol allowing for a dishonest majority was given, in the setting where a single party has a full quantum computer and the other parties can only prepare and send single qubits.

\bigskip

\centerline { $*\:*$ }

\bigskip

Similarly to the classical world, multiparty quantum computation can be a very powerful tool for secure applications, but requires significant overheads and relies on more assumptions. Since such protocols are useful for quantum computations that are (exponentially) harder for classical computers\footnote{Otherwise classical SMPC could be used.}, the quantum technology required to go beyond proof-of-principle experiments \cite{CPE:PRA17} places the majority of proposed protocols well in the fault-tolerant quantum computing era. Given these challenges, and assuming a quantum network connecting the parties involved, the setting with a single (powerful) quantum server that interacts with weak quantum clients might be the most promising approach for ``near-term''  applications.

\section{Outlook and challenges}\label{sec:outlook}

To conclude this review, we present some relevant challenges and open questions in the field of quantum cryptography. Although many of the theoretical and experimental topics discussed in this section are undergoing active research, we also provide some insights that will hopefully trigger new questions and directions.

\subsection{Composability}\label{sec:composability}

As discussed throughout the review, the security of useful applications requires the combination of multiple (quantum) cryptographic primitives. Thus, it is crucial to consider how primitives can be ``composed''. Ideally, one would want to prove the security of a protocol achieving a given functionality, and then be able to use that protocol in any larger, more complicated protocol. This feature is known as \textit{composable security}: one proves the security of a given protocol, and then can use it to implement the ``ideal functionality'' in any composite protocol. Here, we stress that failing to prove a protocol's security in a composable way \emph{does not} imply that the protocol cannot be used as a subroutine in larger protocols. It only means that, in order to do this, one needs to reconsider the overall security, taking into account the details of the protocol that implements the subroutine. Moreover, achieving composability often requires the use of some (standard) assumptions. The most common one is the assumption that parties share a common random reference string (CRS) that cannot be biased by any party. Interestingly, without assuming CRS, even computationally-secure (classical) bit-commitment is impossible \cite{canetti2001universally}.

Composable quantum cryptography is a large field that is beyond the scope of this review\footnote{Even the framework that needs to be used for composable security, suitable for quantum multiparty protocols, is an open area of research.}. Most multiparty protocols cannot be composably secure in the plain model (without assumptions such as the CRS), and this persists in the quantum case. For instance, it was shown that weak coin flipping is not composably secure in the plain model \cite{WHB:Quantum25}, while stand-alone sequential composition is proven secure. The case of relativistic quantum protocols (for bit commitment and coin flipping) were also shown to be non-composable in the plain model \cite{VPL:NJP19}. Introducing assumptions such as CRS or similar quantum assumptions could evade these results. The case of composable security for the verification of multipartite entanglement is more positive; in \cite{YDK:PRA21} composable security was proven against dishonest preparation for GHZ states; while in \cite{CMY:Arx24} this was strengthened by proving composable security for verification of general graph states and allowing malicious parties to collude with the dishonest preparation. \hl{However, limits on existing protocols were shown to be inevitable, in that quantum state verification cannot be both efficient and secure \cite{WCK:arx24}.}

Finally, major results in the field of secure quantum computing were also established: verifiable blind quantum computing schemes such as that of \cite{FK2017unconditionally} were proven composably secure in \cite{DFPR2014Asiacrypt}. On the other hand, classical-client protocols that rely on remote state preparation \cite{GV:IEEE19,CCK:LNCS19} were shown to be non-composable in the plain model \cite{BCC+2020Asiacrypt}, and require extra assumptions (such as the ``measurement buffer'' of \cite{GV:IEEE19}). In the same work however, the stand-alone security of composing classical-client remote state preparation with verifiable quantum computing was demonstrated. Such results highlight that many of the impossibilities of composable security should not be viewed as impossibilities for the use of primitives in larger protocols, but rather as stressing the need for detailed, case-by-case, security analyses.

\subsection{Quantum advantage}

While the notion of quantum advantage for computational speed-up is widely studied and debated \cite{AAB:Nature19,BBC:SciAdv22,PCZ:PRL22,HHY:Science20,LCG:PRL24}, less attention is focused on defining this notion in the context of quantum cryptography. 
 
From a qualitative standpoint, the claim of quantum advantage for primitives such as QKD and unforgeable quantum tokens is somewhat straightforward: they provide i.t.-security for tasks that cannot achieve such a feature classically. For others primitives such as quantum coin flipping, optimal bounds on the performance of classical schemes have even been derived \cite{HW:TCC11}, serving as a useful quantitative benchmark for quantum advantage (Section \ref{sec:SCF}). Arguably, another key advantage in using quantum resources is the fact that they allow for tamper evidence and cheat detection \cite{SR:PRL02,NYC:NatComm23,HK:PRL04,BCH:PRA08,FGH:ACM02,G:QIC03}, which are desirable properties that cannot always be achieved in the classical world.

Despite the simplicity of such arguments, they can lead to discussions and scepticism in practice. Such concerns can be scientifically motivated, but are also intertwined with business considerations: although quantum cryptography provides stronger security guarantees, institutions and companies might still favor the option of insuring data breaches over costly infrastructure changes to implement quantum schemes. Recent recommendations\footnote{A summary of such recommendations, along with links to detailed reports, can be found at https://www.nsa.gov/Press-Room/News-Highlights/Article/Article/2394053/nsa-cybersecurity-perspectives-on-quantum-key-distribution-and-quantum-cryptogr/} from the United States' National Security Agency in favor of post-quantum methods offer another example of practical scepticism towards quantum methods.

This calls for the need to identify use-cases in which quantum cryptography beyond QKD can provide practical benefits over classical methods. In the context of quantum money for instance, such an investigation is ongoing, both for private-key \cite{SKS:NatComm23,CLW:SciAdv24} and public-key use-cases \cite{CS:Quantum20}.

\subsection{Single-photon sources }\label{sec:single}

An ideal source for quantum cryptography should provide high rates of identical single photons on-demand. From a practical standpoint, these requirements translate into high brightness/collection efficiency, low multiphoton noise and high degrees of indistinguishability. Note that the latter property is only required in two-party protocols relying on two-photon interference, or in multipartite quantum cryptography and secure quantum computation where larger entangled states are required. 

For single-photon primitives such as standard BB84 QKD, the decoy-state technique was introduced in order to relax the stringent need for low multiphoton noise  \cite{LMC:PRL05,W:PRL05}. As a consequence, Poisson-distributed sources such as SPDC or attenuated laser states can already provide impressive performances \cite{LZJ:PRL23,BBR:PRL18,Pan:Nat20}. For quantum primitives in which decoy states are not used however, the question of the optimal single-photon source remains open. 

Despite presenting an inherent trade-off between high brightness and low multiphoton noise, Poisson-distributed sources are sufficient for demonstrations of two-partite quantum primitives in somewhat realistic conditions \cite{PJL:NC14,ENG:NatComm14,NS:NatComm12,SKS:NatComm23,CLW:SciAdv24,GHM:Optica14,CAF:SciRep17,CCD:NatComm12}. On the other hand, it was recently shown that quantum dot single-photon sources (QDs), which in principle do not exhibit this trade-off, can provide performance enhancements \hl{for experimental QKD \cite{ZDL:PRL25,YJB:LSA24,MPG:NatComm23} and mistrustful quantum primitives \cite{BVC:npjQI22,VKD:Arx24}}. This is partly thanks to their high single-photon purity, but also thanks to their tunable coherence in the photon-number basis \cite{LAR:NatPhot19,KVK:npjQI24,HBC:AQT25}, which allows to circumvent the need for active phase randomization. Unlike for Poisson-distributed sources, the best performing QDs emit in the infrared \cite{SSW:NatNano17}, thus requiring frequency conversion to telecom wavelengths. On the other hand, QDs emitting directly in the telecom C-band are under active development \cite{YLL:NatNano23,VJN:APL23,YJB:LSA24,HVZ:NatComm24}.

It is worth noting the existence of other single-photon sources such as organic molecules emitting at room temperature \cite{MCH:OptExp23,TGC:NatMat21}, defect centers in diamond \cite{LSF:NJP14,KKB:PRL22} \hl{and atomically-thin single-photon sources \cite{GVA:npj2D23}}. To the best of our knowledge, their role in primitives beyond QKD has not yet been investigated.

\subsection{Loss tolerance}

A key challenge in practical quantum cryptography is to guarantee security over long distances, whether using free space channels or deployed fiber networks. Unfortunately, the no-cloning principle forbids the amplification of quantum signals (Section \ref{sec:cloning}), implying that quantum-cryptographic implementations suffer from the scaling of losses with distance. As discussed throughout the review, this not only affects the rate at which quantum primitives may be carried out, but also their security properties. 

For some functionalities, quantum cryptography is fairly robust to losses, as secure demonstrations over several tens of kilometers are possible thanks to the back-reporting of valid detection time-tags \cite{PJL:NC14,ENG:NatComm14,NS:NatComm12,CAF:SciRep17}. In cases where the quantum part is performed locally between two parties and the long-distance communication only handles classical data, this allows for even larger separations \cite{Zbind:PRL13,Pan:PRL14,JKP:Arxv24}. On the other hand, recent work has shown that most methods used to back-report valid time-tags, including symmetrization of losses, may reveal crucial information about the receiver's measurement basis choice \cite{BCD:PRX21}. Appropriate countermeasures should therefore be designed and implemented in deployed demonstrations of mistrustful quantum cryptography, including for multipartite quantum primitives (Section \ref{sec:multipartite}) and secure quantum computation (Section \ref{sec:computing}).

For protocols in which no back-reporting takes place, the question of whether one can achieve even modest loss tolerance remains open. This is the case for some versions of unforgeable quantum tokens and their use-cases \cite{SKS:NatComm23,BDG:PRA19}, in which the overall loss allowance is limited to $50\%$ to forbid perfect cloning attacks. In some protocols such as quantum position verification, this limit may be overcome by increasing the size of the set of possible encoding/measurement bases \cite{QS:PRA15}. 

Overall, practical security proofs for many types of quantum primitives in this review have not yet been developed (mainly for Sections \ref{sec:computational} and \ref{sec:quantumdata}), especially with respect to loss-dependent attacks. 

\subsection{Device independence}

While some beyond-QKD primitives have been studied in the measurement-device-independent setting \cite{FYC:PRL15,LFL:PRR23,PAW:PRA2016,BDG:PRA19,YWT:PRA2017,RLY:NatCommn2017}, the question of full device-independence (DI) \cite{VV:PRL14}, in which neither quantum sources nor measurement devices can be trusted, remains fairly open. For mistrustful cryptography, DI implies that remote parties must certify the presence of non-locality while displaying competing interests, which differs from the better-studied QKD scenario \cite{ABG:PRL07,PAB:NJP09,WAP:Quantum21}. Although most known mistrustful DI schemes rely on GHZ correlations instead of Bell pairs \cite{SCA:PRL11,SSV:Arx24}, some protocols based on CHSH were proposed for quantum bit commitment and strong coin flipping \cite{AMP:NJP16}. For quantum oblivious transfer, a DI version of the XOR protocol was proposed in \cite{KST:Quantum22}, and security in the bounded storage model was studied under extra post-quantum hardness assumptions \cite{BY:NJP23}. 

Besides raising the question of exactly which quantum-cryptographic primitives admit a DI version and under which assumptions, most existing works are proposed in the ideal setting, where parties do not have to calibrate imperfection parameters related to finite multiphoton emission probability, channel losses, and valid sets of detection time-tags. Thus, in order to develop experimental milestones such as those achieved in DI-QKD \cite{ZLR:Nat22,NDN:Nat22}, security bounds on the tolerable noise and losses of beyond-QKD primitives should be derived.

\section*{Acknowledgments}

We thank Alexandru Gheorghiu and Jarn de Jong for providing some figures in the manuscript, along with the anonymous referees for their very helpful suggestions. M.B. and Ph.W. acknowledge funding in part by the Austrian Science Fund (FWF) [10.55776/F71], the Air Force Office of Scientific Research under award number FA9550-21-1-0355, and the European Union (HORIZON Europe Research and Innovation Programme, QSNP, No 101114043). Views and opinions expressed are however those of the author(s) only and do not necessarily reflect those of the European Union or the European Commission-EU. Neither the European Union nor the granting authority can be held responsible for them. C.C. acknowledges funding from Canada's NSERC and Qu\'ebec's FRQNT. Part of the writing of this paper happened while this author was holding an INRIA international chair at \'Ecole normale sup\'erieure, Paris. Pe.W. acknowledges support from the EPSRC grants EP/T001062/1, EP/X026167/1 and EP/T026715/1 and the STFC grant ST/W006537/1.

\end{document}